\documentclass[longauth]{aa} 

\usepackage{graphicx}
\usepackage{array,multirow,makecell}
\usepackage{txfonts}
\usepackage{xcolor}
\usepackage{hyperref}
\hypersetup{colorlinks,linkcolor={blue},citecolor={blue},urlcolor={blue}} 
\usepackage[capitalise]{cleveref}
\usepackage{natbib}
\usepackage{amsmath}
\usepackage{ulem} 
\bibpunct{(}{)}{;}{a}{}{,} 
\usepackage[normal]{threeparttable}
\usepackage{tabularx}
\usepackage{colortbl}
\usepackage[caption = false]{subfig}
\usepackage{float}  
\usepackage{lscape} 
\usepackage{pdflscape} 
\usepackage{arydshln} 
\usepackage{booktabs} 
\newcommand{\rom}[1]{\uppercase\expandafter{\romannumeral #1\relax}}
\usepackage{pifont}
\usepackage{nccmath} 
\usepackage{gensymb}
\usepackage{soul}

\newcommand{\msun}{\mbox{$M_\odot$}}

\newcommand{\tdust}{\mbox{$T_{\rm dust}$}}
\newcommand{\vlsr}{\mbox{$V_{\rm LSR}$}}
\newcommand{\hii}{H\mbox{\sc ~ii} }

\newcommand{\pc}{\mbox{pc}}
\newcommand{\kpc}{\mbox{kpc}}
\newcommand{\kms}{\mbox{km~s$^{-1}$}}
\newcommand{\bsens}{\texttt{bsens}\xspace}
\newcommand{\cleanest}{\texttt{cleanest}\xspace}

\begin{document} 

\title{ALMA-IMF I - Investigating the origin of stellar masses: Introduction to the Large Program and first results}

\author{F. Motte \inst{1} \and
          S. Bontemps \inst{2} \and
          T. Csengeri \inst{2} \and
          Y. Pouteau \inst{1} \and
          F. Louvet \inst{3,4} \and
          A. M.\ Stutz \inst{5,6} \and
          N. Cunningham \inst{1} \and 
          A. L\'opez-Sepulcre \inst{1,7} \and
          N. Brouillet \inst{2} \and 
          R. Galv\'an-Madrid \inst{8} \and 
          A. Ginsburg \inst{9} \and
          L. Maud \inst{10} \and
          A. Men’shchikov \inst{3} \and
          F. Nakamura \inst{11,12, 13} \and
          T. Nony \inst{8} \and
          P. Sanhueza \inst{11,12} \and
        R. H.\ \'Alvarez-Guti\'errez \inst{5} \and 
        M. Armante \inst{4, 14} \and 
        T. Baug \inst{15} \and 
        M. Bonfand \inst{2} \and 
        G. Busquet \inst{1, 16, 17} \and
        E. Chapillon \inst{2, 7} \and 
        D. D\'iaz-Gonz\'alez \inst{8} \and 
        M. Fern\'andez-L\'opez \inst{18} \and 
        A. E. Guzm\'an \inst{11} \and 
        F. Herpin \inst{2} \and 
        H.-L. Liu \inst{5, 19} \and
        F. Olguin \inst{20} \and 
        A. P.\ M.\ Towner \inst{9} \and
          J. Bally \inst{21} \and
          C. Battersby \inst{22} \and
          J. Braine \inst{2} \and
          L. Bronfman \inst{23} \and
          H.-R. V. Chen \inst{20} \and
          P. Dell'Ova \inst{4} \and
          J. Di Francesco \inst{24} \and
          M. Gonz\'alez \inst{3} \and
          A. Gusdorf \inst{4} \and
          P. Hennebelle \inst{3} \and
          N. Izumi \inst{11, 25, 26} \and
          I. Joncour \inst{1} \and
          Y.-N.\ Lee \inst{27} \and
          B. Lefloch \inst{1} \and
          P. Lesaffre \inst{4} \and
          X. Lu \inst{11} \and
          K. M.\ Menten \inst{28} \and
          R. Mignon-Risse \inst{3} \and
          J. Molet \inst{2} \and
          E. Moraux \inst{1} \and
          L. Mundy \inst{29}
          Q.~Nguy$\tilde{\hat{\rm e}}$n~Lu{\hskip-0.65mm\small'{}\hskip-0.5mm}o{\hskip-0.65mm\small'{}\hskip-0.5mm}ng \inst{30} \and
          N. Reyes \inst{28, 31} \and   
          S. D.\ Reyes Reyes \inst{5} \and
          J.-F. Robitaille \inst{1} \and
          E. Rosolowsky \inst{32} \and    
          N. A.\ Sandoval-Garrido \inst{5} \and
          F. Schuller \inst{28, 33} \and    
          B. Svoboda \inst{34} \and 
          K. Tatematsu \inst{11} \and
          B. Thomasson \inst{1} \and
          D. Walker \inst{35} \and  
          B. Wu \inst{11, 36} \and  
          A. P.\ Whitworth \inst{37} \and  
          F. Wyrowski \inst{28}
          }

\institute{Univ. Grenoble Alpes, CNRS, IPAG, 38000 Grenoble, France  
    \and Laboratoire d'astrophysique de Bordeaux, Univ. Bordeaux, CNRS, B18N, all\'ee Geoffroy Saint-Hilaire, 33615 Pessac, France  
    \and AIM, CEA, CNRS, Universit{\'e} Paris-Saclay, Universit{\'e} de Paris, F-91191 Gif-sur-Yvette, France 
    \and Laboratoire de Physique de l'\'Ecole Normale Sup\'erieure, ENS, Universit\'e PSL, CNRS, Sorbonne Universit\'e, Universit\'e de Paris, Paris, France  
    \and Departamento de Astronom\'{i}a, Universidad de Concepci\'{o}n, Casilla 160-C, 4030000 Concepci\'{o}n, Chile    
    \and Max-Planck-Institute for Astronomy, K\"{o}nigstuhl 17, 69117 Heidelberg, Germany   
    \and Institut de RadioAstronomie Millim\'etrique (IRAM), Grenoble, France    
    \and Instituto de Radioastronom\'ia y Astrof\'isica, Universidad Nacional Aut\'onoma de M\'exico, Morelia, Michoac\'an 58089, M\'exico 
    \and Department of Astronomy, University of Florida, PO Box 112055, USA 
    \and European Southern Observatory, Karl-Schwarzschild-Strasse 2, 85748 Garching bei M\"unchen, Germany     
    \and National Astronomical Observatory of Japan, National Institutes of Natural Sciences, 2-21-1 Osawa, Mitaka, Tokyo 181-8588, Japan 
    \and Department of Astronomical Science, SOKENDAI (The Graduate University for Advanced Studies), 2-21-1 Osawa, Mitaka, Tokyo 181-8588, Japan   
    \and The Graduate University for Advanced Studies (SOKENDAI), 2-21-1 Osawa, Mitaka, Tokyo 181-0015, Japan 
    \and Observatoire de Paris, PSL University, Sorbonne Universit\'e, LERMA, 75014, Paris, France 
    \and S. N. Bose National Centre for Basic Sciences, Block JD, Sector III, Salt Lake, Kolkata 700106, India    
    \and Institut de Ci\`encies de l’Espai (ICE, CSIC), Can Magrans, s/n, 08193, Cerdanyola del Vall\`es, Catalonia, Spain   
    \and Institut d’Estudis Espacials de Catalunya (IEEC), 08340, Barcelona, Catalonia, Spain 
    \and Instituto Argentino de Radioastronom\'\i a (CCT-La Plata, CONICET; CICPBA), C.C. No. 5, 1894, Villa Elisa, Buenos Aires, Argentina 
    \and Department of Astronomy, Yunnan University, Kunming, 650091, PR China    
    \and Institute of Astronomy, National Tsing Hua University, Hsinchu 30013, Taiwan   
    \and Department of Astrophysical and Planetary Sciences, University of Colorado, Boulder, Colorado 80389, USA    
    \and University of Connecticut, Department of Physics, 196A Auditorium Road, Unit 3046, Storrs, CT 06269 USA    
    \and Departamento de Astronomía, Universidad de Chile, Casilla 36-D, Santiago, Chile   
    \and Herzberg Astronomy and Astrophysics Research Centre, National Research Council of Canada, 5071 West Saanich Road, Victoria, BC CANADA V9E 2E7    
    \and College of Science, Ibaraki University, 2-1-1 Bunkyo, Mito, Ibaraki 310-8512, Japan   
    \and Institute of Astronomy and Astrophysics, Academia Sinica, No. 1, Section 4, Roosevelt Road, Taipei 10617, Taiwan  
    \and Department of Earth Sciences, National Taiwan Normal University, Taipei 116, Taiwan 
    \and Max Planck Institute for Radio Astronomy, Auf dem H\"{u}gel 69, 53121 Bonn,  Germany  
    \and Department of Astronomy, University of Maryland, College Park, MD 20742, USA 
    \and CSMES, The American University of Paris, 2bis, Passage Landrieu 75007 Paris, France 
    \and Departamento de Ingenier\'ia El\'ectrica, Universidad de Chile, Santiago, Chile  
    \and 4-183 CCIS, University of Alberta, Edmonton, Alberta, Canada 
    \and Leibniz-Institut für Astrophysik Potsdam (AIP), An der Sternwarte 16, D-14482 Potsdam, Germany 
    \and National Radio Astronomy Observatory, PO Box O, Socorro, NM 87801 USA
    \and University of Connecticut, Department of Physics, 196A Auditorium Road, Unit 3046, Storrs, CT 06269 USA 
    \and NVIDIA Research, 2788 San Tomas Expy, Santa Clara, CA 95051, USA  
    \and School of Physics and Astronomy, Cardiff University, Cardiff, UK 
}

\date{Received June 30, 2021; accepted \today}
 
\abstract
{}
{Thanks to the high angular resolution, sensitivity, image fidelity, and frequency coverage of ALMA, we aim to improve our understanding of star formation.
One of the breakthroughs expected from ALMA, which is the basis of our Cycle~5 ALMA-IMF Large Program, is the question of the origin of the initial mass function (IMF) of stars. Here we present the ALMA-IMF protocluster selection, first results, and scientific prospects.}
%
{ALMA-IMF imaged a total noncontiguous area of $\sim$53~pc$^2$, covering extreme, nearby protoclusters of the Milky Way.
We observed 15 massive ($2.5-33\times 10^3~\msun$), nearby ($2-5.5$~kpc) protoclusters that were selected to span relevant early protocluster evolutionary stages. Our 1.3~mm and 3~mm observations provide continuum images that are homogeneously sensitive to point-like cores with masses of $\sim$0.2~$\msun$ and $\sim$0.6~$\msun$, respectively, with a matched spatial resolution of $\sim$2\,000~au across the sample at both wavelengths. Moreover, with the broad spectral coverage provided by ALMA, we detect lines that probe the ionized and molecular gas, as well as complex molecules. Taken together, these data probe the protocluster structure, kinematics, chemistry, and feedback over scales from clouds to filaments to cores.}
%
{We classify ALMA-IMF protoclusters as Young (six protoclusters), Intermediate (five protoclusters), or Evolved (four protoclusters) based on the amount of dense gas in the cloud that has potentially been impacted by \hii region(s). The ALMA-IMF catalog contains $\sim$700 cores that span a mass range of $\sim$0.15~$\msun$ to $\sim$250~$\msun$ at a typical size of $\sim$2\,100~au. We show that this core sample has no significant distance bias and can be used to build core mass functions (CMFs) at similar physical scales.
Significant gas motions, which we highlight here in the G353.41 region, are traced down to core scales
and can be used to look for inflowing gas streamers and to quantify the impact of the possible associated core mass growth on the shape of the CMF with time. Our first analysis does not reveal any significant evolution of the matter concentration from clouds to cores (i.e., from 1~pc to 0.01~pc scales) or from the youngest to more evolved protoclusters, indicating that cloud dynamical evolution
and stellar feedback have for the moment only had a slight effect on the structure of high-density gas in our sample.
Furthermore, the first-look analysis of the line richness toward bright cores indicates that the survey encompasses several tens of hot cores, of which we highlight the most massive in the G351.77 cloud. 
Their homogeneous characterization can be used to constrain the emerging molecular complexity in protostars of high to intermediate masses.}
%
{The ALMA-IMF Large Program is uniquely designed to transform our understanding of the IMF origin, taking the effects of cloud characteristics and evolution into account. It will provide the community with an unprecedented database with a high legacy value for protocluster clouds, filaments, cores, hot cores, outflows, inflows, and stellar clusters studies.}


\maketitle
\section{Introduction}
\label{s:intro}

The relative number of stars born with masses between $0.01~\msun$ and $>$100~$\msun$, the so-called initial mass function (IMF), is among the very few key parameters transcending astrophysical fields. For example, it is critically important for cosmology and stellar physics \citep{madau&dickinson2014, hopkins18}.
In studies of both the Galactic and cosmic history of star formation, the IMF is often considered to be universal \citep[e.g.,][]{bastian10, kroupa13}. A few studies of young massive stellar clusters in the Milky Way \citep{lu13, maia16, hosek19} or in nearby galaxies \citep{schneiderF18} and indirect constraints at high redshift \citep{smithR14, zhangZ18} suggest, however, incidences of IMFs with noncanonical, top-heavy shapes \citep[see the recent review by][]{hopkins18}. The IMF also varies with metallicity, becoming top-heavy or bottom-heavy in low- or high-metallicity environments, respectively \citep[e.g.,][]{marks12, 
martin-navarro15}.
Overall, the IMF may not be as universal as once thought, but may vary with galactic environment and evolve over time. Therefore, the central astrophysical importance of the IMF motivates a vigorous investigation into the question of its origin.

In the star-formation community, both the IMF origin and its dependence on environment remain the subject of heated debate \citep[see reviews by][]{offner14, krumholz15, ballesteros20, lee20}. In the star-forming regions studied in the last two decades, the mass distribution of cores, the core mass function (CMF), is strikingly similar to the IMF \citep[e.g.,][]{motte98, motte01, TeSa98, alves07, enoch08, konyves15}. These studies, which were conducted in Gould Belt clouds,
 star-forming regions in the solar neighborhood that form solar-type stars, led to the interpretation that the shape of the IMF may simply be inherited from the CMF. These nearby regions are, however, unrepresentative of the larger Milky Way. For instance, they do not capture clouds that form stars more massive than $5~\msun$, high-mass cloud environments, or the vast extent and range of conditions in the Galaxy. Our current understanding of the origin of stellar masses is therefore biased. 
Massive protoclusters are key laboratories for the study of the emergence of the IMF because these clusters of cores are the gas-dominated cradles of rich star clusters,
probing substantially different, and cosmically important, environments.
A\ detailed scrutiny and study of statistical samples of massive protoclusters is mandatory to test observationally 
whether the IMF origin is in fact independent of cloud characteristics or not. The ALMA-IMF\footnote{
    ALMA project \#2017.1.01355.L; see \url{http://www.almaimf.com}.}
Large Program (PIs: Motte, Ginsburg, Louvet, Sanhueza) is a survey of 15 nearby Galactic protoclusters observed at matched sensitivity and physical resolution
that aims for statistically meaningful results on the origin of the IMF (see below).

Even before studying the relationship between the IMF and the CMF, it is important to realize that how the IMF originates from the observed CMF depends directly on the definition of the cores, {assumed to be the} gas mass reservoir used for the formation of each star or binary system. As shown by \cite{louvet21}, defining this mass reservoir may seem obvious in the observed map of a cloud, but core characteristics (size, mass) depend heavily on the spatial scales probed by the observations. In addition, the theoretical definition of cores also depends on whether the star-formation scenario is quasi-static or dynamic. In the former scenario, cores are gas condensations sufficiently dense to be on the verge of gravitational collapse, and they convert the core gas into stars \citep{shu87, chabrier03, MKOs07, andre14}. After a quasi-static phase of concentration of the cloud gas into cores, cores become distinct from their surrounding cloud and start to collapse, and their future stellar content becomes independent of the properties of the parental cloud. In the latter scenario, dynamics play a major role during all phases of the star-formation process  \citep[e.g.,][]{ballesteros07, HeFa12, padoan14}. In particular, global infall of filament networks and gas inflow toward cores are expected to be important drivers of star formation \citep[e.g.,][]{smith09, vazquez19, padoan20}. In this framework, filaments, cores, and stellar embryos simultaneously accrete gas, and the gas reservoir associated with star formation largely exceeds the extent of the observed cores. This so-called clump-fed scenario was proposed in various recent papers and described in detail in the review by \citet[][see references therein]{motte18a}. One of the main objectives of the ALMA-IMF Large Program is to discriminate between 
the quasi-static and dynamic scenarios by quantifying the role of cloud kinematics in defining core mass and in possibly changing it over time. 

In the ALMA-IMF pilot study, \cite{motte18b} identified the first definitive observation of a CMF whose shape differs from that of the IMF. The authors derived this CMF in W43-MM1, which is a dense cloud efficiently forming stars at the tip of the Galactic bar \citep{nguyen13, louvet14}. Fitted by a single power-law relation in both the solar-type and high-mass regimes ($2-100~\msun$ cores, thus $\sim$1$-50~\msun$ stars with a $50\%$ conversion efficiency; \citealt{motte18b}), this CMF is flatter than those of reference CMF studies from nearby, low-mass star-forming regions \citep[e.g.,][]{motte98, konyves15, diFrancesco20}. It is also quantitatively flatter than the CMF derived from a one-to-one mapping of the high-mass end, $m\ge1~\msun$, of the stellar IMF, ${\rm d} N / {\rm d}\log(m) \propto m^{-1.35}$ \citep{salpeter55, kroupa01}. Such an excess of high-mass cores with respect to their solar-type counterparts indicates a top-heavy CMF. This was previously suggested by single-pointing observations \citep[e.g.,][]{bontemps10, zhang15} but could not be substantiated further than a mass segregation effect. Top-heavy CMFs were also observed in combined CMFs, built from the combination of cores extracted in a dozen to several dozen massive clumps \citep{csengeri17b, liu18, sanhueza19, sadaghiani20, lu20, oneill21}. 
To date, the only two statistically significant studies carried out on single massive clouds are those of \cite{motte18b} and \cite{kong19}. If confirmed, these results challenge either the direct relation between the CMF and the IMF or the IMF universality, and most probably both.

To achieve the objectives of ALMA-IMF, we must tackle both individual cores and their connection to the larger-scale cloud environment, which is most immediately accessible via kinematics. Following the nomenclature in \cite{motte18a}, but adapting the definitions to gas structures containing massive protoclusters, clouds are a few parsecs in size, clumps are structures on intermediate scales on the order of a few times $\sim$0.1 pc, and cores are $\sim$0.01 pc in size. Cores could subfragment, making them the precursors of either single star or multiples stars, but they will not form stellar clusters. The large spectral coverage of ALMA 
makes it possible to simultaneously image molecular lines that characterize both cores and clouds. The presence of outflows and infall, often traced by CO, SiO, CS, and HCO$^+$ lines, typically provides the first indication of the evolutionary nature of the cores {that can be pre-stellar or protostellar}. In addition, as the luminosity of the protostars increases, they further interact with their immediate surroundings, creating hot cores and, for the most massive, \hii regions. Hot cores classically correspond to high-mass protostellar objects \citep[e.g.,][]{kurtz00, cesaroni05}, which are dominated by radiatively heated gas above 100~K. At these temperatures, the ice mantles of the dust grains formed in the cold, high-density medium of cores evaporate \citep[e.g.][]{HerVD09}. 
Hot cores are therefore associated with a rich molecular content observed by a large number of lines from complex organic molecules \citep[COMs; e.g.,][]{Gibb2000,Schilke06}. Considering the formation of dense cloud structures (filaments and cores $>$10$^5$~cm$^{-3}$), 
the most abundant, light molecules (such as CO, N$_2$H$^+$, and CS) are typically used to probe the gas density and kinematics. The kinematics of dominant filaments in low- and high-mass star-forming regions have already been studied in some detail \citep[e.g.,][]{schneider10, peretto13, fernandez14, battersby14, stutz18, hacar18, jackson19};  
however, very little is known about the gas feeding of cores \citep[]{galvan09,csengeri11b, olguin21, sanhueza21}. This process must now be a priority for CMF studies because constraints on any hierarchical inflow of gas could link cloud kinematics to the growth of core mass.

To deepen our understanding of the IMF origin and quantify the CMF dependence, if any, with respect to the properties of clouds over their lifetimes, various cloud environments must be sampled. Observational limitations lead to initiating such work by targeting the most massive and closest protoclusters in our Milky Way, as was done by, for example, \cite{motte18b} and \cite{sanhueza19}. 
During the past decade, the APEX/ATLASGAL\footnote{
    The APEX Telescope Large Area Survey of the Galaxy; see \url{https://atlasgal.mpifr-bonn.mpg.de}.},
CSO/BGPS, and \textsl{Herschel}/HiGAL surveys have covered the inner Galactic plane at (sub)millimeter and far-infrared wavelengths, providing complete samples of $0.1-1$~pc clumps up to distances of at least 8~kpc \citep{ginsburg13, csengeri14, koenig17, elia21}. 
From the CSO/BGPS catalog, \cite{ginsburg12} identified 18 particularly massive protocluster clouds, the most well known of which are Sgr~B2, W49, W51-E, W51-IRS2, W43-MM1, and W43-MM2 \citep{sanchez17, galvan13, ginsburg15, nguyen13, motte18b}. 
From the APEX/ATLASGAL catalog of \cite{csengeri14}, \cite{csengeri17b} identified a sample containing the 200 most massive clumps covered by ATLASGAL. As these clumps represent the early stages of massive cluster formation, this sample is the ideal choice for selecting the best targets for {a Large Program with ALMA}.

In the present paper, we provide an introduction of the ALMA-IMF Large Program. Section~\ref{s:intro} presents the main scientific objectives, and Sect.~\ref{s:target} describes the selection criteria that led to the targeting of 15 massive protoclusters. Section \ref{s:obs} presents the Large Program data set, whose data reduction and continuum images are fully described in a companion paper, Paper II  \citep{ginsburg22}. Section \ref{s:highlights} details the evolutionary stages of the ALMA-IMF protocluster clouds and investigates their core content from catalogs presented in a second companion paper, Paper III \citep[][]{louvet22}. In Sect.~\ref{s:highlights}, we also present the preliminary line data cubes used to illustrate the potential of the ALMA-IMF data set to constrain the kinematics and chemical complexity of clouds. Finally, Sect.~\ref{s:conc} summarizes our initial conclusions.

\setcounter{figure}{0}
\begin{figure*}[htbp!]
    \centering
    \includegraphics[width=0.85\textwidth]{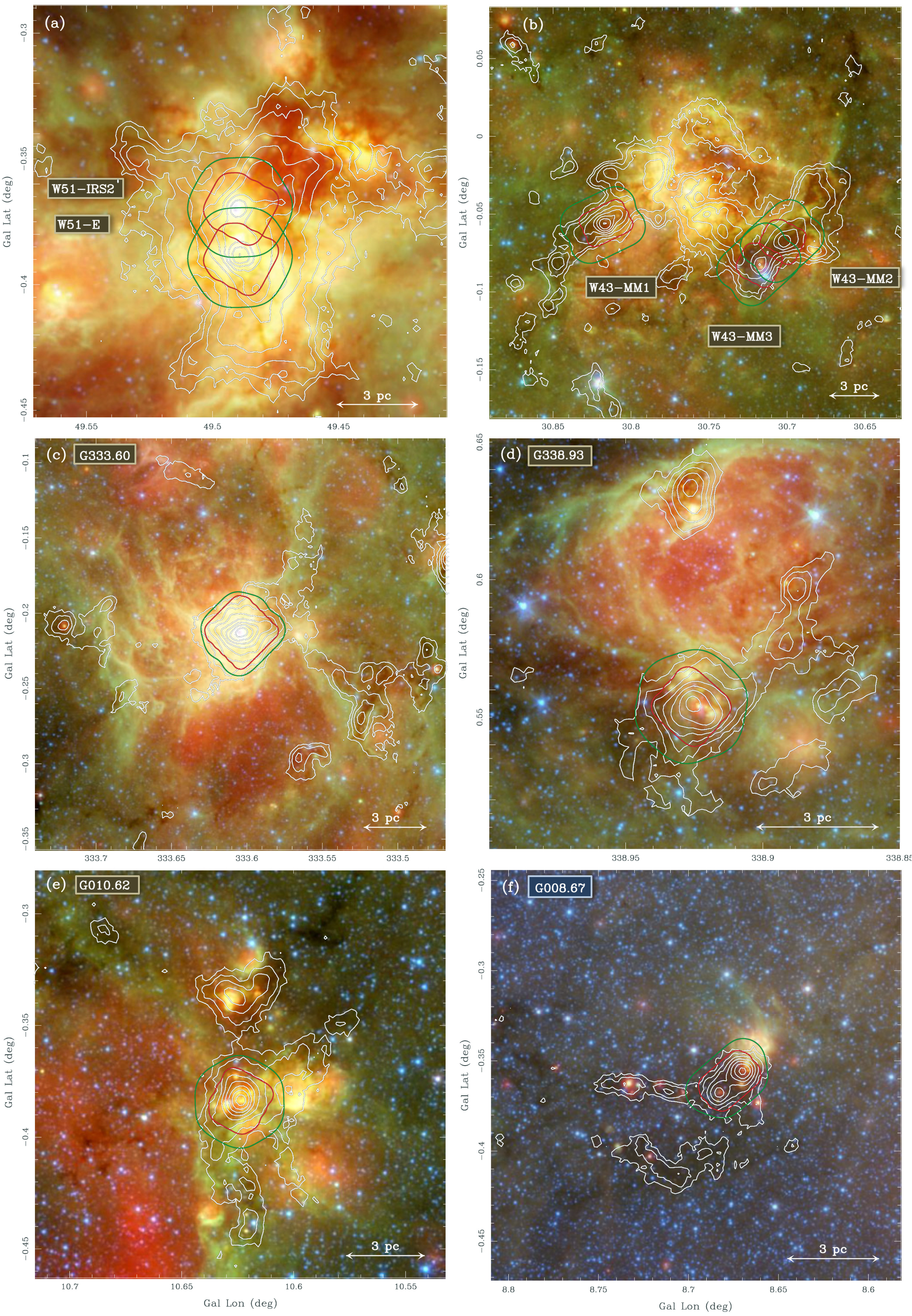}
    \caption{Overview of the surroundings of the ALMA-IMF protocluster clouds, ordered by decreasing mass of their central clump (Col.~9 of Table~\ref{tab:sample}): W51-E and W51-IRS2 (in \textsl{a}), W43-MM1, W43-MM2, and W43-MM3 (in \textsl{b}), G333.60 (in \textsl{c}), G338.93 (in \textsl{d}), G010.62 (in \textsl{e}), and G008.67 (in \textsl{f}). ATLASGAL $870\;\mu$m emission contours (logarithmically spaced from 0.45~Jy/beam to 140~Jy/beam with a $19.2\arcsec$ beam) are overlaid on \textsl{Spitzer} three-color images (red represents MIPS $24\;\mu$m, green GLIMPSE $8\;\mu$m, and blue GLIMPSE $3.6\;\mu$m). The green and red contours outline the primary beam response of the ALMA 12~m array mosaics down to $15\%$ at 3~mm and 1.3~mm, respectively. A 3 pc scale bar is shown in the lower-right corner of each image.
    }
    \label{fig:overview}
\end{figure*}

\setcounter{figure}{0}
\begin{figure*}[htbp!]
    \centering
    \includegraphics[width=0.85\textwidth]{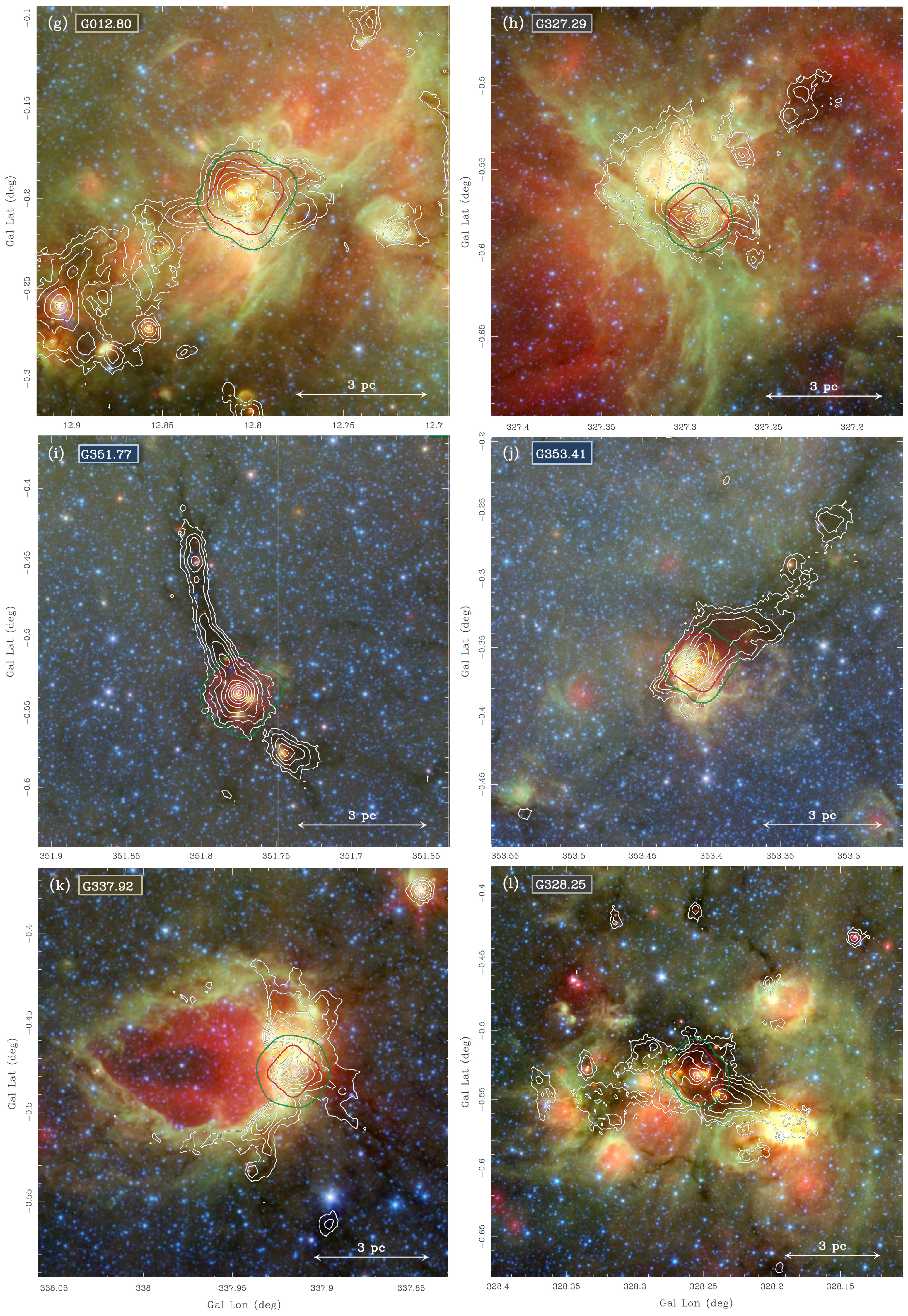}
       \caption{\textbf{(Continued)} Overview of the surroundings of the protocluster clouds G012.80 (in \textsl{g}), G327.29 (in \textsl{h}), G351.77 (in \textsl{i}), G353.41 (in \textsl{j}), G337.92 (in \textsl{k}), and G328.25 (in \textsl{l}).
       }
\end{figure*}

\begin{table*}[htbp!]
\centering
\begin{threeparttable}[c]
\caption{ALMA-IMF targets, a representative sample of massive protoclusters in the Milky Way.}
\label{tab:sample}
\begin{tabular}{lccccccccc}
\hline \noalign {\smallskip}
Protocluster  & RA\tnote{1}    & Dec\tnote{1}    & \vlsr\tnote{1} & $d$   & Ref.  
    & Evolutionary & FWHM$_{\rm 870\,\mu m}^{\rm clump}$\tnote{4} & $M_{\rm 870\,\mu m}^{\rm clump}$\tnote{4}  \\ 
cloud name\tnote{1}    & \multicolumn{2}{c}{[ICRS]}   & [$\kms$]      & [kpc] &  papers\tnote{2} &  stage\tnote{3}  & [pc] & [$\times 10^3~\msun$]  \\
\hline \noalign {\smallskip}
    
W51-E     
    & 19:23:44.18 & $+$14:30:29.5 & $+55$   & 5.4$\pm 0.3$ & (1) 
    & IR-bright &  0.41 & $10.4$ \\     
    
W43-MM1    
    & 18:47:47.00 &  $-$01:54:26.0 & $+97$    & 5.5$\pm 0.4$ & (2) 
    & IR-quiet & 0.47 & {$7.0$} \\      
        
G333.60     
    & 16:22:09.36 & $-$50:05:58.9 & $-47$   & 4.2$\pm 0.7$ & (3) 
    &  IR-bright & 0.58 & $5.4$         \\
    
W51-IRS2     
    & 19:23:39.81 & $+$14:31:03.5 & $+55$   & 5.4$\pm 0.3$ & (1) 
    & IR-bright & 0.39 & $4.8$ \\       

G338.93     
    & 16:40:34.42 & $-$45:41:40.6 & $-62$   & 3.9$\pm 1.0$ & (3) 
    & IR-quiet & 0.58 & $4.5$\\ 

G010.62        
    & 18:10:28.84 & $-$19:55:48.3 & $-2$     & 4.95$\pm 0.5$ & (4) 
    & IR-bright &  0.42 & $4.2$ \\      

W43-MM2         
    & 18:47:36.61 & $-$02:00:51.1 & $+97$   & 5.5$\pm 0.4$ & (2) 
    & IR-quiet & 0.57 & $4.2$ \\  

G008.67     
    & 18:06:21.12 & $-$21:37:16.7 & $+37.6$   & 3.4$\pm 0.3$ & (3)  
    & IR-quiet & 0.33 & $1.9$ \\ 

G012.80      
    & 18:14:13.37 & $-$17:55:45.2 & $+37$   & 2.4$\pm 0.2$ & (5)  
    & IR-bright & 0.32 & {$1.7$} \\
    
G327.29     
    & 15:53:08.13 & $-$54:37:08.6 & $-45$   & 2.5$\pm 0.5$ & (3) 
    & IR-bright & 0.16 & $1.4$ \\       

W43-MM3         
    & 18:47:41.46 & $-$02:00:27.6 & $+97$   & 5.5$\pm 0.4$ & (2) 
    & IR-bright & 0.47 & $1.1$ \\

G351.77     
    & 17:26:42.62 & $-$36:09:20.5 &   $-3$  & 2.0$\pm 0.7$ & (6) 
    & IR-bright & 0.16 & $1.0$ \\ 

G353.41     
    & 17:30:26.28 & $-$34:41:49.7 & $-17$   & 2.0$\pm 0.7$ & (6) 
    &  IR-bright & 0.30 & $0.9$ \\ 

G337.92         
    & 16:41:10.62 & $-$47:08:02.9 & $-40$   & 2.7$\pm 0.7$ & (6) 
    & {IR-bright} & 0.20 & {$0.8$} \\ 

G328.25     
    & 15:57:59.68 & $-$53:58:00.2 & $-43$   & 2.5$\pm 0.5$ & (3) 
    & IR-quiet & 0.21 & $0.5$ \\ 

\hline \noalign {\smallskip}
\end{tabular}
\begin{tablenotes}
\item[1] Protocluster name, central position of the mosaics, and velocity at rest used for the ALMA-IMF observations.
The Galactic coordinates of the associated ATLASGAL clumps are given in \cref{appendixtab:timea-table}. The
$\vlsr$ values are taken from the high-density gas studies by \cite{wienen15} and \cite{ginsburg15} for W51, \cite{nguyen13} for W43, and \cite{immer14} for G012.80. The phase center of W43-MM1 in the pilot study is 18:47:46.50, -01:54:29.5. 
\item[2] 
References for the distance to the Sun: (1) \cite{sato10}; (2) \cite{zhang14}; (3) \cite{csengeri17b};  (4) \cite{sanna14}; (5) \cite{immer13}; (6) this paper.
\item[3] Evolutionary stage of the ATLASGAL clump, as classified by \cite{csengeri17b}: IR-bright or IR-quiet (see Sect.~\ref{s:evol}).
\item[4] Size and mass of the ATLASGAL clump, as defined by \cite{csengeri17b}, located at the center of the ALMA-IMF clouds, whose area and mass are given in \cref{tab:evol}.
Mass estimates assume $\tdust=20$~K and 30~K for the IR-quiet and IR-bright sources, respectively. Sizes correspond to the geometric mean of the beam-deconvolved major and minor FWHM axes of Gaussian fits made in \citet{csengeri14}.
\end{tablenotes}
\end{threeparttable}
\end{table*}

\section{ALMA-IMF targets}
\label{s:target}

In an effort to investigate statistically the richest protoclusters of the Milky Way, we compiled a list of 15 massive clouds. \cref{tab:sample} lists their adopted names, the coordinates used as phase center, and velocity in the kinematic Local Standard of Rest. Figure~\ref{fig:overview} illustrates their surroundings with overlays of the mid-infrared \textsl{Spitzer} emission associated with the heating of luminous sources \citep{benjamin03, carey09} and the ATLASGAL submillimeter emission tracing the cloud gas \citep{schuller09}. As one of the main goals of the ALMA-IMF Large Program is to create large catalogs of protocluster cores, we focused on massive clouds selected from \cite{csengeri17b}, which have sizes of a few parsecs and can be properly imaged by ALMA with an angular resolution down to a few thousand au (see Sect.~\ref{s:select}). We then selected a representative sample of half (i.e., 15) of these protoclusters spanning a range of evolutionary stages (see Sect.~\ref{s:evol}).

\subsection{The most massive protocluster clouds of the Milky Way}
\label{s:select}

From the catalog of \cite{csengeri17b}, which contains the 200 most massive APEX/ATLASGAL clumps, we identified the most massive protocluster clouds of the Milky Way, whose core content can be characterized by ALMA (see \cref{appendixtab:timea-table} and \cref{fig:sample}). In order to reach an angular resolution of a couple of thousand au and a subsolar mass sensitivity with reasonable ALMA integration times, while reaching the exceptional W43 mini-starburst region at 5.5 kpc \citep{nguyen11a}, we applied a distance-limited criterion of $d\leq 5.5~\kpc$. With this upper limit for the distance and avoiding Galactic longitudes toward the Galactic center ($355\degree < l < 5\degree$)\footnote{
    At these longitudes, kinematic distances are very uncertain. For this reason, we chose to exclude two of the brightest ATLASGAL sources, whose radial velocities would locate them below 5.5~kpc but that could well be within the Galactic center region.
}, we further excluded regions at larger distances  
that would require long integration times and are already the focus of dedicated ALMA studies \citep[e.g.,][]{galvan13,sanchez17}. 
On the other hand, setting a lower distance limit of 2~kpc allows us to more easily observe the entire extent of the parsec-size clouds with ALMA mosaics. Furthermore, massive cloud complexes at lower distances (<2~kpc), including Cygnus~X, NGC~6334, M~17, and Orion, have already been extensively studied by, for example, the \textsl{Herschel}/HOBYS key program \citep{motte10} and already revealed the nearest sites of high-mass star and cluster formation \citep[e.g.,][]{bontemps10, ohashi16, louvet19, sadaghiani20, fischer20}.

As our aim is to focus on the densest $\sim$1~pc size clouds hosting the ATLASGAL clumps, we selected the ATLASGAL sources of \cite{csengeri17b} that have an integrated $870\;\mu$m flux larger than 25~Jy. This threshold, which is used to define the most massive protoclusters, corresponds to five times that used in \citet{csengeri17b} and leads to a list of 28 potential targets. This flux threshold ensures a minimum mass of $\sim$400$-3\,000~\msun$ for the densest regions in the massive protoclusters at distances of 2-5.5~kpc, and assuming appropriate values for our sample of $\tdust=25$~K and $\kappa_{\rm 870\,\mu m} = \rm 0.0185~cm^2\,g^{-1}$ in \cref{eq:mass870} (see Sect.~\ref{s:evol}). 
Among these 28 massive protoclusters, we find W51-E, W51-IRS2, W43-MM1, and W43-MM2 previously identified as extremely massive and active mini-starburst clumps \citep{motte03, ginsburg12}. Furthermore, we add two sources to this list: the ATLASGAL source W43-MM3 to cover the W43-MM2\&MM3 mini-starburst ridge, which is suspected to host an extreme protocluster \citep{nguyen13}, and G328.25-0.58, which is the most massive, young protocluster of \citet{csengeri17b} {that} exhibits at its center a single high-mass protostar \citep{csengeri18}. The catalog of the 30 selected massive clumps is given in  \cref{appendixtab:timea-table}. Their Galactic coordinates, evolutionary stage, and integrated flux at $870\;\mu$m are taken from \cite{csengeri17b}. We also include a list of ALMA projects that previously targeted these protoclusters, the name of the molecular complex hosting them, and their distance to the Sun in \cref{appendixtab:timea-table}.

Ten of the massive clumps of \cref{appendixtab:timea-table} have a distance measured by trigonometric parallaxes using masers \citep{sato10, immer13, sanna14, reid14, zhang14}. The distances for the remaining sources were estimated by \cite{csengeri17b} using kinematic distance estimates, associations with cloud complexes, and mid-infrared absorption features \citep[as done by][]{moises11,wienen15}. These distance estimates are subject to uncertainties, such as the Galactic rotation curve and association with the near or far kinematic distance solutions. Improvements can only be expected when parallax distances are available using either weaker masers or, for the closest clumps, \textsl{Gaia} measurements. We here modify the distances of three ATLASGAL clumps of \cref{appendixtab:timea-table} using recent improvements made by the BeSSeL\footnote{
    The Bar and Spiral Structure Legacy (BeSSeL) Survey; see \url{http://bessel.vlbi-astrometry.org}.}
project \citep[see][]{reid14}.
With its revised kinematic distance calculator, the two relatively nearby clumps, G351.77 and G353.41, both have almost equally probable distance solutions of $\sim$1.3~kpc and $\sim$2.7~kpc.
Given that the cloud gas between these two clumps presents a velocity continuity, G353.41-0.36 and G351.41-0.54 are probably part of the same complex and we adopted the average distance of $\sim$2.0~kpc, with a dispersion of $\pm0.7$~kpc, for both clumps. Moreover, we updated the distance of  the G337.92-0.48 clump to $2.7\pm0.3$~kpc, following that given by the BeSSeL calculator.

\subsection{A representative sample of 15 massive protoclusters at various evolutionary stages}
\label{s:evol}

We extracted from \cref{appendixtab:timea-table} a smaller sample of clouds, covering a range of evolutionary stages. \cite{csengeri17b} classified ATLASGAL clumps as either IR-bright or IR-quiet, based on their fluxes at mid-IR wavelengths. Initially proposed by \cite{motte07}, this classification has been adapted, in \cite{csengeri17b}, to use a flux threshold of 289~Jy at $22\;\mu$m and $d=1$~kpc, scaling it to the distances of the sources and extrapolating fluxes from {the Wide-field Infrared Survey Explorer} (\textsl{WISE}) or {the Multiband Imaging Photometer for} \textsl{Spitzer} (\textsl{Spitzer}/MIPS) {observatories}. This classification is expected to distinguish between clumps hosting faint infrared (IR-quiet) sources corresponding to deeply embedded, Class-0-like, high-mass protostars  \citep[e.g.,][]{bontemps10} and those with luminous infrared (IR-bright) objects. The latter are either ultra-compact \hii regions or clumps hosting evolved protostars, referred to as high-mass protostellar objects ({HMPOs}, see \citealt{beuther02a}) or massive young stellar objects ({MYSOs}, see \citealt{lumsden13}). The IR-quiet/IR-bright classification makes it possible to follow the evolution from cold to warm cloud structures \citep[e.g.,][]{motte18a}, where cold stages are sometimes referred to as infrared dark clouds {(IRDCs)}, even if they host various stages of low- and high-mass star formation \citep[e.g.,][]{peretto13}.

\cref{appendixtab:timea-table} contains only seven IR-quiet clumps, which exclusively populate the low $870\,\mu$m flux end of the sample distribution. IR-quiet protoclusters, however, as they are not yet significantly impacted by stellar feedback, probably represent the early stage during which it should be easier to study the CMF and its variation with cloud properties. We therefore rebalanced the sample of massive protocluster clouds to be used for the ALMA-IMF Large Program by systematically selecting the top seven, IR-bright, clumps of \cref{appendixtab:timea-table}, but favoring IR-quiet clumps among the remaining 21 clumps. 

To complement this selection, we first chose to cover all of the extreme protoclusters, which lie in the two exceptional, and relatively distant, cloud complexes W51 and W43: W51E, W51-IRS2, W43-MM1, W43-MM2, and W43-MM3. For the remaining molecular cloud complexes, we instead chose to observe only one protocluster per complex to sample various parts of the Milky Way. In the RCW106 and G327 complexes, we therefore only selected the brightest ATLASGAL clump, which happened to be IR-bright. In total, we selected for the ALMA-IMF survey five (out of seven, $71\%$) and ten (out of 23, $43\%$) of the IR-quiet and IR-bright clumps from \cref{appendixtab:timea-table}, respectively. 

\cref{tab:sample} lists the 15 protocluster clouds selected for the ALMA-IMF Large Program, which constitute a representative and well-balanced sample of the most massive protoclusters in the Milky Way. It gives their distance to the Sun, evolutionary stage, and the size and mass of their central clump, FWHM$_{\rm 870\,\mu m}^{\rm clump}$ and $M_{\rm 870\,\mu m}^{\rm clump}$, the latter being used here to order the cloud sample. In this final selection, the 15 massive protoclusters are located at $2-5.5$~kpc with a mean distance of 3.9~kpc. Since the $870\,\mu$m fluxes mainly correspond to thermal dust emission, which is largely optically thin, the clump mass is computed from the integrated fluxes, $S^{\rm int}_{\rm 870\,\mu m}$ listed in \cref{appendixtab:timea-table}, assuming a mass-averaged dust temperature and a distance to the Sun. We used the following equation, and provide here a numerical application whose dependence on each physical variable is given, for simplicity, in the Rayleigh-Jeans approximation:
\begin{eqnarray}
 \label{eq:mass870}
 M^{\rm clump}_{\rm 870\,\mu m} & =    & \frac{S^{\rm int}_{\rm 870\,\mu m}\; d^2}{ \kappa_{\rm 870\,\mu m}\; B_{\rm 870\,\mu m}(\tdust)}\\
      &\approx& 84 \ \msun \times 
              \left(\frac {S^{\rm int}_{\rm 870\,\mu m}}{\mbox{1~Jy}}\right) 
              \left(\frac {d}{\mbox{3.9~kpc}} \right)^2 \nonumber\\
      &      & \times 
\left(\frac {\kappa_{\rm 870\,\mu m}}{\rm 0.0185\,cm^2\, g^{-1}}\right)^{-1}
                     \left(\frac {\tdust}{\rm 20~K}\right)^{-1},
\nonumber
\end{eqnarray}
where $B_{\rm 870\,\mu m}(\tdust)$ is the Planck function for a dust temperature $\tdust$, $d$ is the distance of ALMA-IMF protoclusters, 
and $\kappa_{\rm 870\,\mu m}$ is the dust opacity per unit (gas~$+$~dust) mass at $870\,\mu$m. The adopted dust opacity, $\kappa_{\rm 870\,\mu m}=0.0185\,\rm cm^2\, g^{-1}$, follows the prescriptions by \cite{OsHe94} and assumes a gas-to-dust mass ratio of 100.  We adopted dust temperatures of $\tdust=20$~K and 30~K for the IR-quiet and IR-bright regions, respectively (see also Sect.~\ref{s:revised} for more discussion). These are the mean temperatures of the brightest ATLASGAL sources as measured in NH$_3$ \citep{wienen12, wienen18} in agreement with dust temperatures of \cite{koenig17}. In these 15 massive protoclusters, the mass of the central clump ranges from $500~\msun$ to $1\times 10^4~\msun$.

Figure~\ref{fig:overview} illustrates the IR-quiet versus IR-bright evolutionary stage of these protocluster clouds. The five IR-quiet protocluster clouds are all observed as strong extended ATLASGAL cloud structures associated with extinction patterns at mid-infrared wavelengths (see Figs.~\ref{fig:overview}b, d, f, l). In contrast, the ten IR-bright protoclusters emit at near- to mid-infrared wavelengths, either weakly (see Figs.~\ref{fig:overview}b, i--j) or more strongly (see Figs.~\ref{fig:overview}a, c, e, g--h, k).

Beyond their different evolutionary stages, the targeted protoclusters may represent different conditions for cluster formation in the galactic disk of the Milky Way. The five protoclusters in the W43 and W51 cloud complexes are among the most active star-forming regions of the Milky Way. W43 is located at the end of the Galactic bar and W51 could be a massive cloud compressed by Galactic motions along the Perseus arm \citep{nguyen11a, ginsburg15}. Seven other ALMA-IMF clouds could be under the influence of massive stellar clusters. Each of the G010.62, G337.92, and G338.93 clouds is indeed located at the periphery of a large bubble, presumably excited by OB stars, and the G333.60, G327.29, G328.25, and G012.80 clouds are found in complex networks of such bubbles (see \cref{fig:overview}). In contrast, the G008.67, G351.77, and G353.41 clouds seem more isolated, without obvious interaction with massive stellar clusters or Galactic motions.

\begin{table*}[htbp!]
\centering
\caption{Observational data summary of the 12~m array continuum images of ALMA-IMF protoclusters.}\label{tab:sensitivity_scale_overview}
\resizebox{\textwidth}{!}{
\begin{threeparttable}[c]
\begin{tabular}{l|ccllc|ccllc}
\hline
 & \multicolumn{5}{c|}{1.3~mm (Band 6)} & \multicolumn{5}{c}{3~mm (Band 3)}\\
Protocluster & Imaged FOV\tnote{1} & Resolution\tnote{2} & $\sigma$(\cleanest)\tnote{3} & $\sigma$(\bsens)\tnote{3} & {LAS$_{10\%}$}\tnote{4} & Imaged FOV\tnote{1} & Resolution\tnote{2} & $\sigma$(\cleanest)\tnote{3} & $\sigma$(\bsens)\tnote{3} & {LAS$_{10\%}$}\tnote{4} \\
cloud name & [$\arcsec\times\arcsec$] & [$\arcsec\times\arcsec$] & 
\multicolumn{2}{c}{[$\mathrm{mJy\,beam^{-1}}$]} 
& [$\arcsec$] & [$\arcsec\times\arcsec$] & [$\arcsec\times\arcsec]$ & \multicolumn{2}{c}{[$\mathrm{mJy\,beam^{-1}}$]} 
& [$\arcsec$] \\
\hline
W51-E    & $100\times90$  & $0.35\times0.27$ & 0.17  & 0.16  & 5.0 & $150\times160$ & $0.29\times0.26$ & 0.055 & 0.035 & 6.3 \\
W43-MM1  & $117\times53$  & $0.50\times0.35$ & 0.19  & 0.18  & 4.6 & $190\times150$ & $0.56\times0.33$ & 0.051 & 0.038 & 4.8 \\
G333.60  & $143\times143$ & $0.56\times0.51$ & 0.11  & 0.12  & 5.8 & $190\times180$ & $0.46\times0.44$ & 0.070 & 0.047 & 10  \\
W51-IRS2 & $92\times98$   & $0.50\times0.44$ & 0.097 & 0.076 & 6.1 & $160\times150$ & $0.28\times0.27$ & 0.061 & 0.061 & 9.6 \\
G338.93  & $86\times92$   & $0.56\times0.51$ & 0.17  & 0.16  & 5.5 & $152\times160$ & $0.40\times0.38$ & 0.068 & 0.044 & 11  \\
G010.62  & $98\times90$   & $0.53\times0.41$ & 0.083 & 0.082 & 5.2 & $150\times160$ & $0.39\times0.32$ & 0.051 & 0.051 & 9.4 \\
W43-MM2  & $90\times98$   & $0.52\times0.41$ & 0.075 & 0.063 & 5.5 & $190\times150$ & $0.30\times0.24$ & 0.037 & 0.024 & 8.1 \\
G008.67  & $132\times87$  & $0.72\times0.59$ & 0.37  & 0.20  & 5.9 & $190\times125$ & $0.51\times0.40$ & 0.094 & 0.080 & 10  \\
G012.80  & $132\times132$ & $1.09\times0.70$ & 0.65  & 0.74  & 6.6 & $190\times180$ & $1.48\times1.26$ & 0.21  & 0.24  & 9.9 \\
G327.29  & $105\times109$ & $0.69\times0.62$ & 0.36  & 0.32  & 5.5 & $160\times152$ & $0.43\times0.37$ & 0.13  & 0.075 & 10  \\
W43-MM3  & $100\times90$  & $0.51\times0.43$ & 0.061 & 0.063 & 5.7 & $190\times150$ & $0.41\times0.29$ & 0.031 & 0.028 & 8.2 \\
G351.77  & $132\times132$ & $0.89\times0.67$ & 0.42  & 0.31  & 6.2 & $190\times180$ & $1.52\times1.30$ & 0.26  & 0.12  & 15  \\
G353.41  & $131\times131$ & $0.93\times0.66$ & 0.42  & 0.40  & 6.2 & $190\times180$ & $1.46\times1.27$ & 0.18  & 0.17  & 10  \\
G337.92  & $92\times86$   & $0.61\times0.48$ & 0.22  & 0.23  & 5.6 & $160\times152$ & $0.45\times0.38$ & 0.070 & 0.051 & 11  \\
G328.25  & $120\times120$ & $0.62\times0.47$ & 0.37  & 0.29  & 4.9 & $160\times180$ & $0.62\times0.44$ & 0.087 & 0.076 & 7.5 \\
\hline
\end{tabular}
\begin{tablenotes}
\item[1] Field of view (FOV) corresponding to the combined primary beam of the 1.3~mm and 3~mm mosaics, down to $15\%$ of the peak sensitivity.
\item[2] Angular resolution resulting from a \texttt{tclean} process with the Briggs robust parameter \texttt{robust}=0. 
\item[3] Noise level measured in the \bsens and \cleanest 12~m array images at 1.3~mm and 3~mm.
\item[4] The maximum recoverable scale of the 12~m array, often called the largest angular scale (LAS), is estimated for each 1.3~m and 3~mm images as the 10$^{\rm th}$ percentile of the baseline lengths of 12~m array data (see Figs.~5--7 of \citealt{ginsburg22}).
\end{tablenotes}
\end{threeparttable}
}
\end{table*}

\begin{table*}[htbp!]
\centering
\small
\begin{threeparttable}[c]
\caption{Spectral setup of the ALMA-IMF Large Program.}
\label{tab:lines}
 \begin{tabular}{lcccccl}
\hline \noalign {\smallskip}
ALMA  & Spectral  & Frequency & Bandwith & \multicolumn{2}{c}{Resolution}  & Main spectral lines \\
band & window & [GHz] & [MHz] & [kHz] & [\kms] &  \\
\hline \noalign {\smallskip}
Band 6  & SPW0 & 216.200  & 234 & 244 & 0.34 & DCO$^+$~(3-2), CH$_{3}$OCHO, OC$^{33}$S~(18-17), HCOOH \\
        & SPW1 & 217.150  & 234 & 282 & 0.39  &
        SiO~(5-4), DCN~(3-2), $^{13}$CH$_{3}$OH, CH$_{3}$OCH$_{3}$ \\
        & SPW2 & 219.945  & 117 & 282 & 0.38  & SO~(6-5), H$_{2}^{13}$CO~(3$_{1,2}$-2$_{1, 1}$), CH$_{3}$OH \\ 
        & SPW3 & 218.230  & 234 & 244 & 0.33  & H$_{2}$CO~(3-2), O$^{13}$CS~(18-17), HC$_{3}$N~(24-23), CH$_{3}$OCHO \\
        & SPW4 & 219.560  & 117 & 244 & 0.33  & C$^{18}$O~(2-1), C$_{2}$H$_{5}$CN \\ 
        & SPW5 & 230.530  & 469 & 969 & 1.3  & CO~(2-1), CH$_{3}$CHO, CH$_{3}$OH, C$_{2}$H$_{3}$CN, C$_{2}$H$_{5}$OH \\
        & SPW6 & 231.280  & 469 & 488 & 0.63 & $^{13}$CS~(5-4), N$_{2}$D$^{+}$~(3-2), OCS~(19-18), CH$_{3}$CHO, CH$_{3}$OH,\\
        & & & & & &     CH$_{3}^{18}$OH, C$_{2}$H$_{5}$CN \\
        & SPW7 & 232.450  & 1875   & 1130 & 1.5 & H30$\alpha$, CH$_{3}$CHO, CH$_{3}$OH, CH$_{3}$OCHO, C$_{2}$H$_{5}$OH, C$_{2}$H$_{5}$CN, \\
        & & & & & & CH$_{3}$OCH$_{3}$, CH$_{3}$COCH$_{3}$, $^{13}$CH$_{3}$CN~(13-12), H$_{2}$C$^{34}$S~(7$_{1,7}$-6$_{1,6}$), HC(O)NH$_{2}$  \\
        \hline
Band 3  & SPW0 & 93.1734 & 117 & 71 & 0.23 & N$_{2}$H$^{+}$~(1-0), CH$_{3}$OH \\
        & SPW1 & 92.2000 & 938 & 564 & 1.8 &  CH$_3$CN~(5-4), H41$\alpha$, CH$_{3}$$^{13}$CN, $^{13}$CS~(2-1), $^{13}$CH$_{3}$OH, CH$_{3}$OCHO\\
        & SPW2 & 102.600 & 938 & 564 & 1.6 &  CH$_3$CCH~(6-5), CH$_{3}$OH, H$_{2}$CS,C$_{2}$H$_{5}$CN, C$_{2}$H$_{5}$OH, CH$_{3}$NCO \\
        & SPW3 & 105.000 & 938 & 564 & 1.6 & H$_{2}$CS, CH$_{3}$OH, C$_{2}$H$_{3}$CN, C$_{2}$H$_{5}$OH, CH$_{3}$OCH$_{3}$ \\
\hline \noalign {\smallskip}

\end{tabular}
\end{threeparttable}
\end{table*}

\begin{table*}[htbp!]
\centering
\begin{threeparttable}[c]
\caption{Main characteristics (size and mass) of the massive protocluster clouds imaged by ALMA-IMF and their evolutionary stage.}
\label{tab:evol}
 \begin{tabular}{lccrrcll}
\hline \noalign {\smallskip}
Protocluster   & \multicolumn{2}{c}{Imaged areas\tnote{1}}   
    & $S^{\rm cloud}_{\rm 870\,\mu m}$\tnote{2} & $M^{\rm cloud}_{\rm 870\,\mu m}$\tnote{3} 
    & $S^{\rm cloud}_{\rm 1.3\,mm}$
    & $\Sigma^{\rm free-free}_{\rm H41\alpha}$~\tnote{4}  & Refined \\
cloud name    & \multicolumn{2}{c}{[$\pc \times\pc$]} 
    & \multicolumn{2}{c}{within $A_{\rm 1.3\,mm}$}
    & $/S^{\rm cloud}_{\rm 3\,mm}$  & over $A_{\rm 1.3\,mm}$ & evolutionary \\
& $A_{\rm 1.3\,mm}$ & $A_{\rm 3\,mm}$ & [Jy]  & [$\times 10^3~\msun$] & over $A_{\rm 1.3\,mm}$ & [$\rm Jy\,pc^{-2}$] & stage\tnote{5}\\
\hline \noalign {\smallskip}

W43-MM1 & $3.1\times2.3$        & $5.1 \times 4.0$      & {80.3}        & {13.4}  
    & 13        & 0.005 & IR-quiet $=$ Y \\ 
    
W43-MM2 & $2.6 \times 2.4$      & $5.1 \times 4.0$  & ~~69.6    & 11.6  
    &   15 & 0.009      & IR-quiet $=$ Y \\
    
G338.93          & $1.6 \times 1.6$     & $2.9 \times 2.8$      &  ~~84.6 & 7.1   
    & 7.2       & 0.02  & IR-quiet $=$ Y \\
    
G328.25          & $1.4 \times 1.4$     & $2.2 \times 1.9$        &  ~~73.2 & 2.5
    & 8.3       &   0.03 & IR-quiet $=$ Y \\
    
G337.92             & $1.2 \times 1.1$  & $2.1 \times 2.0$      &  ~~63.3 & {2.5}   
    & 5.8       & 0.04, faint \hii      & {IR-bright $\rightarrow$} Y \\
    
G327.29          & $1.3 \times 1.3$     & $1.9 \times 1.8$      &  147.9 & 5.1   
    & 11        & 0.1, faint \hii       & IR-bright $\rightarrow$ Y \\
       
\hline

G351.77  & $1.3 \times 1.3$    & $1.8 \times 1.7$ &  158.7 & 2.5        
    & 8.3       &  0.2, UC\hii  & IR-bright $\rightarrow$ I \\
        
G008.67  & $2.2 \times 1.4$     & $3.1 \times 2.1$  & ~~66.5 & 3.1          
    & 3.2       &       0.6, UC\hii  &  IR-quiet $\rightarrow$ I \\
    
W43-MM3  & $2.7 \times 2.4$     & $5.1 \times 4.0$  & ~~43.2 & 5.2      
    & 2.2       & 0.2, UC\hii & IR-bright $\rightarrow$ I \\
    
W51-E  & $2.6 \times 2.4$       & $4.2 \times 3.9$      &  278.9 & 32.7 
    & 2.2       & 1, two HC\hii $+$ \hii        & IR-bright $\rightarrow$ I \\
    
G353.41          & $1.3 \times 1.3$ &  $1.8 \times 1.7$ & {153.4}  & {2.5}     
    & 1.9     & 0.7, UC\hii $+$ \hii &  IR-bright $\rightarrow$ I \\

\hline    

G010.62    & $2.3 \times 2.2$   & $3.8 \times 3.6$ & ~~87.3  & 6.7      
    & 1.8       & 2, two \hii   & IR-bright $=$ E \\ 
    
W51-IRS2    & $2.6 \times 2.4$  & $4.2 \times 3.9$      &  224.4 & 20.6 
    & 1.4       & 2, two \hii   & IR-bright $=$ E \\
    
G012.80     & $1.5 \times 1.5$    & $2.2 \times 2.1$ & 255.4    & 4.6   
    & 1.1       & 7, \hii & IR-bright $=$ E\\
    
G333.60             & $2.9 \times 2.9$    & $3.9 \times 3.7$ & 216.3  & 12.0    
    & 0.8   & 5, \hii & IR-bright $=$ E \\ 
    
\hline \noalign {\smallskip}
\end{tabular}
\begin{tablenotes}
\item[1] Physical areas encompassing the combined primary beam of the 1.3~mm and 3~mm mosaics, down to $15\%$. 
These areas define the cloud sizes.
\item[2] Integrated flux density at 870$\,\mu$m measured on the ATLASGAL images \citep{schuller09} within the area of the 1.3\,mm ALMA images (Col.~2).
\item[3] Cloud mass computed from the $870\,\mu$m integrated flux (Col.~4) assuming $\kappa_{\rm 870\,\mu m} = \rm 0.0185~cm^2\,g^{-1}$ and using $\tdust=20$~K, 25~K and 30~K for the Young, Intermediate, and Evolved regions (see Col.~8), respectively.
\item[4] Flux surface density of the free-free emission at 92.034~GHz, estimated from the integrated flux of the H41$\alpha$ recombination line. 
Spatial distribution of the ionized gas: HCH$\mbox{\sc ~ii}$ and UCH$\mbox{\sc ~ii}$ consist of $<$0.05~pc and $\sim$0.1~pc bubbles of ionized gas, respectively; H$\mbox{\sc ~ii}$ are larger regions of ionized gas with a non-spherical structure, some are faint H$\mbox{\sc ~ii}$ regions
(see Sect.~\ref{s:revised} and \citealt{motte18a}).
\item[5] Classification of the ALMA-IMF protocluster clouds: Young (Y), Intermediate (I), and Evolved (E). Their evolutionary stage is refined from that of \cite{csengeri17b} by measuring their 1.3~mm to 3~mm flux ratio (Col.~6) and their estimated free-free emission flux density (Col.~7). We assume that IR-quiet and IR-bright ATLASGAL clumps are associated with Young and Evolved clouds, respectively. Any evolution of this classification is discussed in Sect.~\ref{s:revised} and indicated here by an arrow.
\end{tablenotes}
\end{threeparttable}
\end{table*}

\section{Observations and data reduction}
\label{s:obs}

The ALMA-IMF Large Program (\#2017.1.01355.L, PIs: Motte, Ginsburg, Louvet, Sanhueza) was set up following the pilot program \#2013.1.01365.S. 
The ALMA-IMF Large Program images each of the 15 massive protocluster clouds of \cref{tab:sample} both at 1.3~mm (ALMA Band 6)\footnote{
    The 1.3~mm observations of W43-MM1 are part of the pilot study \#2013.1.01365.S and \#2015.1.01273.S.}
and 3~mm (Band 3). We here explain our observation strategy (see Sect.~\ref{s:strategy}) and briefly discuss the resulting data set (see Sect.~\ref{s:dataset}), which is described in more detail in Paper II \citep{ginsburg22}. Tables~\ref{tab:sensitivity_scale_overview}-\ref{tab:lines} give the  mapping and spectral  setups  of  the  ALMA-IMF Large Program.

\subsection{Observing strategy}
\label{s:strategy}
 
To resolve the $\sim$2\,000~au typical diameter of cores \citep{zhang09,bontemps10, palau13} and image the $\sim$1$-8~\pc^2$ protocluster cloud extent, 1.3~mm and 3~mm mosaics (shown in Figs.~\ref{fig:overview}a--l and whose extent is listed in \cref{tab:sensitivity_scale_overview}) were requested with $0.37\arcsec-0.95\arcsec$ synthesized beams depending on their distance (see \cref{tab:sensitivity_scale_overview}).

We chose the 1.3~mm and 3~mm spectral bands primarily for their mostly optically thin emission in (massive) cores and their relatively well-defined dust opacities. The central frequencies of the ALMA-IMF bands are $\nu_{\textrm{B6}}=228.965$~GHz (1.3~mm) and $\nu_{\textrm{B3}}=100.713$~GHz (3~mm) \citep[see Table~D1 of][]{ginsburg22}, assuming a spectral index of $\alpha=3.5$ that corresponds to optically thin dust emission with an emissivity index of $\beta=1.5$, well suited for protostars \citep{AWB93, juvela15}. According to \cite{OsHe94} and assuming a gas-to-dust mass ratio of 100, the dust opacities per unit (gas $+$ dust) mass, recommended for cores, are $\kappa_{\rm 1.3\,mm}=\rm 0.01~cm^2\,g^{-1}$ and $\kappa_{\rm 3\,mm}=\kappa_{\rm 1.3\,mm} \times \left(\frac{\nu_{\textrm{B6}}} {\nu_{\textrm{B3}}}\right)^{-1.5} \simeq \rm 0.0034~cm^2\,g^{-1}$. The ALMA-IMF Large Program was designed to reach, for point-like cores, a gas mass sensitivity of $0.15~\msun$ ($3\,\sigma$) at 1.3~mm, for all protocluster clouds of \cref{tab:sample} and over their whole extents (see Figs.~\ref{fig:overview}a--l). Assuming optically thin dust emission, the above dust opacity, and a dust temperature of $20$~K, this requirement led to a large range of continuum sensitivity requests: $1\,\sigma = 0.1-0.6$~mJy\,beam$^{-1}$ at 1.3~mm. To complete these 1.3~mm detections and correct them for a few optically thick (massive) cores, we aimed to reach a point mass sensitivity of $0.6~\msun$ ($3\,\sigma$) at 3~mm, corresponding to $1\,\sigma = 0.03-0.18$~mJy\,beam$^{-1}$. 
As we show in Sect.~\ref{s:revised}, comparing the 1.3~mm and 3~mm continuum images allows us to distinguish thermal dust emitting sources, such as cloud filaments and cores, from free-free emitting sources associated with ionized gas of \hii regions.

The spectral setup chosen for the ALMA-IMF Large Program contains eight spectral windows at 228.4~GHz (1.3~mm) and four at 99.66~GHz (3~mm)
(see \cref{tab:lines}). 
The 228.4~GHz setup is exactly the one used for the ALMA-IMF pilot project that targeted the W43-MM1 protocluster cloud \citep{nony18, nony20}. The main characteristics of these 12 spectral windows are given in \cref{tab:lines}, including the main lines they cover. ALMA-IMF has a particular focus on the N$_2$H$^+$~(1-0), DCO$^+$~(3-2), DCN~(3-2), and C$^{18}$O~(2-1) lines intended to be used to trace gas mass inflows from the cloud to its cores \citep[e.g.,][]{csengeri11a, peretto13, henshaw14, chen19, alvarez21}. The $^{12}$CO~(2-1), SiO~(5-4), and SO~(6-5) lines were chosen to trace protostellar outflows and shocks associated with protostellar accretion or cloud formation \citep[e.g.,][]{gusdorf08, 
sanhueza13, duarte14, louvet14, li20}. Additionally, the $^{13}$CS~(5-4) and N$_2$D$^+$~(3-2) lines were chosen to estimate the core turbulence levels \citep[e.g.,][]{tan13, nony18}. As for the H41$\alpha$ and H30$\alpha$ recombination lines, they pinpoint \hii regions and allow for more robust gas mass estimates by accounting for free-free contamination of the millimeter fluxes \citep[e.g.,][]{liu19}. Molecules such as CH$_3$CN and CH$_3$CCH, can themselves be used to probe the gas temperature of hot cores and their envelopes {or} host cores, respectively. The lines of H$_{2}$CO, OCS, and other COMs (see \cref{tab:lines}) emitting in the 6.4~GHz noncontinuous bandwidth of the ALMA-IMF setup probe the physical and chemical conditions of hot cores, 
protostellar outflows, and shocks \citep[e.g.,][]{giannetti17, Lefloch2017, csengeri19, molet19, bonfand19}.

The ALMA-IMF Large Program was designed to provide sensitive continuum estimates through wide spectral windows: one $\sim$2~GHz window at 1.3~mm and three $\sim$1~GHz windows at 3~mm, with a velocity resolution of $\sim$1.5--1.8~$\kms$, also allowing the detection of broad hot core lines from COMs with confidence \citep[e.g.,][]{molet19, belloche20, olguin21}. The narrow spectral windows, except two at 1.3~mm, have a spectral resolution corresponding to $\sim$0.3~$\kms$ (see \cref{tab:lines}), suitable to follow the gas kinematics at the spectral resolution of the sonic line width \citep[e.g.,][]{henshaw14, chen19}.
The two narrow spectral windows with a lower spectral resolution, $\sim$1.3~$\kms$ and $\sim$0.6~$\kms$, are customized to detect CO~(2-1) outflows and to cover both the $^{13}$CS~(5-4) and N$_2$D$^+$~(3-2) lines, respectively. 

The line sensitivity of the ALMA-IMF Large Program at 1.3~mm is driven by the need to detect $>$1~$\kms$ lines, such as $^{13}$CS~(5-4) or N$_2$D$^+$~(3-2), toward dust cores. With the requested 1.3~mm continuum sensitivity, $1\,\sigma = 0.1-0.6$~mJy\,beam$^{-1}$, the expected noise level is $0.6-0.8$~K averaged within a spectral resolution element of 1~$\kms$. The targeted lines generally are ten to hundred times brighter \citep{tan13, nony18}, thus allowing their correct characterization in terms of line width. This sensitivity is also enough to detect outflowing material in CO~(2-1) and SiO~(5-4) around candidate protostellar sources \citep[e.g.,][]{nisini07, duarte13, plunkett13, nony20} and COMs tracing hot cores and shocks \citep[e.g.,][]{Lefloch2017,molet19, csengeri19, olguin21}. At 3~mm, the $1\sigma = 0.03-0.18$~mJy\,beam$^{-1}$ continuum sensitivity leads to a sensitivity of $2-3$~K at 1~$\kms$ resolution. This allows N$_2$H$^+$~(1-0) cubes to be sensitive down to the weakest filaments crossing the ALMA-IMF protoclusters (see, e.g., \cref{fig:dynamics}).

While compact structures, such as cores traced by their thermal continuum emission or hot cores identified by line forests, are marginally affected by interferometric artifacts, a proper analysis of molecular outflows and gas inflows would require combining ALMA 12~m array mosaics with 7~m array ({Atacama Compact Array}, ACA) data and, if possible, Total Power Array data \citep{zhang16, li20, hara21}. Therefore, the ALMA-IMF Large Program  observed all the appropriate short spacing data that will allow correct analyses and provide the community with a complete and uniformly produced data set.

\setcounter{figure}{1}

\begin{figure*}[htbp!]
    \centering
    \includegraphics[width=0.54\textwidth]{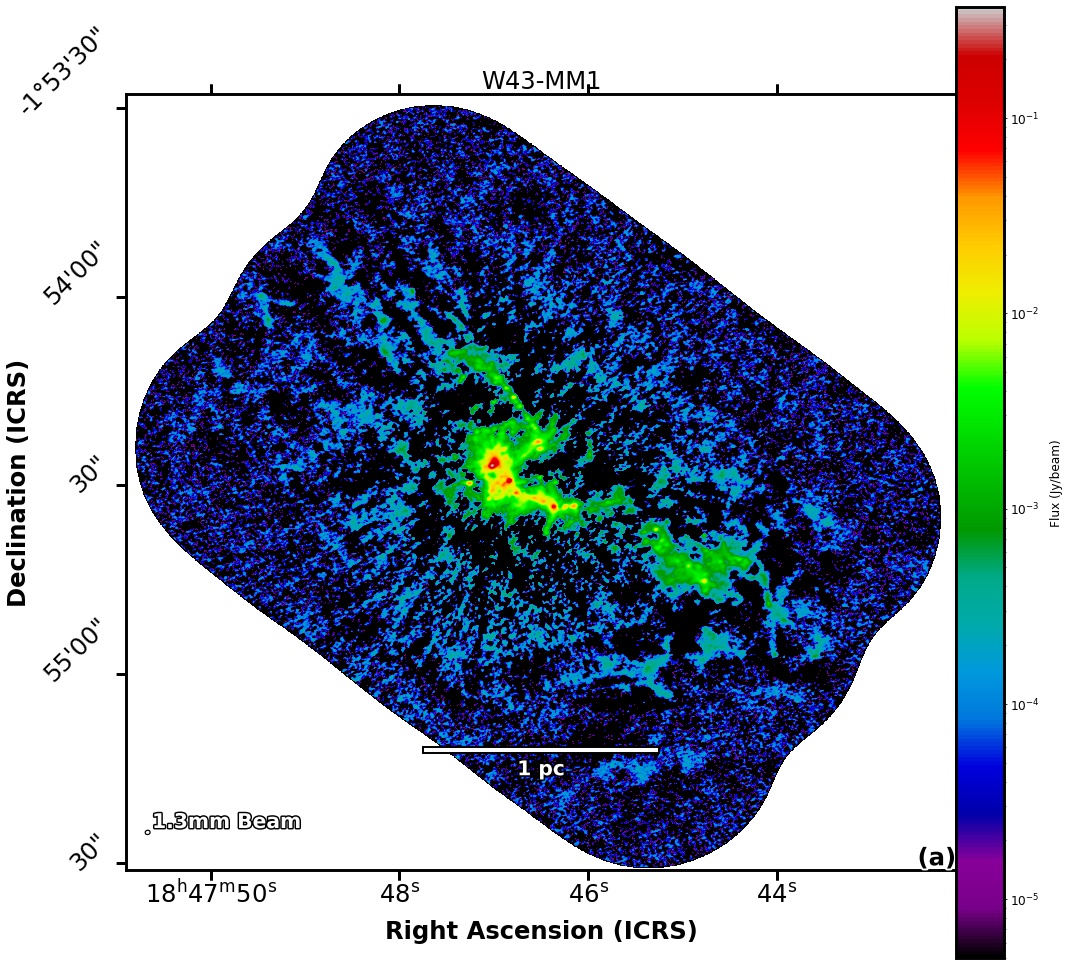}\hskip 0.3cm\includegraphics[width=0.41\textwidth]{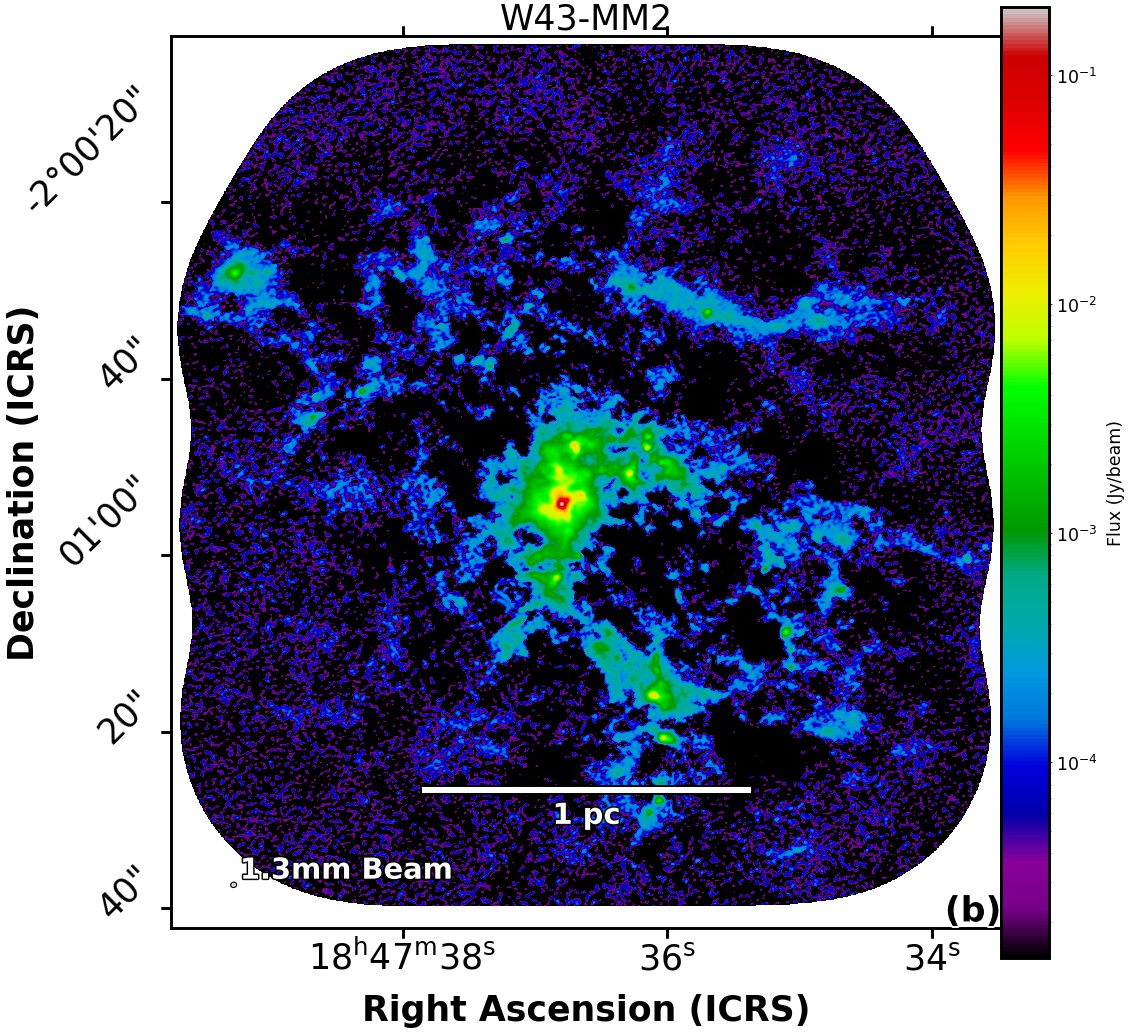}
    \vskip 0.5cm
    \includegraphics[width=0.295\textwidth]{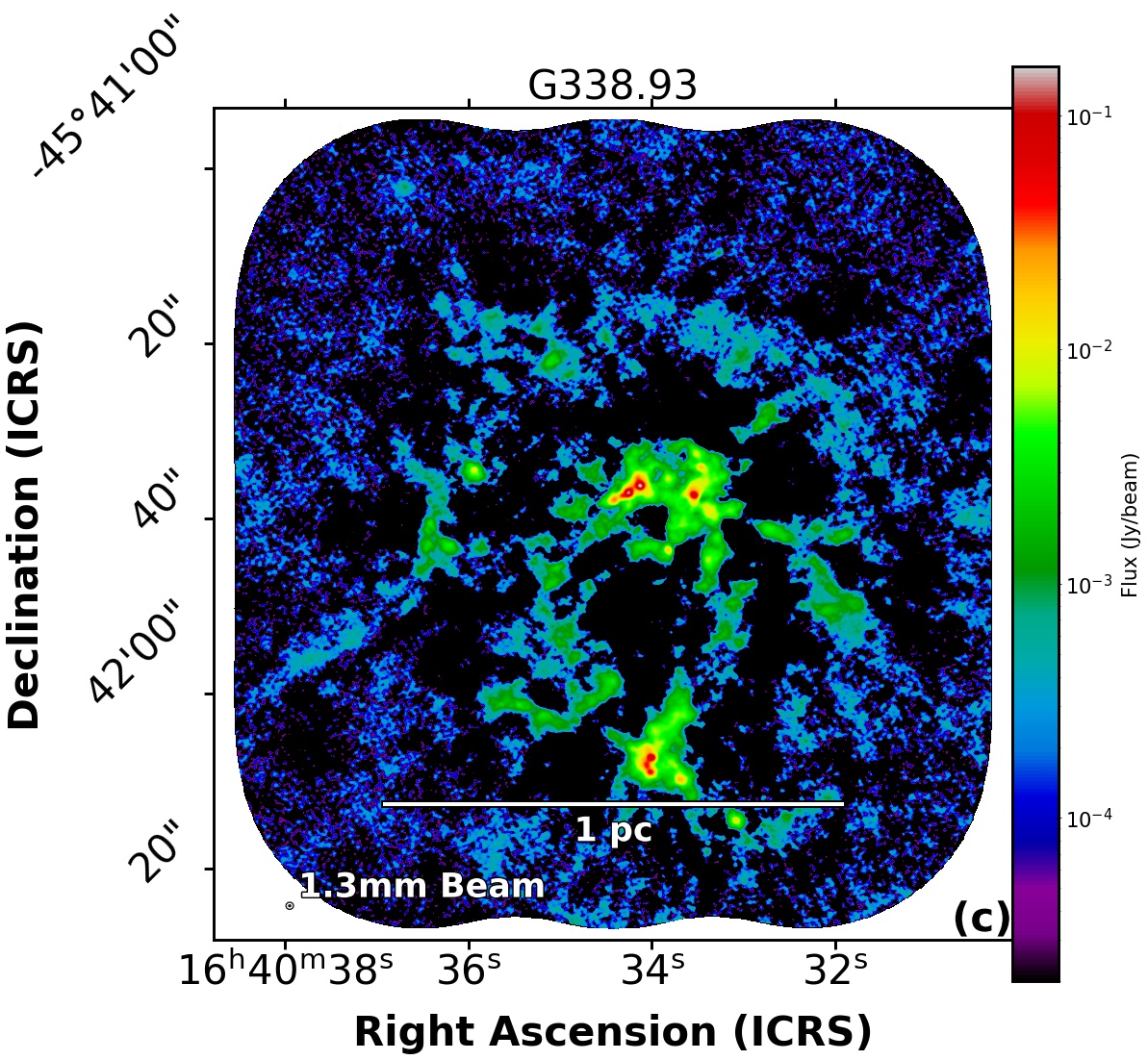}\hskip 1cm\includegraphics[width=0.28\textwidth]{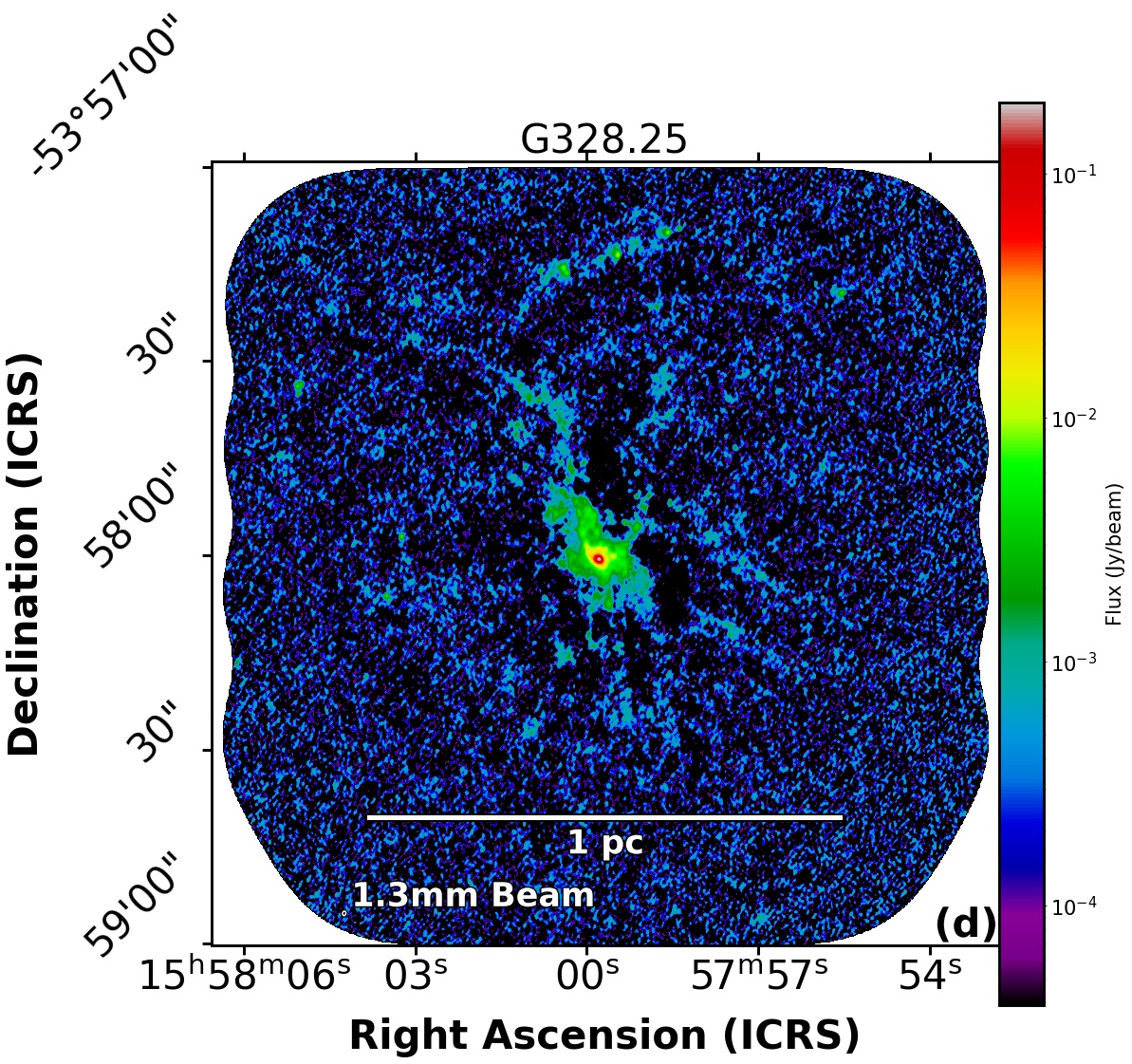}\\
    \vskip 0.5cm
    \includegraphics[width=0.28\textwidth]{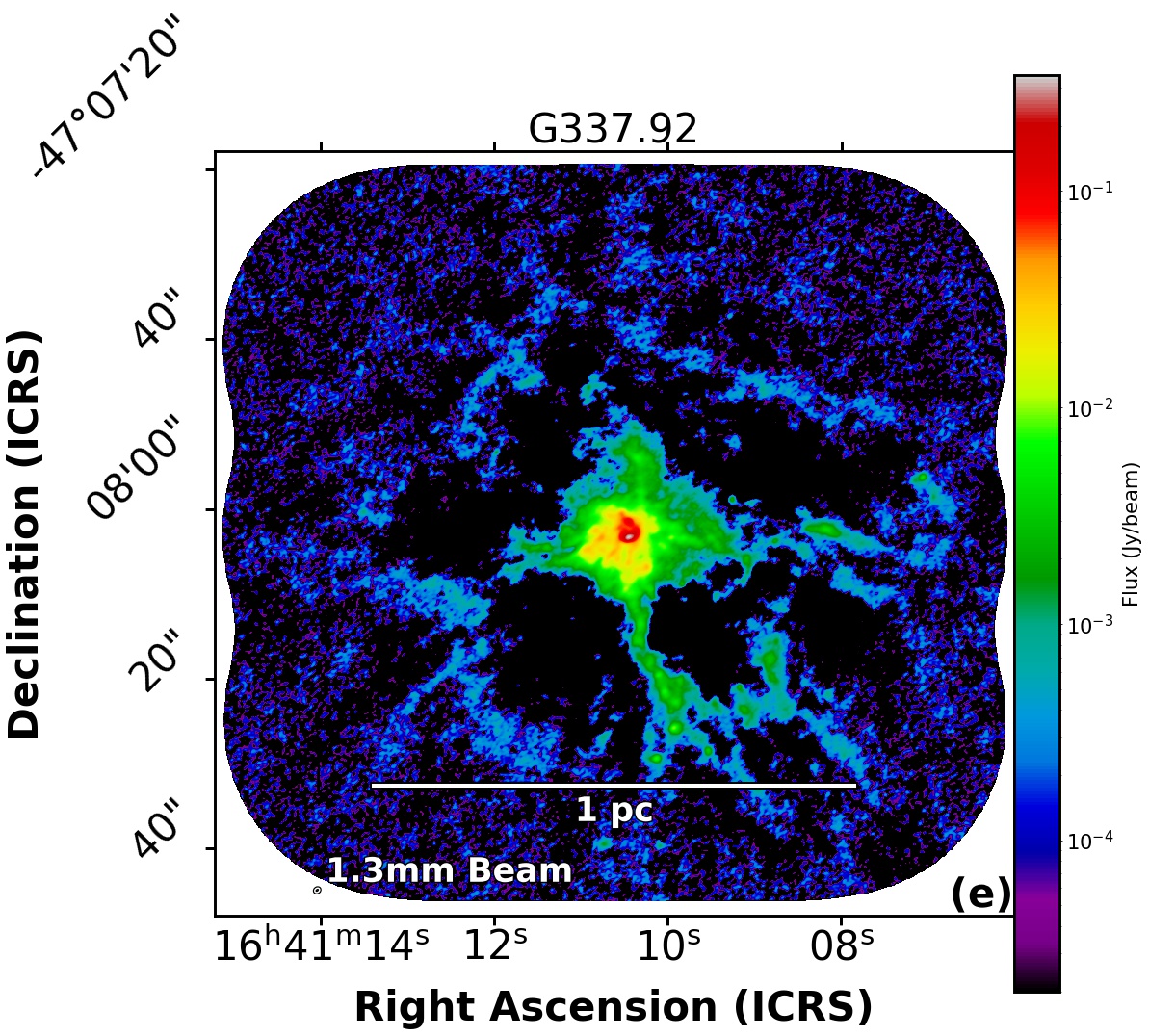}\hskip 1cm\includegraphics[width=0.245\textwidth]{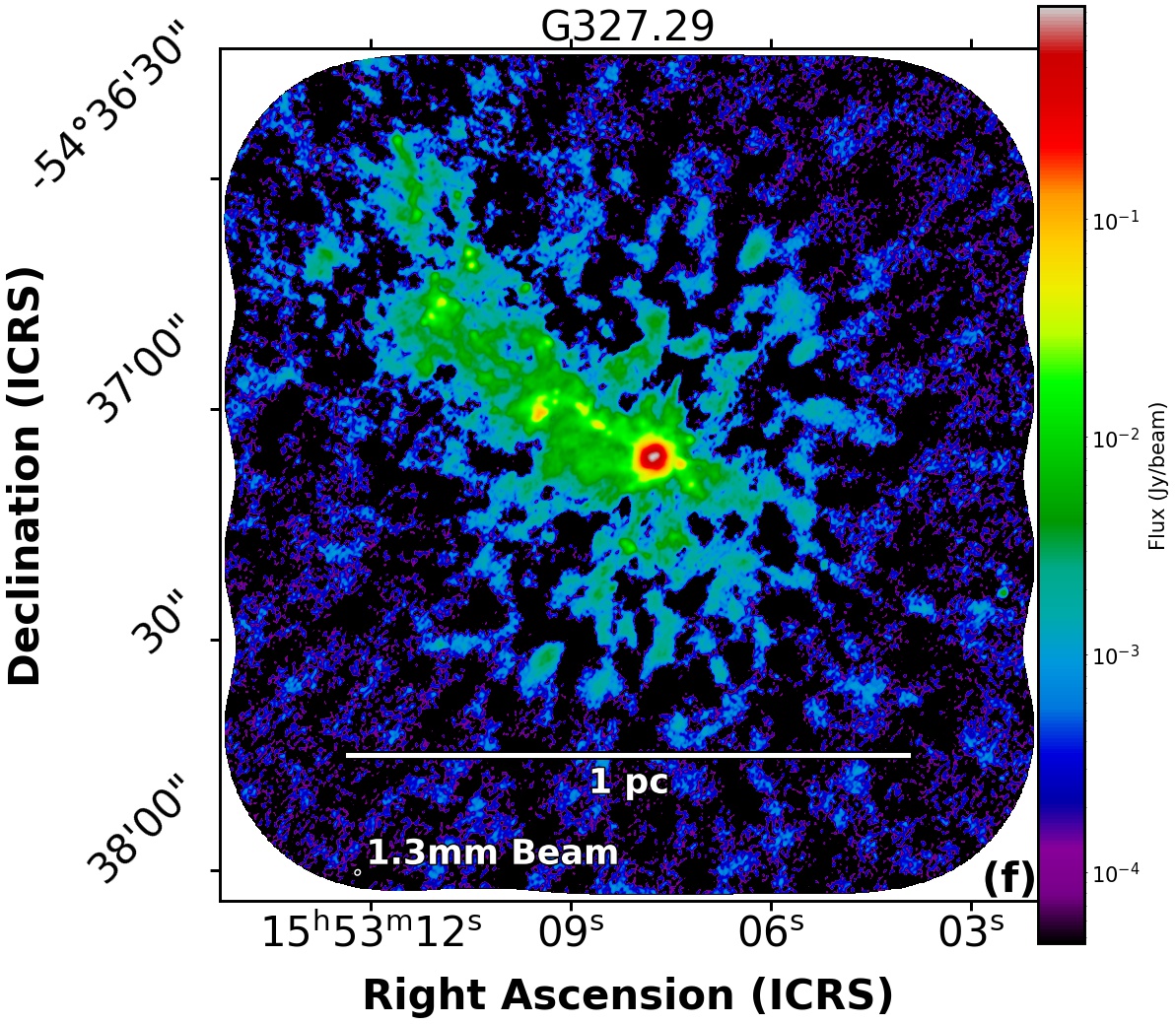}
    \vskip 0.3cm
    \caption{1.3~mm continuum ALMA 12~m array images of the Young clouds: W43-MM1 (in \textsl{a}), W43-MM2 (in \textsl{b}), G338.93 (in \textsl{c}), G328.25 (in \textsl{d}), G337.92 (in \textsl{e}), and G327.29 (in \textsl{f}), plotted at the same physical scale. 
    }
    \label{fig:ALMA1mm}
\end{figure*}

\setcounter{figure}{1}
\begin{figure*}[htbp!]
    \centering
     \vskip 0.2cm
    \includegraphics[width=0.24\textwidth]{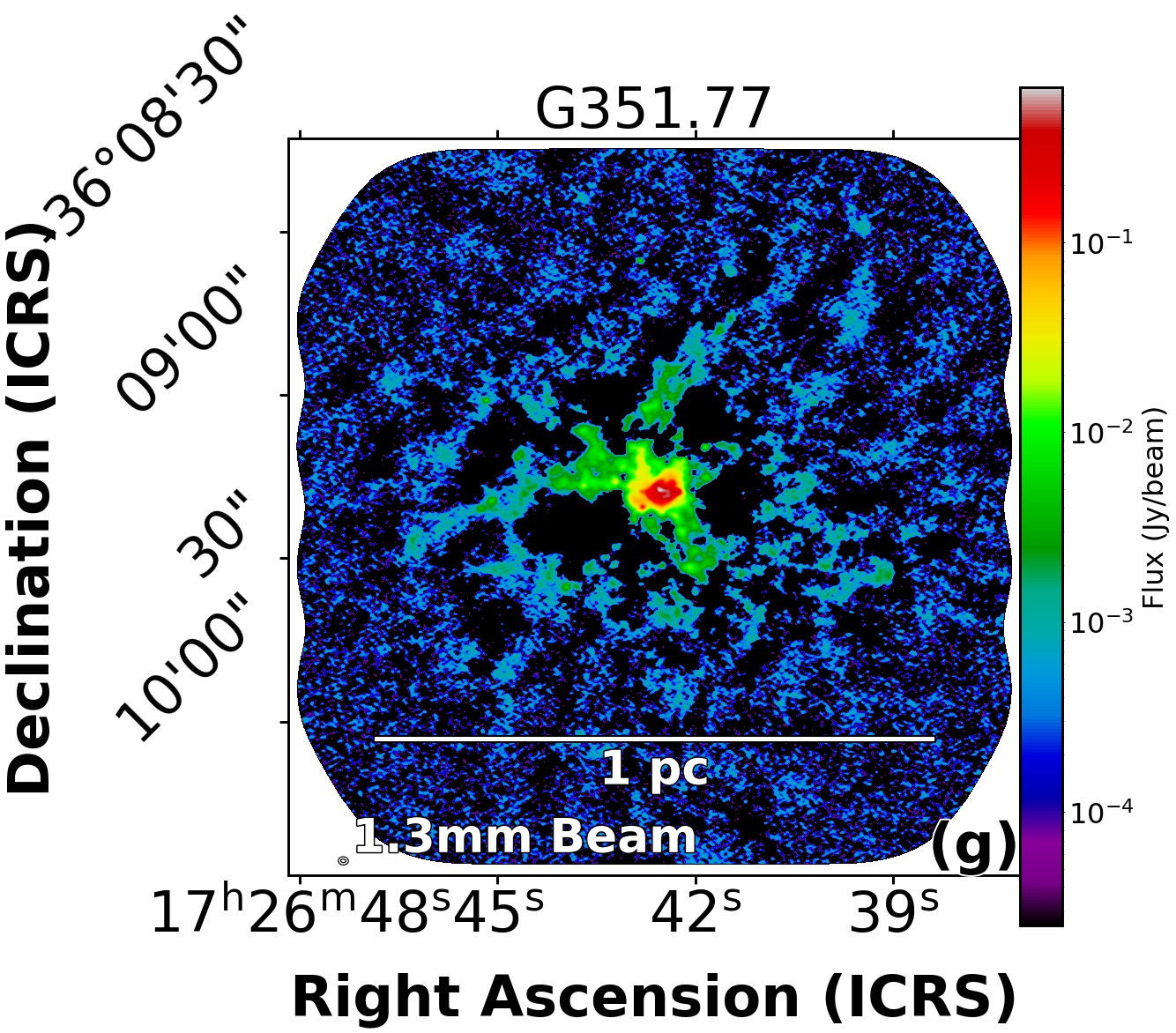}\hskip 1cm\includegraphics[width=0.38\textwidth]{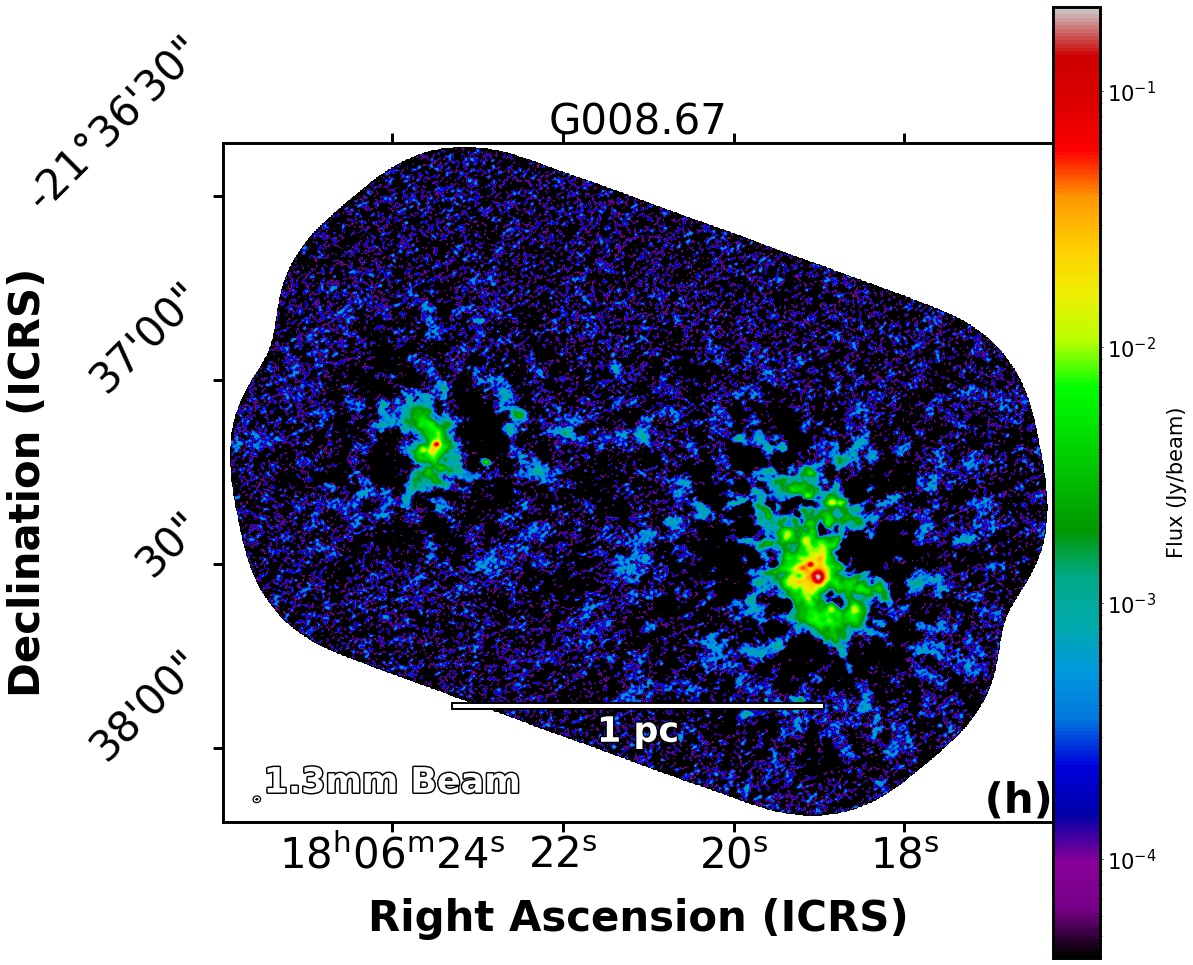}\hskip 1cm\includegraphics[width=0.25\textwidth]{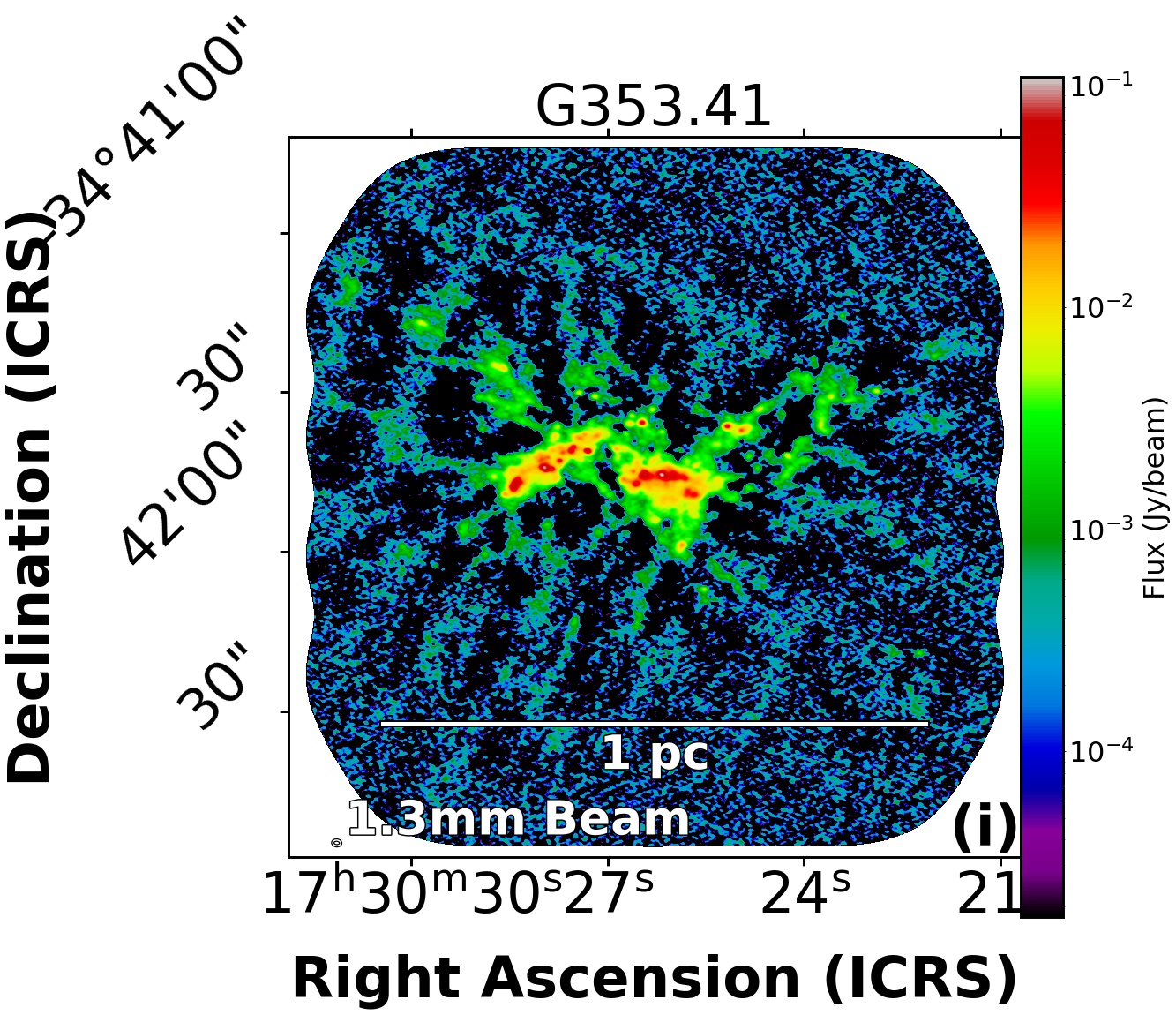}
    \vskip 0.5cm
    \includegraphics[width=0.42\textwidth]{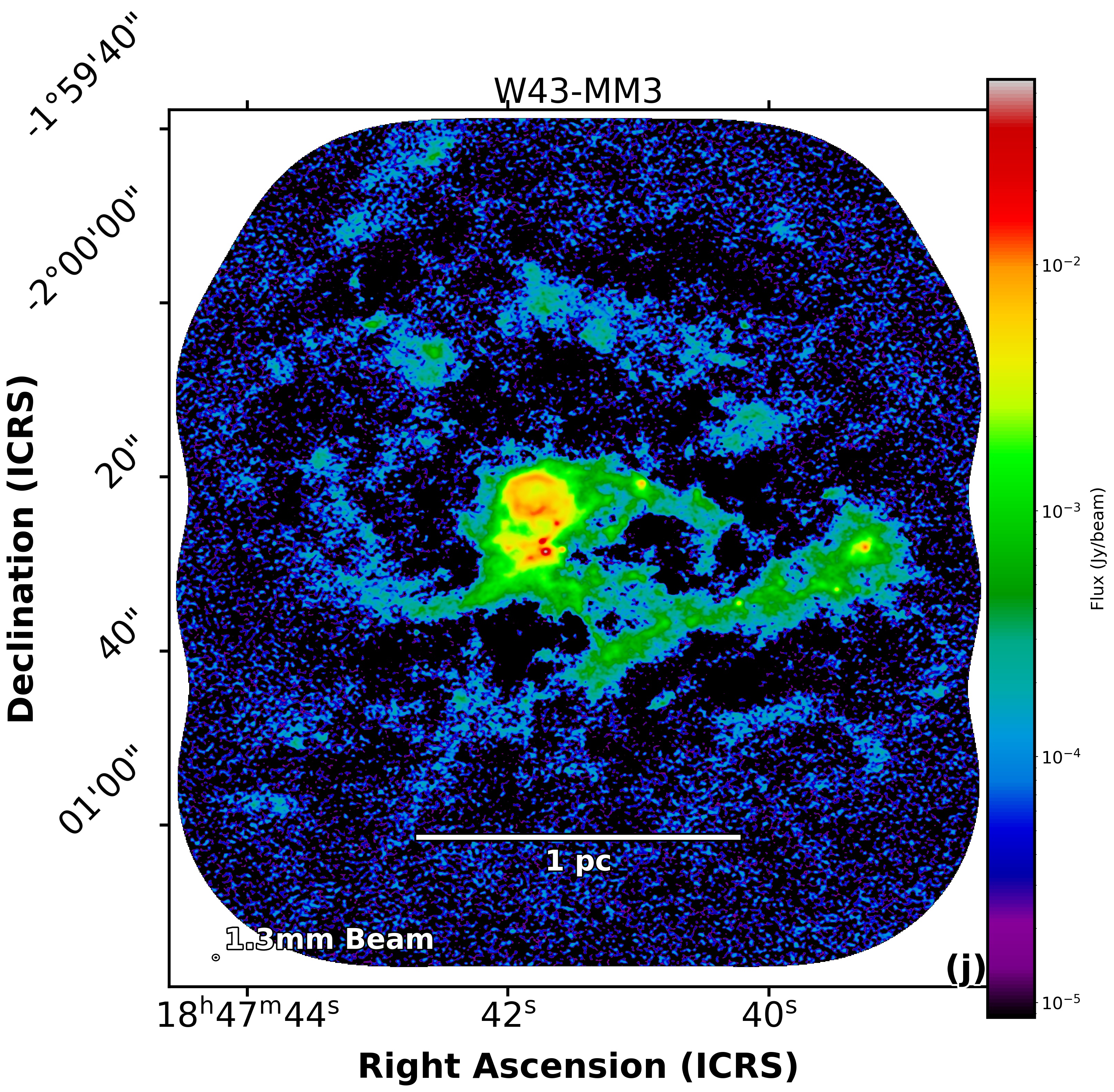}\hskip 1cm\includegraphics[width=0.41\textwidth]{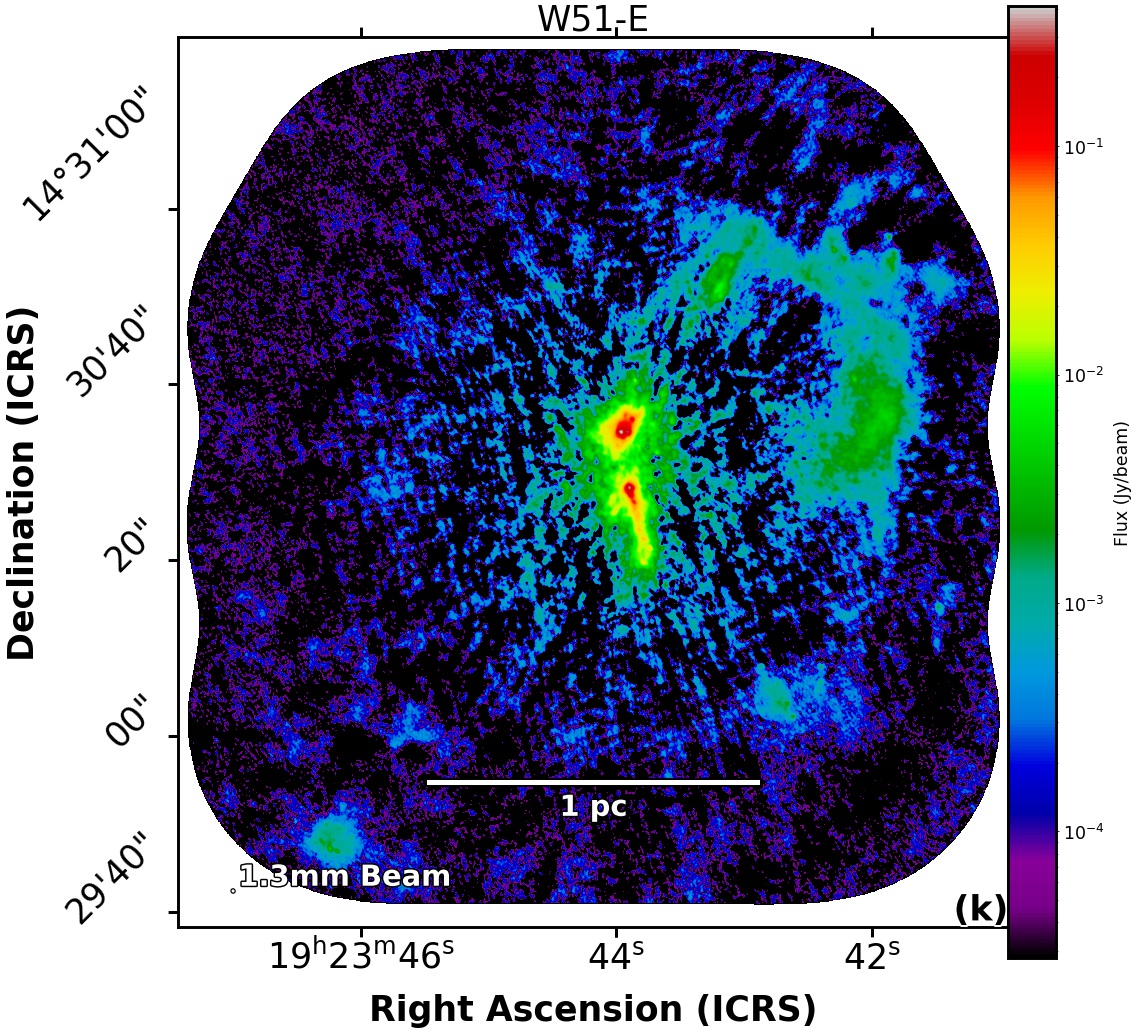}
    \vskip 0.2cm
    \caption{\textbf{(Continued)} 1.3~mm continuum ALMA 12~m array images of the Intermediate clouds: G351.77 (in \textsl{g}), G008.67 (in \textsl{h}), G353.41 (in \textsl{i}), W43-MM3 (in \textsl{j}), and W51-e (in \textsl{k}). 
    }
\end{figure*}

\setcounter{figure}{1}
\begin{figure*}[htbp!]
    \centering
    \vskip 0.2cm
    \includegraphics[width=0.395\textwidth]{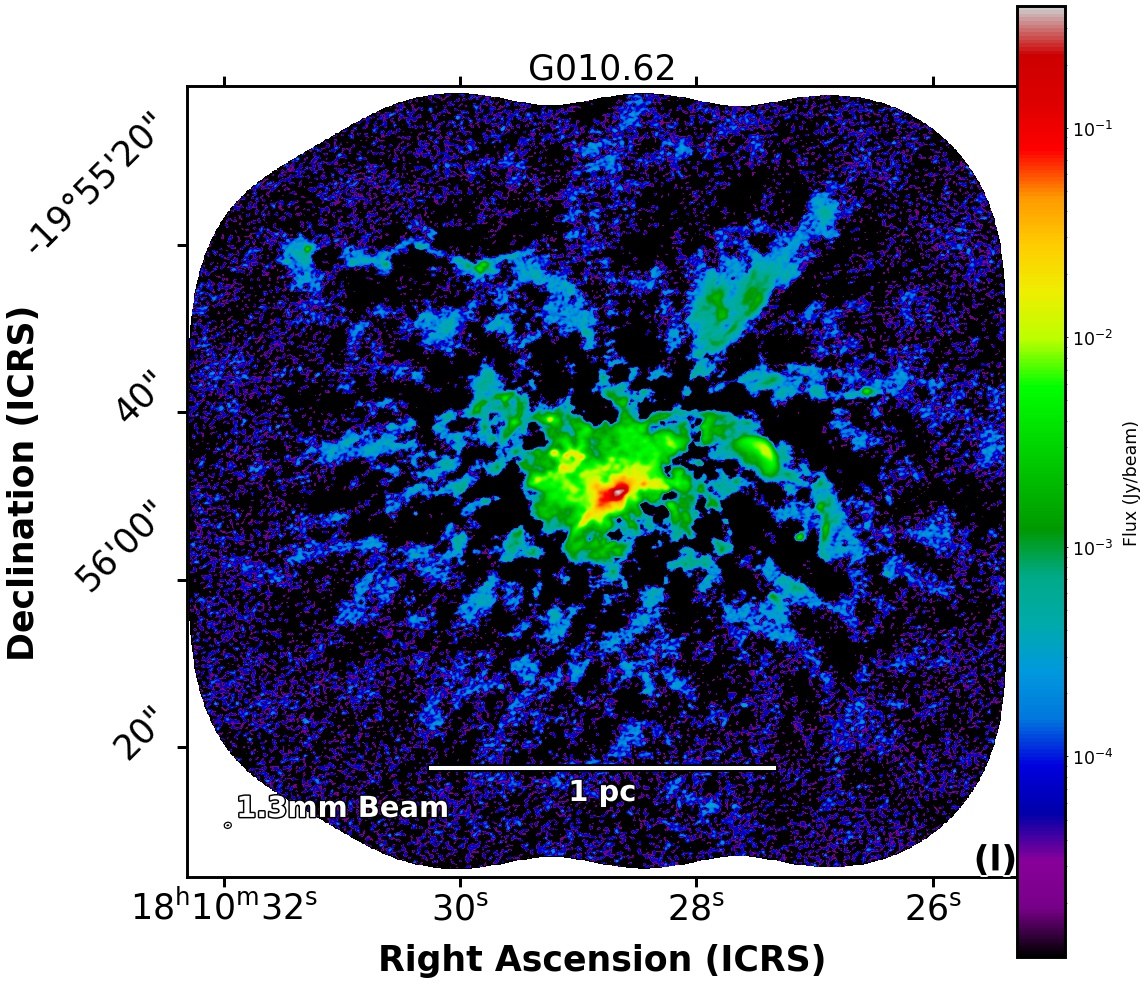}\hskip 1cm\includegraphics[width=0.41\textwidth]{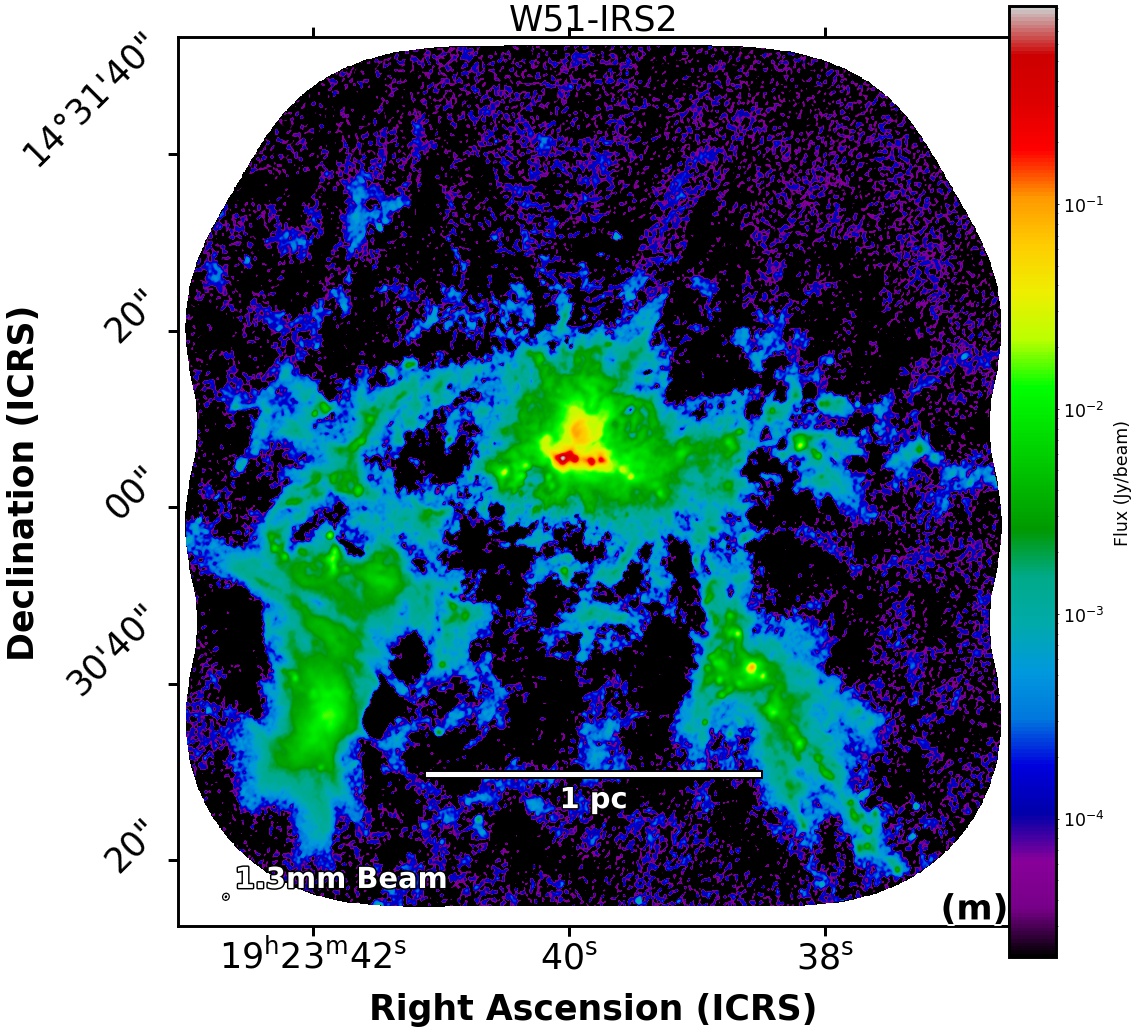}
    \vskip 0.5cm
    \includegraphics[width=0.29\textwidth]{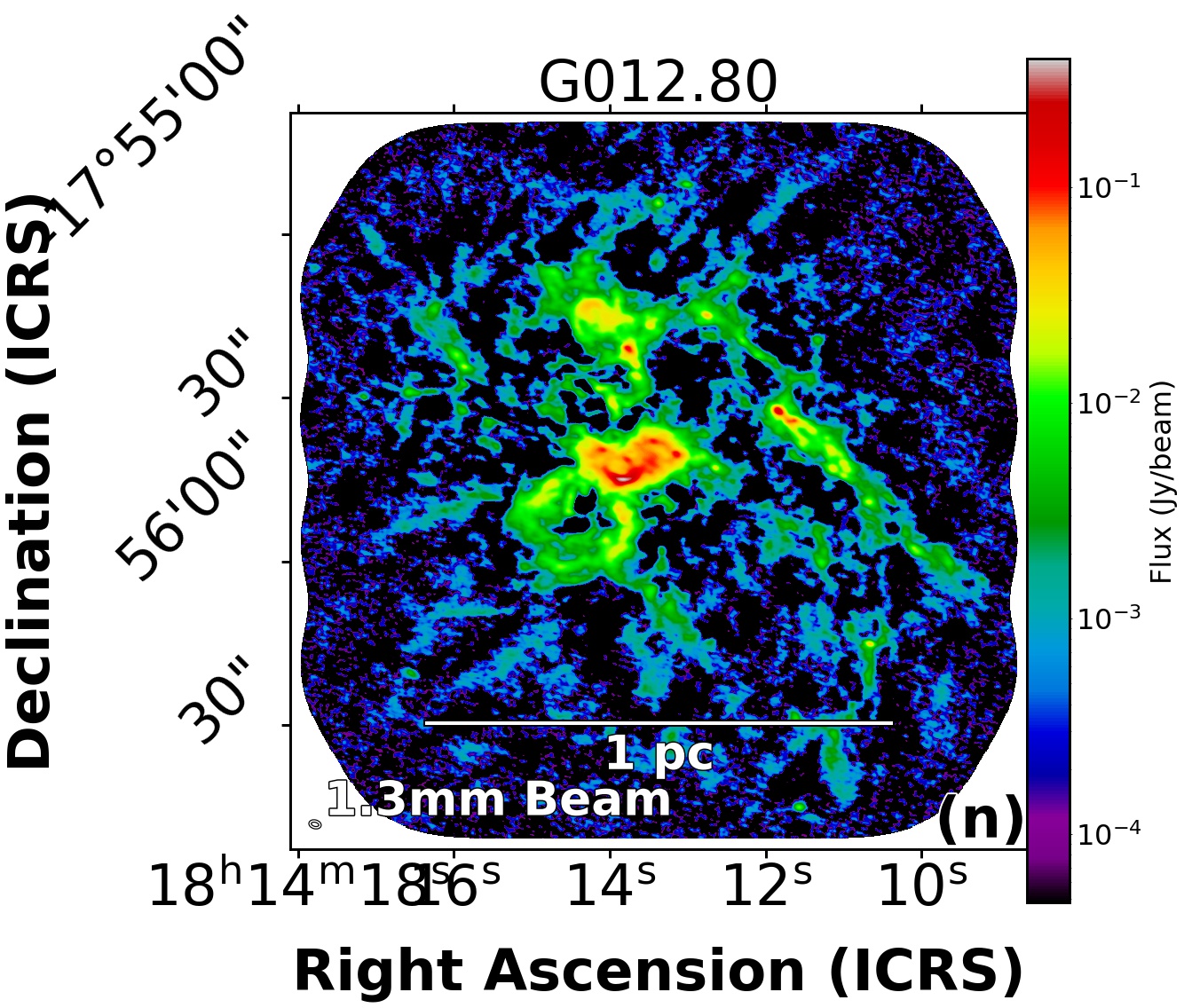} \hskip 1cm\includegraphics[width=0.49\textwidth]{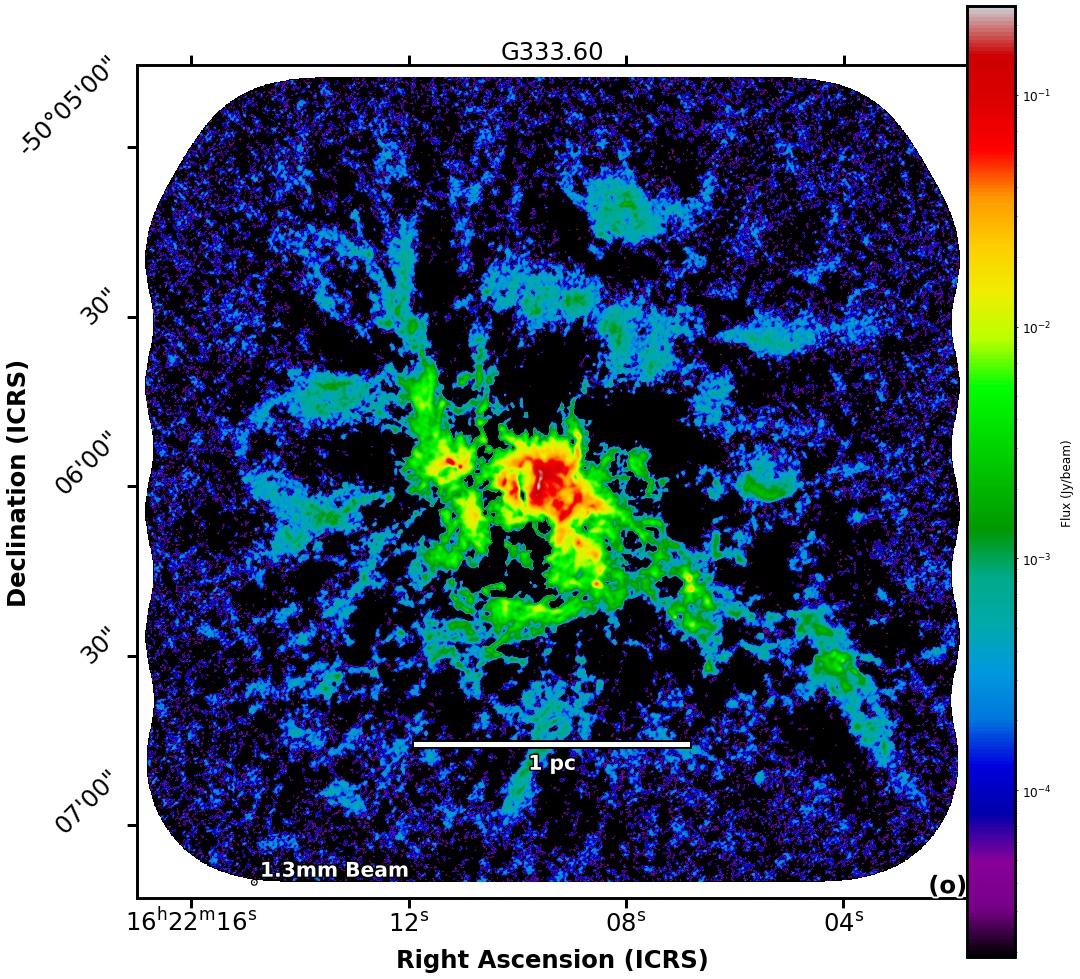}
    \vskip 0.2cm
    \caption{\textbf{(Continued)} 1.3~mm continuum ALMA 12~m array images of the Evolved clouds: G010.62 (in \textsl{l}), W51-IRS2 (in \textsl{m}), G012.80 (in \textsl{n}), and G333.60 (in \textsl{o}). 
    }
\end{figure*}

\subsection{ALMA-IMF data set}
\label{s:dataset}

The ALMA-IMF Large Program was observed from October 2017 to August 2018 with a total observation time of 69~hours and 172~hours with the 12~m and 7~m (ACA) arrays, respectively, and 595~hours with the Total Power Array. In this paper, we concentrate on the 1.3~mm and 3~mm mosaics done with the 12~m array (outlines are shown in \cref{fig:overview}), for the protocluster clouds of \cref{tab:sample}.
\cref{tab:sensitivity_scale_overview} shows a summary of the imaged fields of view, angular resolutions, and sensitivities of the 12~m array continuum images at 1.3~mm and 3~mm. \cref{tab:evol} lists the imaged 1.3~mm and 3~mm areas, which cover the whole extent of the protocluster clouds and their surroundings, respectively. The 1.3~mm areas were defined to systematically cover a 1\,pc~$\times$~1\,pc area around the targeted ATLASGAL clumps of \cref{tab:sample}, with larger areas for W43 and W51 clumps. The 3~mm areas aim to cover the filaments converging toward the ATLASGAL clumps (see \cref{fig:overview}).
On average, these 15 massive protoclusters cover $\sim$3.5$\pm{2.0}~\pc^2$ each, and sum up to $\sim $53$~\pc^2$ (see \cref{tab:evol}).

The ALMA-IMF consortium built a pipeline
that allows a homogeneous, repeatable, and high-quality reduction of its data set, starting with the 12~m array continuum mosaics \citep{ginsburg22}. 
Two sets of continuum images have been produced for each protocluster cloud and band. The first set of continuum images, called  \bsens, are derived using all observed spectral channels of the ALMA band 6 (1.3~mm or 228.965~GHz with a spectral index of $\alpha=3.5$) or band 3 (3~mm or 100.713~GHz with $\alpha=3.5$), summing up to 3.7~GHz and 2.9~GHz, respectively. The second set, constituting the \cleanest continuum images, is built using only the line-free channels
\citep[see][their Figs.~3--4]{ginsburg22}. 
The  \bsens continuum images are up to two times more sensitive than the \cleanest maps (see \cref{tab:sensitivity_scale_overview}) but their continuum emission is contaminated by line emission. 
The 1.3~mm  \bsens images allow the detection of point-like cores down to the $3\,\sigma$ sensitivity level of $0.09~\msun$ to $0.5~\msun$, with a median of $\sim$0.18$~\msun$,
corresponding to a point-source sensitivity of $\sim$7$\times10^{22}$~cm$^{-2}$ in column density.
Figure~\ref{fig:ALMA1mm} presents the 1.3~mm images of the 15 ALMA-IMF clouds obtained from the 12~m array \bsens data set of the ALMA-IMF Large Program.
The 12~m array \bsens and \cleanest continuum images at 1.3~mm and 3~mm are provided to the community, along with the present paper and Paper II.
The ALMA-IMF data processing pipelines and analysis are made public as described in \citet{ginsburg22}.
The code is available on the github repository, \url{https://github.com/ALMA-IMF/reduction},
and ongoing work and data release updates can be found there and on the  ALMA-IMF website (\url{https://almaimf.com}).

The angular resolutions achieved by the 12~m array ALMA continuum images at 1.3~mm and 3~mm, using the Briggs robust parameter \texttt{robust}=0, 
are within 30\% of those requested, with the exception of a couple of outliers
at 3~mm (see \cref{tab:sensitivity_scale_overview} 
of \citealt{ginsburg22}).
Taking the distances of the ALMA-IMF clouds from \cref{tab:sample}, we computed the linear resolutions of the  1.3~mm and 3~mm  \bsens images. Listed in \cref{tab:concentration}, the 1.3~mm spatial resolution ranges from 1\,350~au to 2\,690~au, with a median value and standard deviation of $2\,100\pm{400}$~au. As for the 3~mm images, they have on average a slightly better spatial resolution of $1\,800\pm{500}$~au. The dynamic range in angular scales (i.e., $DR = \theta_{\rm LAS}/\theta_{\rm beam}$, where $\theta_{\rm LAS}$ is the largest angular scale) ranges from $DR=6$ to 14 at 1.3~mm and $DR=7$ to 34 at 3~mm.
The largest angular scale, also called maximum recoverable scale, spans ranges of $4.6\arcsec-6.6\arcsec$ at 1.3~mm and $4.8\arcsec-11\arcsec$ at 3~mm (see \cref{tab:sensitivity_scale_overview}). They correspond to linear scales with mean values and $1\,\sigma$ dispersions of $\sim$0.1~pc and $\pm0.03$~pc at 1.3~mm and $\sim$0.16~pc and $\pm0.05$~pc at 3~mm, respectively, with maximum variation factors of $2.7-2.8$. The ratio of the 3~mm to 1.3~mm largest angular scales has a mean value of 1.7, close to the inverse ratio of the observed frequencies.

The line data cubes were processed within the framework of the ALMA-IMF data pipeline.
The different 12~m array configurations were combined following the same procedure as for the continuum data, but cleaning and imaging is adapted to the line data cubes. They however first need to be corrected for the system temperature and the spectral data normalization\footnote{
    ALMA ticket: \url{https://help.almascience.org/kb/articles/607}, \url{https://almascience.nao.ac.jp/news/amplitude-calibration-issue-affecting-some-alma-data}} \citep[see also Sect. ~2 of][]{ginsburg22}. 
The ALMA-IMF data were indeed affected by a systematic error in the spectral data normalization and returned to the Joint ALMA Observatory (JAO) for further processing in November 2020. Any data downloaded from the archive before this time are therefore affected by these issues. When processing is complete, line data cubes will be used to discuss the ionized component of the ALMA-IMF clouds, the cloud kinematics, outflows, and chemical enrichment. The sensitivities measured in the preliminary data cubes used here are $1\,\sigma=0.6-1$~K at 1.3~mm and  $1\,\sigma=1.4-3.0$~K
at 3~mm with a $1~\kms$ resolution. 

As shown in \cite{ginsburg22}, the products combining ALMA 12~m array data with 7~m array data are of inconsistent quality across the ALMA-IMF sample. For several of the target fields, incorporating the 7~m array data resulted in increased noise levels and/or imaging artifacts. The increased noise levels is particularly problematic for source extraction on the $\sim$2\,100~au scales. The 7~m array data are therefore not used for the present analysis.

\setcounter{figure}{2}

\begin{figure*}[htbp!]
    \centering
    \includegraphics[width=0.54\textwidth]{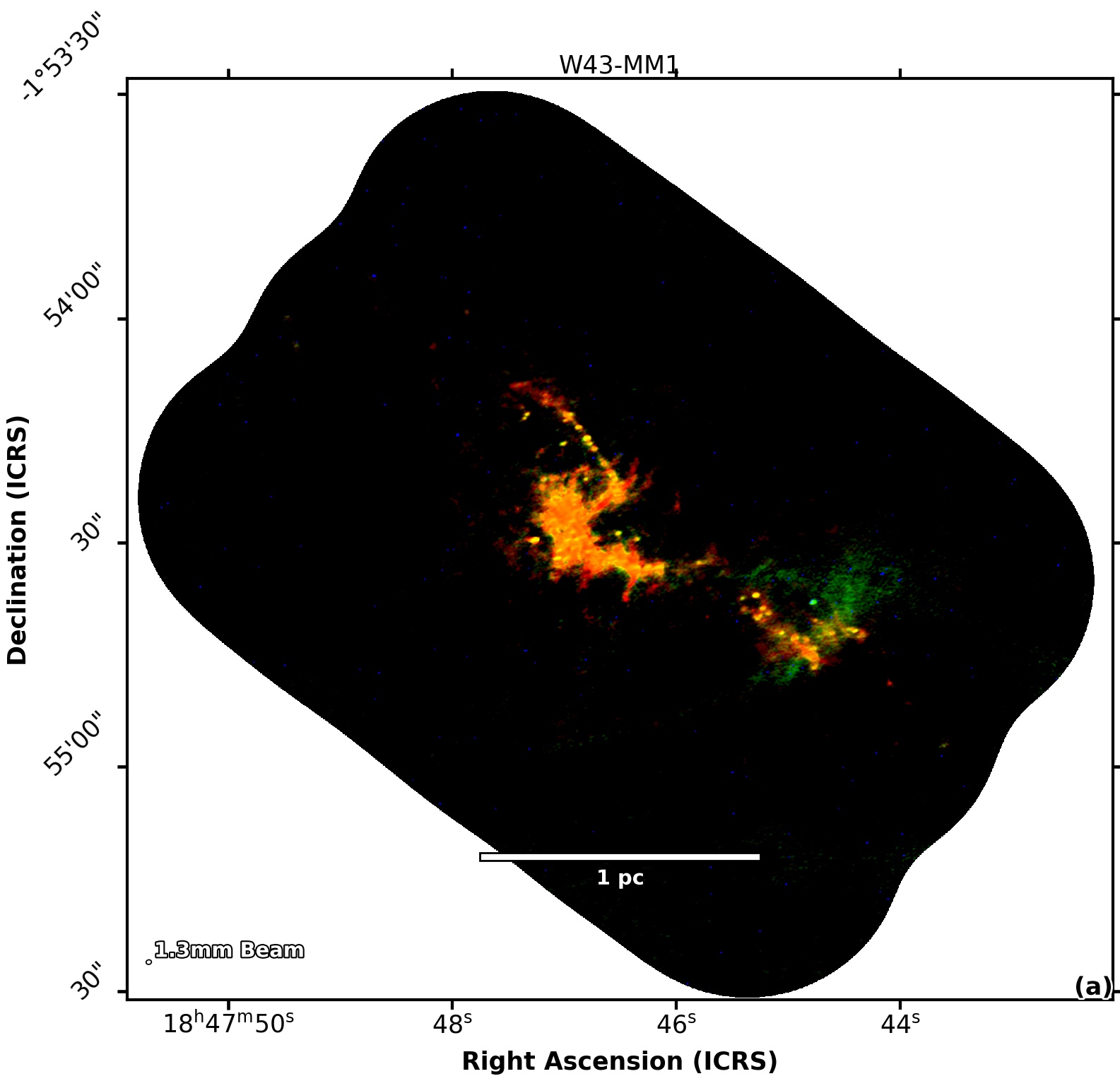}\hskip 0.3cm\includegraphics[width=0.41\textwidth]{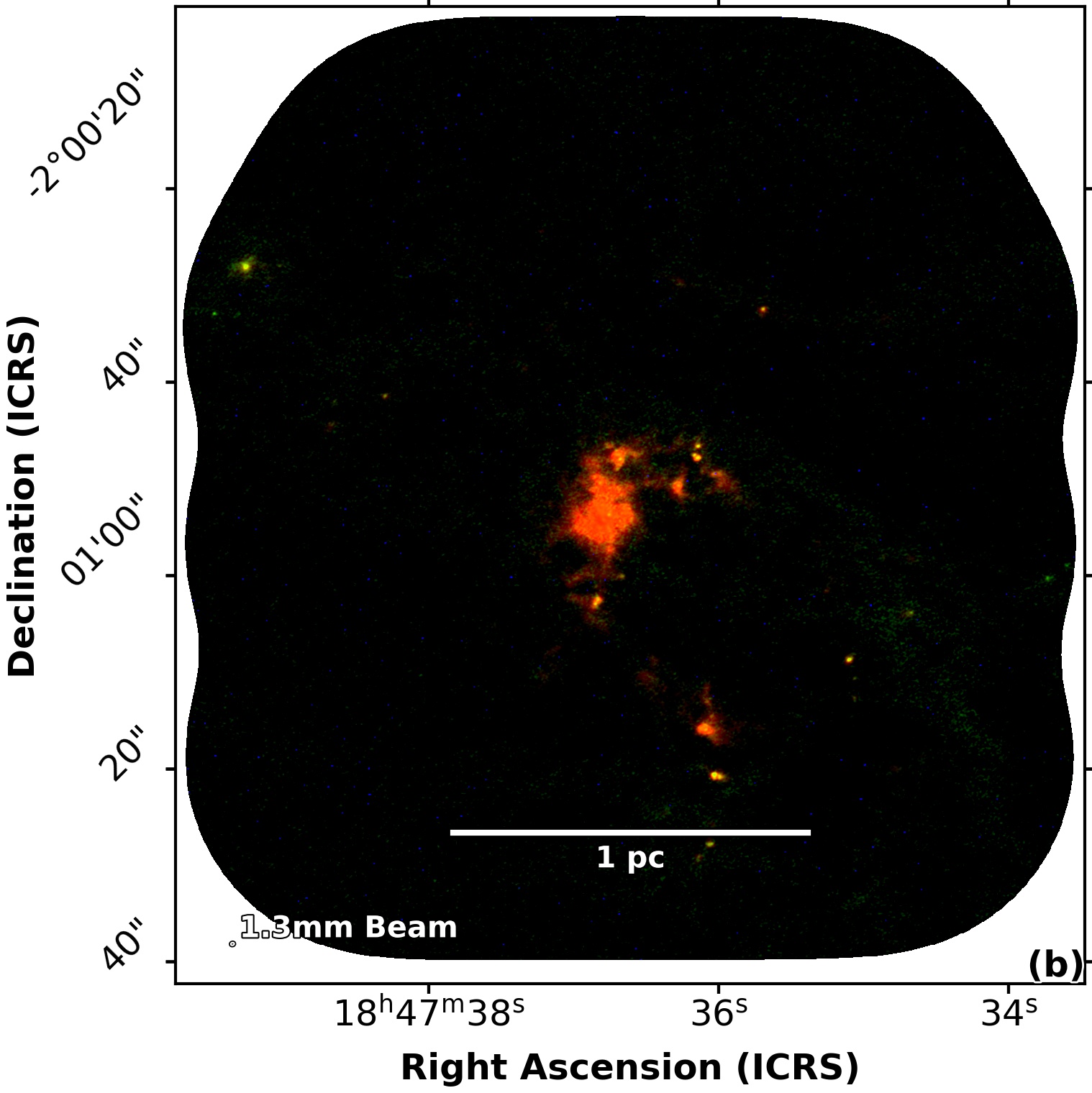}
   \vskip 0.5cm
    \includegraphics[width=0.295\textwidth]{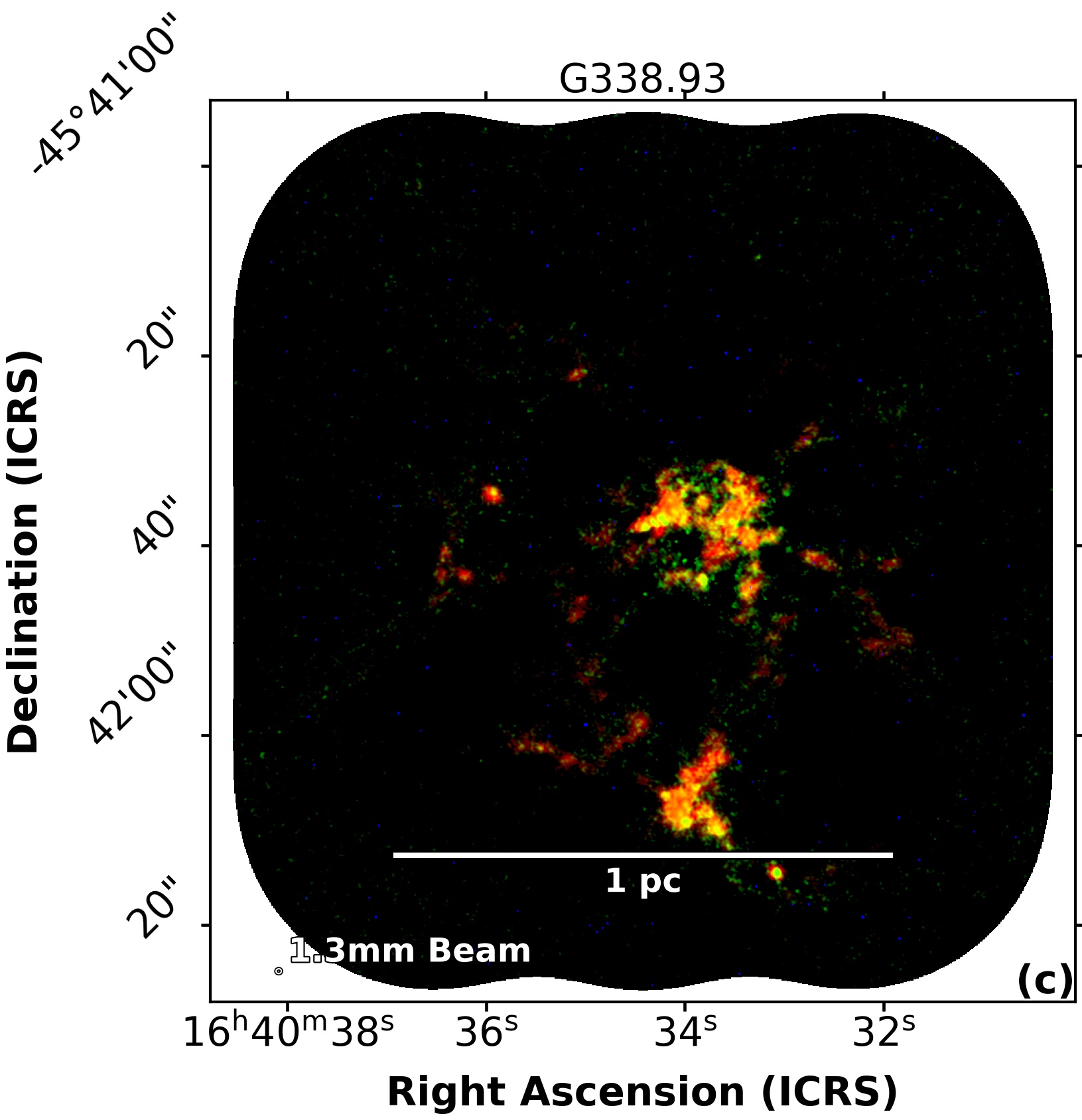}\hskip 1cm\includegraphics[width=0.28\textwidth]{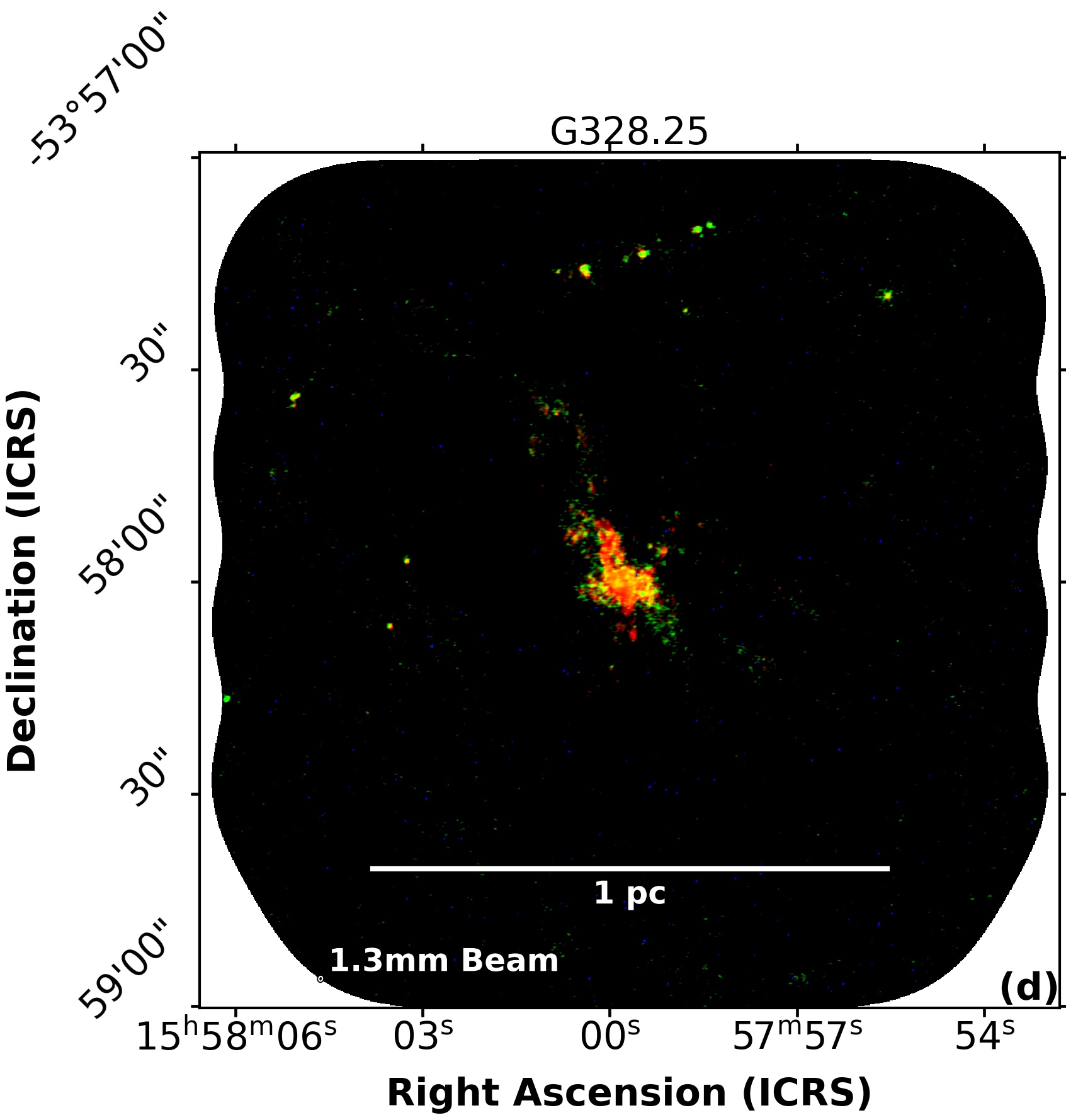}\\
    \vskip 0.5cm
    \includegraphics[width=0.28\textwidth]{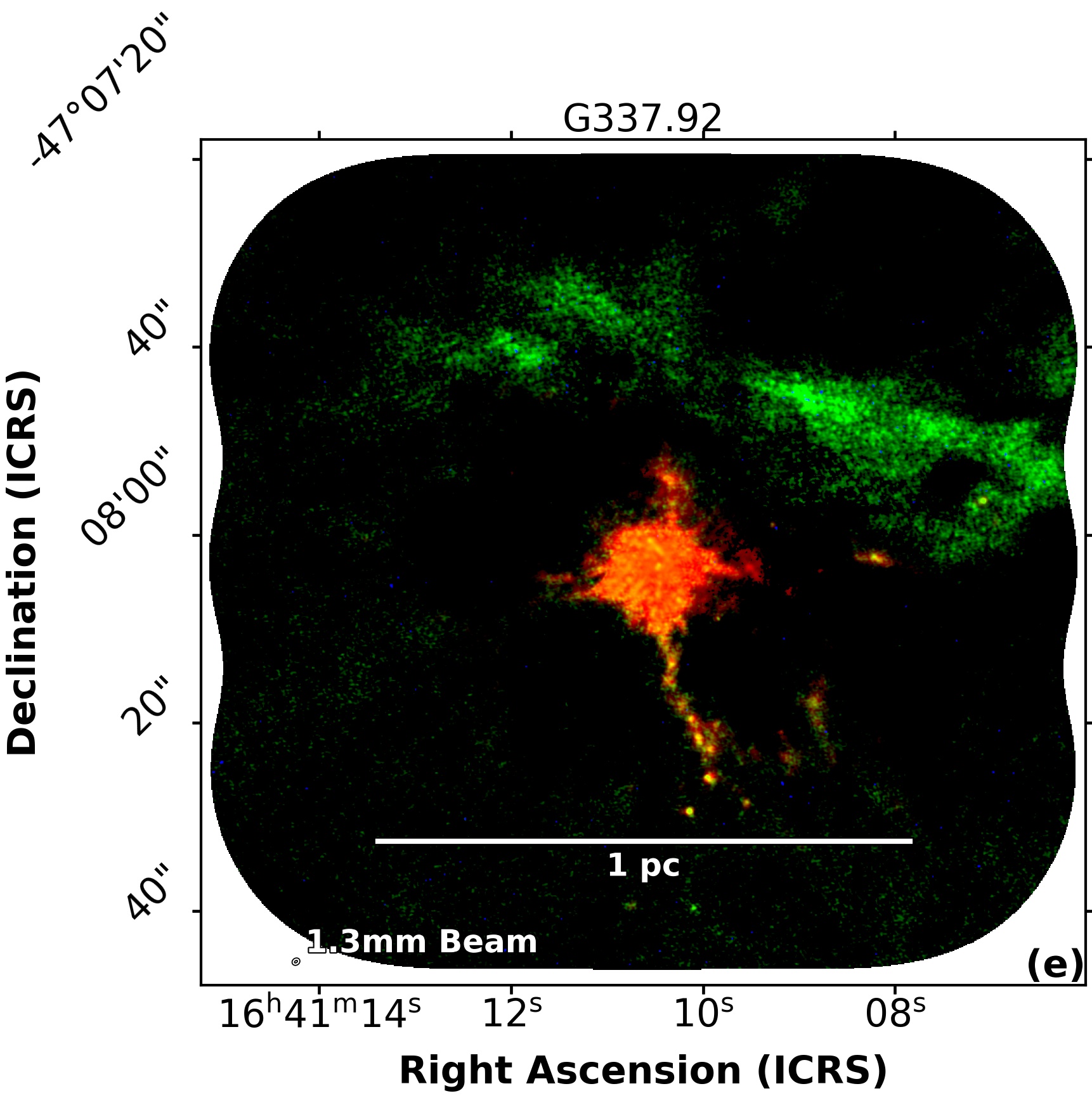}\hskip 1cm\includegraphics[width=0.245\textwidth]{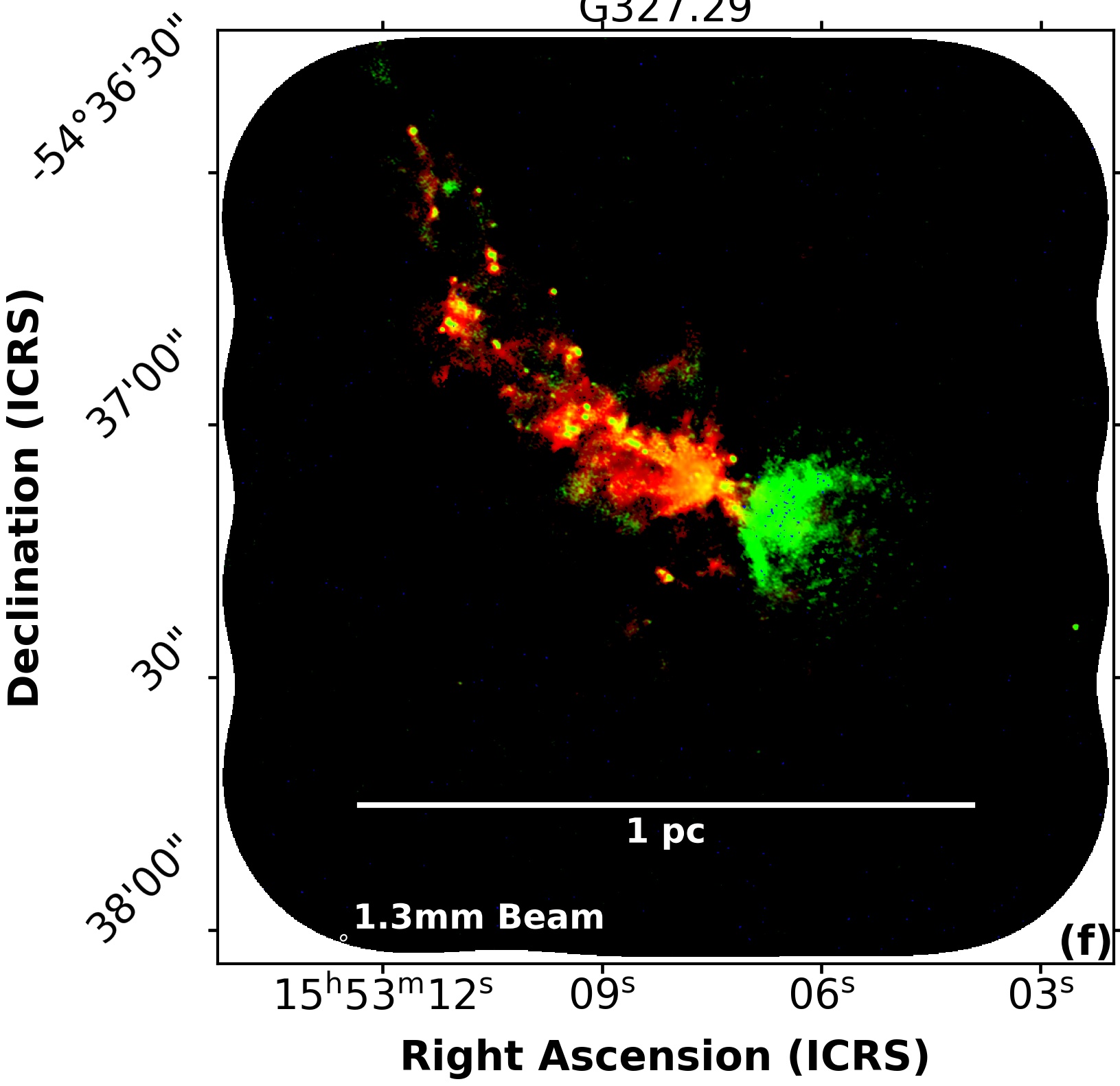}
    \vskip 0.3cm
    \caption{Three-color ALMA 12~m array images of the Young clouds: W43-MM1 (in \textsl{a}), W43-MM2 (in \textsl{b}), G338.93 (in \textsl{c}), G328.25 (in \textsl{d}), G337.92 (in \textsl{e}), and G327.29 (in \textsl{f}), plotted at the same physical scale. Red and green display the continuum images at 1.3~mm and 3~mm, respectively, with the green color for the longer, 3~mm, wavelength because the 3~mm emission is partly associated with hot ionized gas.
    Blue corresponds to the free-free emission at the frequency of the H$41\alpha$ recombination line. Thermal dust emission of filaments and cores is shown in orange shades, while diffuse green features locate weak free-free emission from faint \hii regions in \textsl{e} and \textsl{f}.
    }
    \label{fig:3col}
\end{figure*}

\setcounter{figure}{2}
\begin{figure*}[htbp!]
    \centering
     \vskip 0.2cm
    \includegraphics[width=0.24\textwidth]{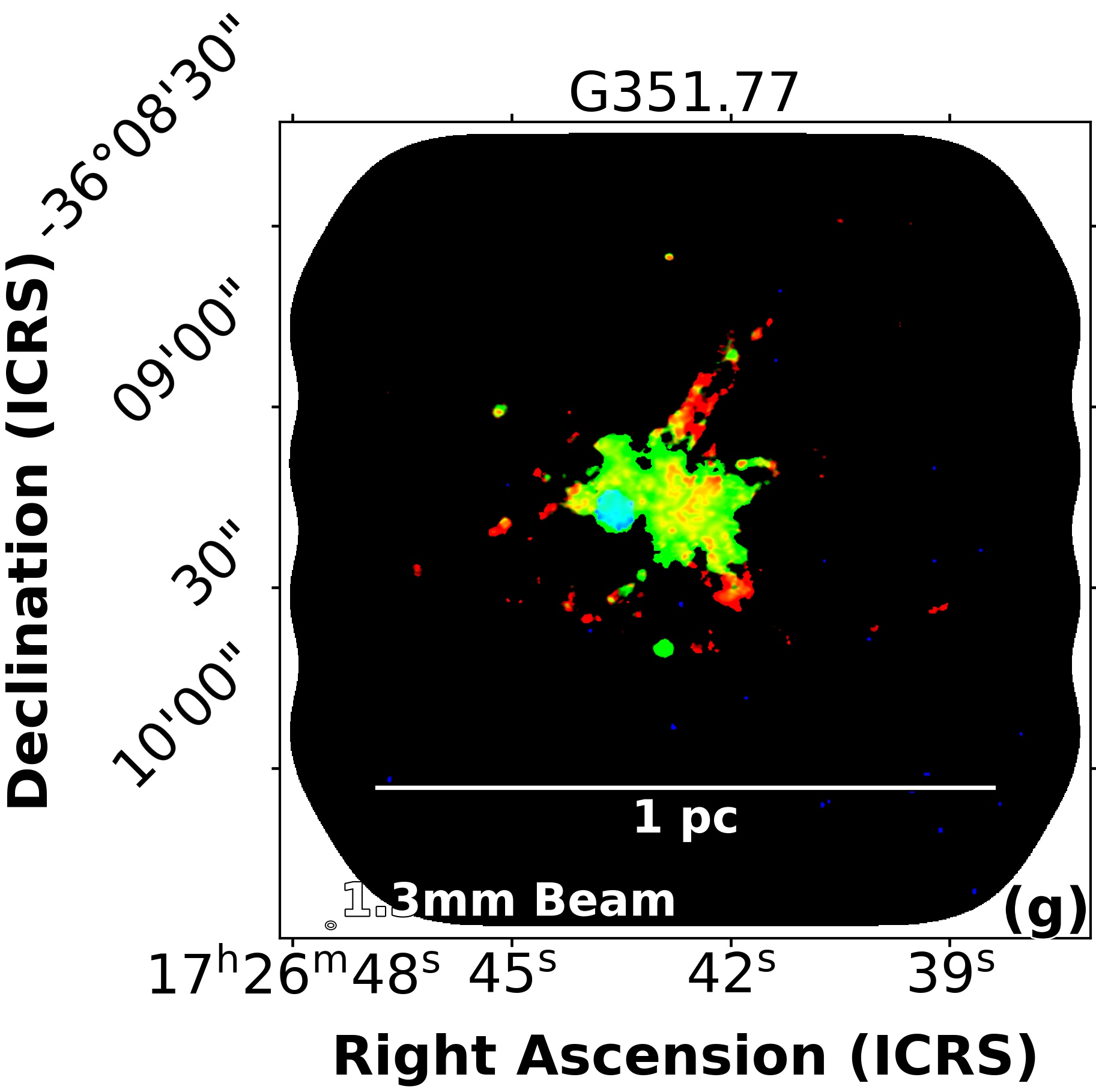}\hskip 1cm\includegraphics[width=0.38\textwidth]{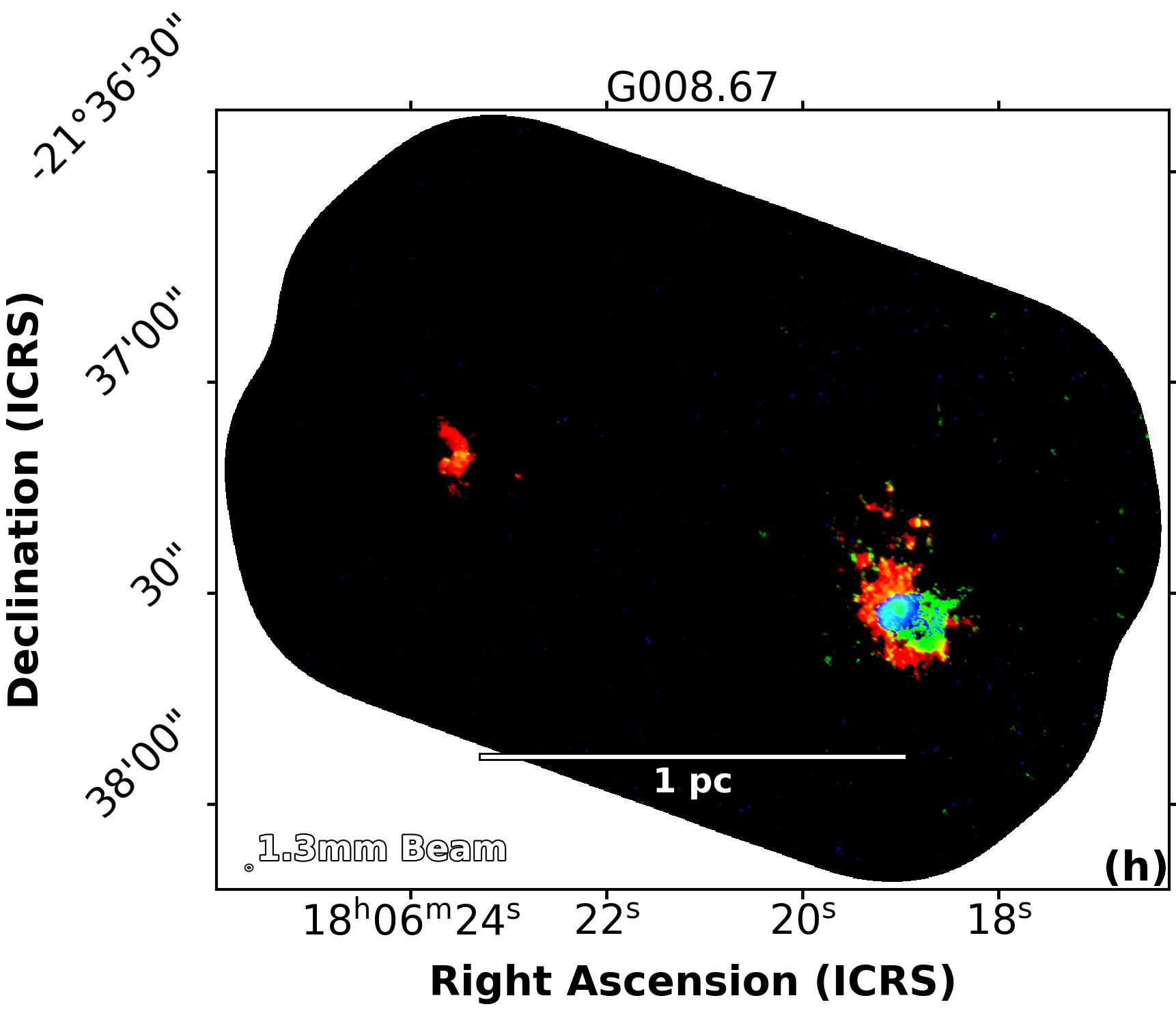}\hskip 1cm\includegraphics[width=0.25\textwidth]{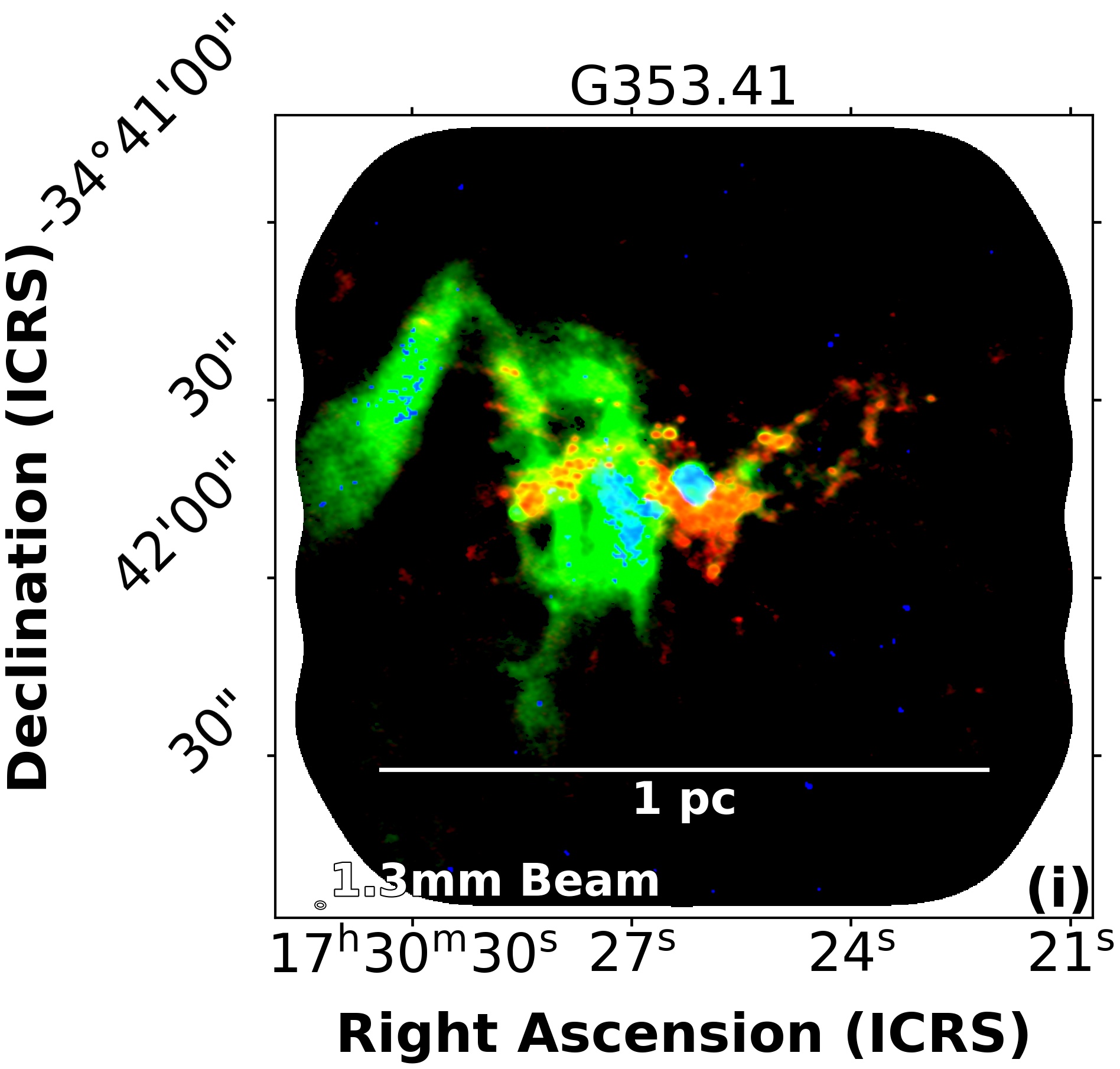}
    \vskip 0.5cm
    \includegraphics[width=0.42\textwidth]{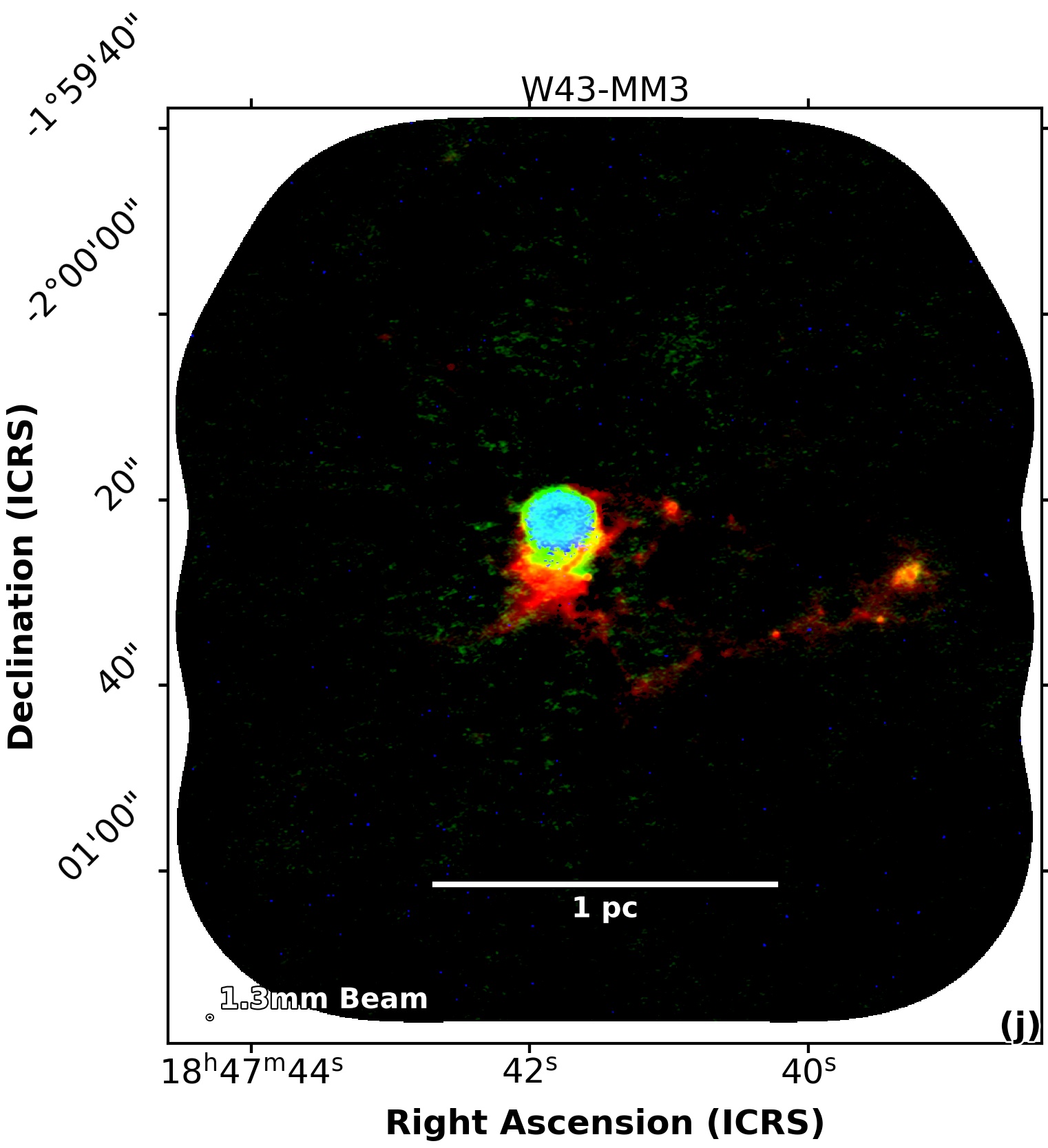}\hskip 1cm\includegraphics[width=0.41\textwidth]{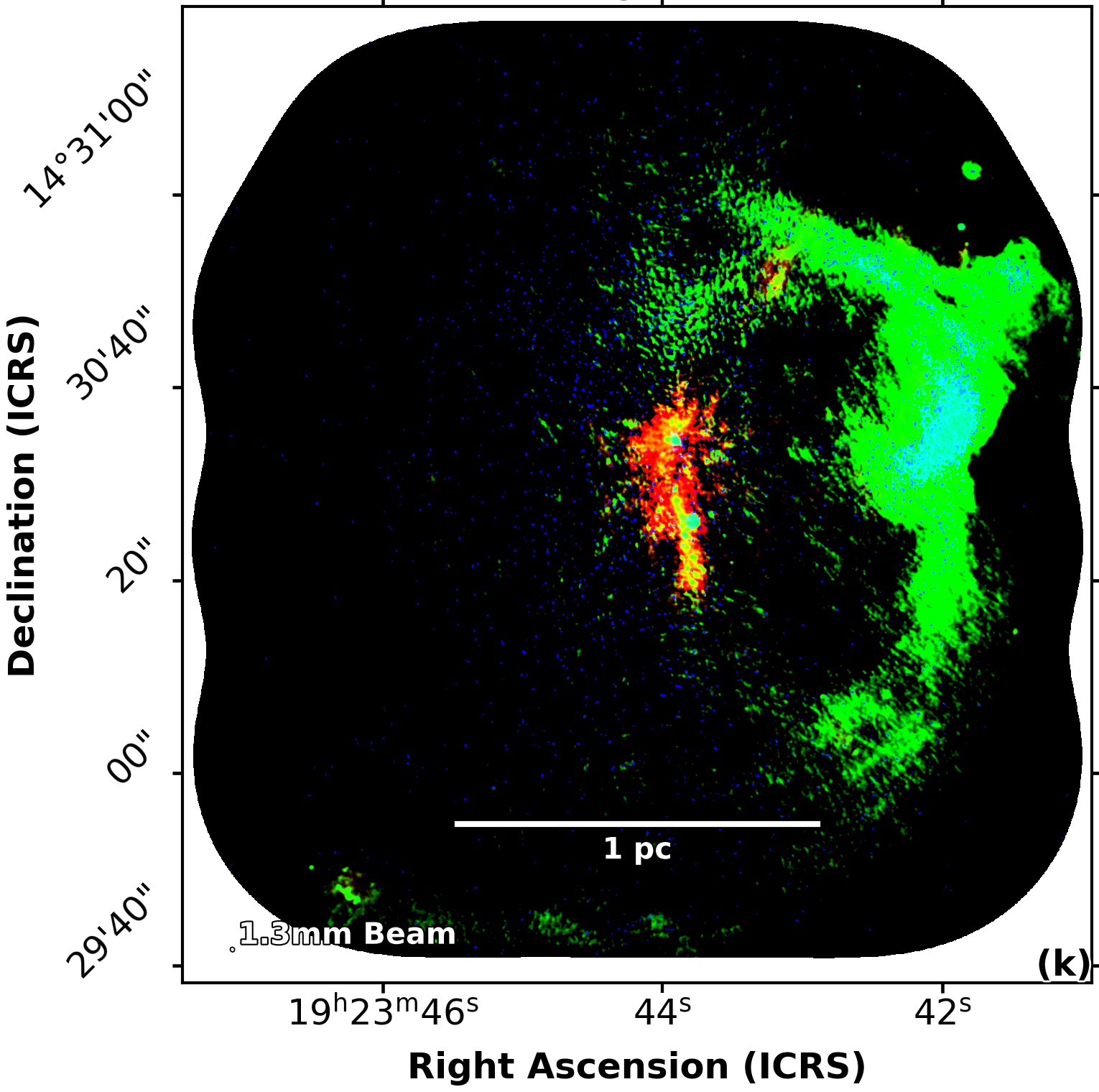}
    \vskip 0.2cm
    \caption{\textbf{(Continued)} Three-color ALMA  12~m array images of the Intermediate clouds: G351.77 (in \textsl{g}), G008.67 (in \textsl{h}), G353.41 (in \textsl{i}), W43-MM3 (in \textsl{j}), and W51-e (in \textsl{k}). The coincidence of blue and green emission locates six HC or UC\hii region bubbles (in \textsl{g--k}) and diffuse arc-like emission of compact to extended \hii regions (in \textsl{i} and \textsl{k}).}
\end{figure*}

\setcounter{figure}{2}
\begin{figure*}[htbp!]
    \centering
    \vskip 0.2cm
    \includegraphics[width=0.395\textwidth]{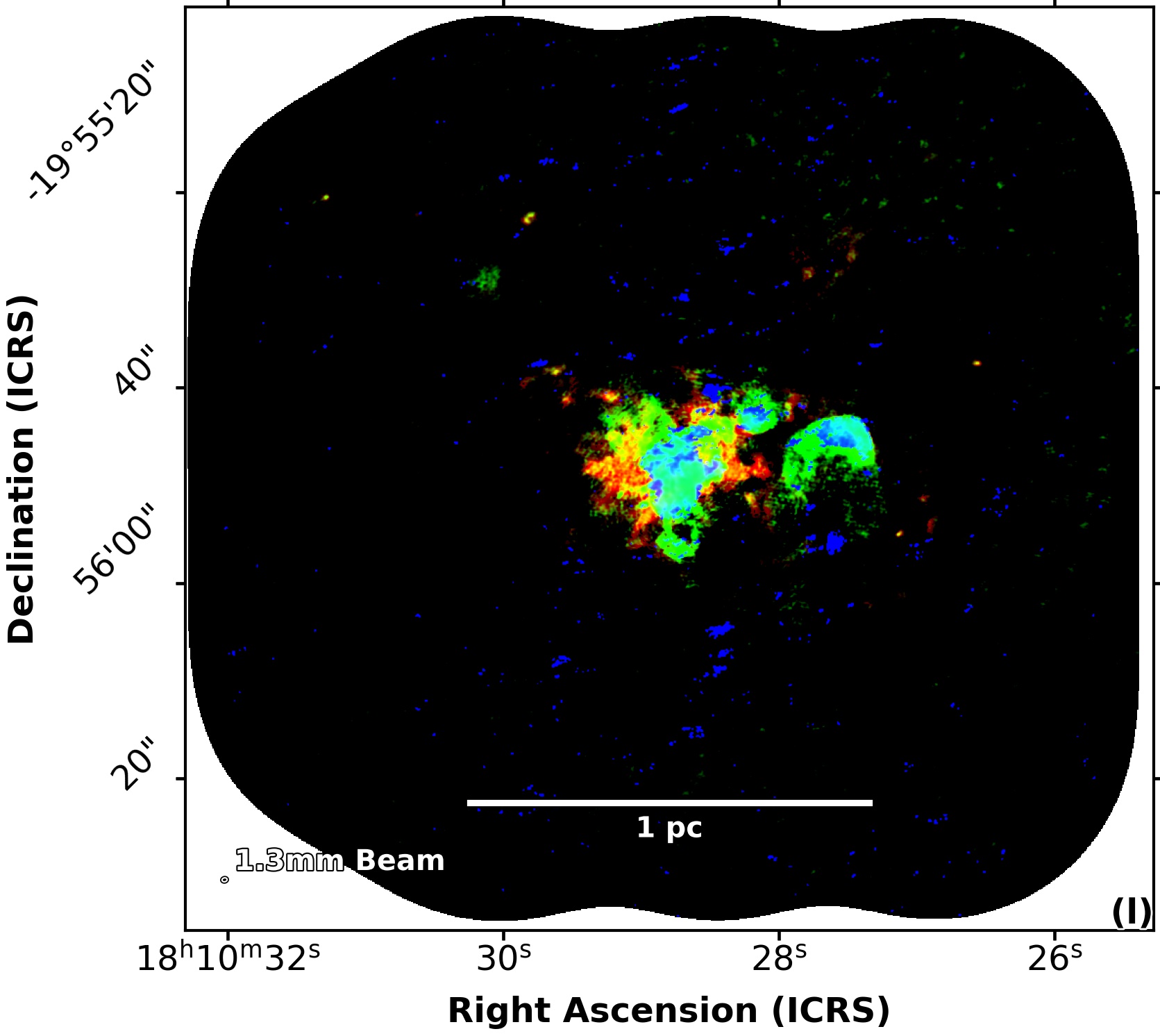}\hskip 1cm\includegraphics[width=0.41\textwidth]{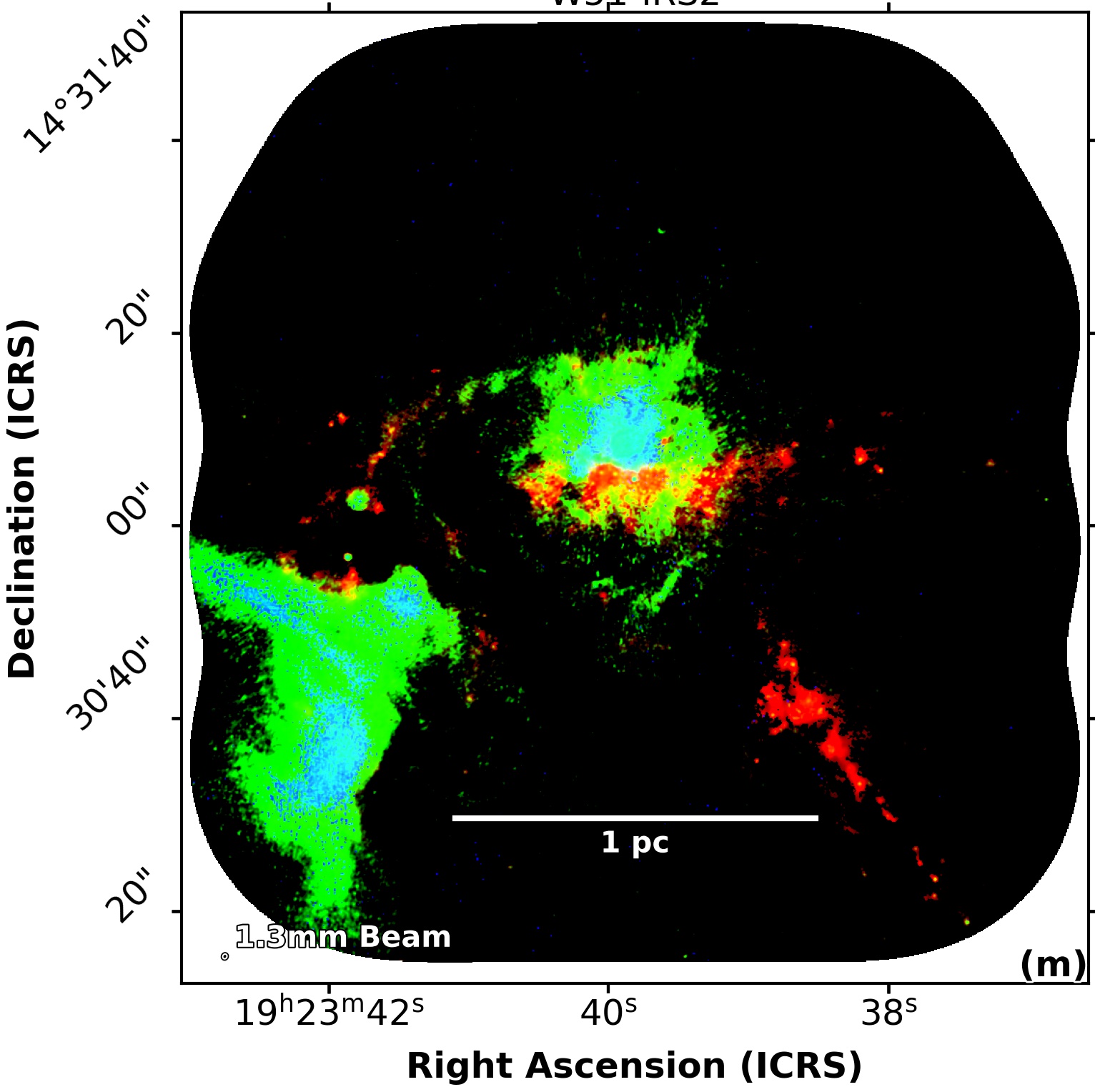}
    \vskip 0.5cm
    \includegraphics[width=0.29\textwidth]{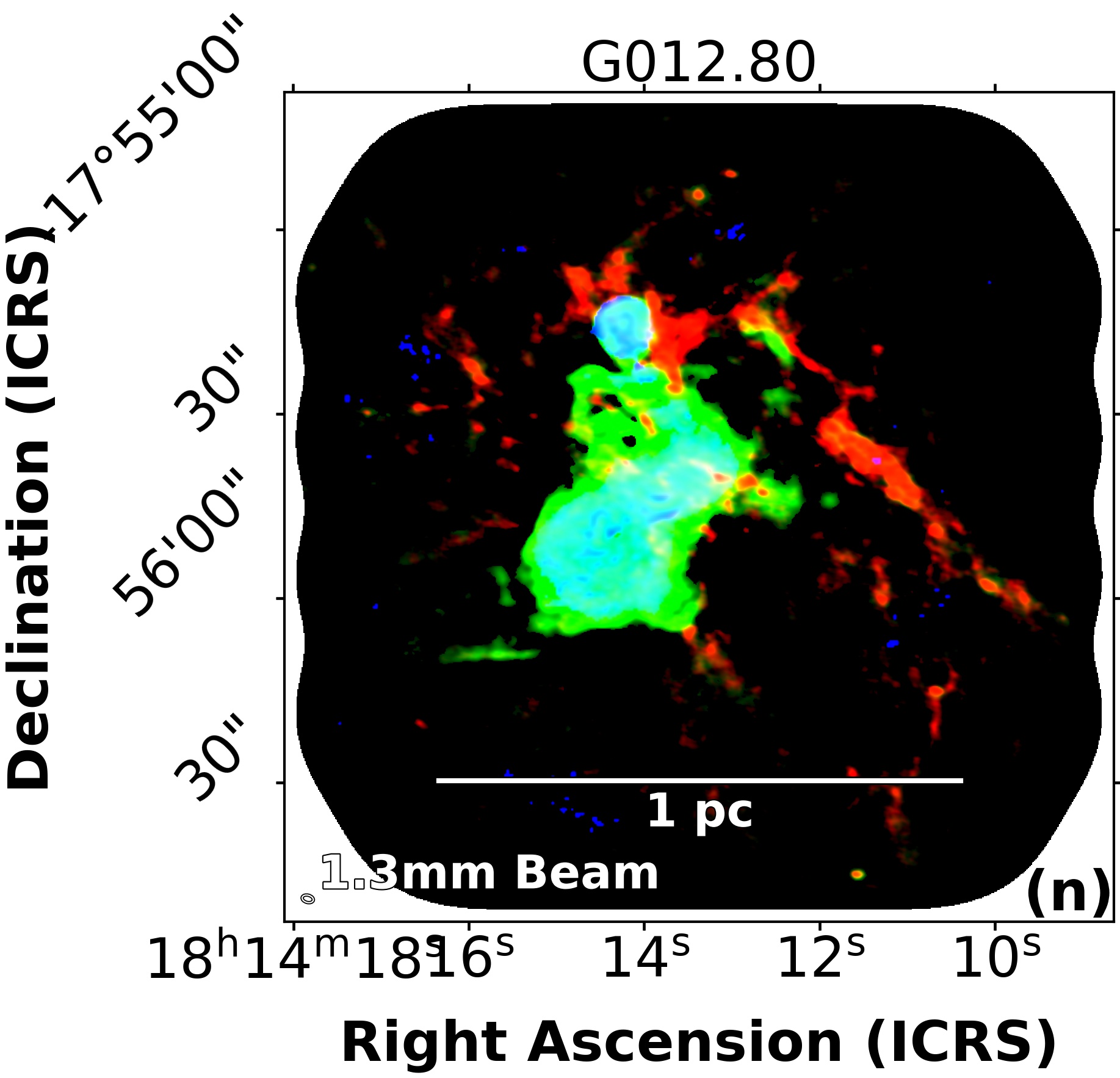} \hskip 1cm\includegraphics[width=0.49\textwidth]{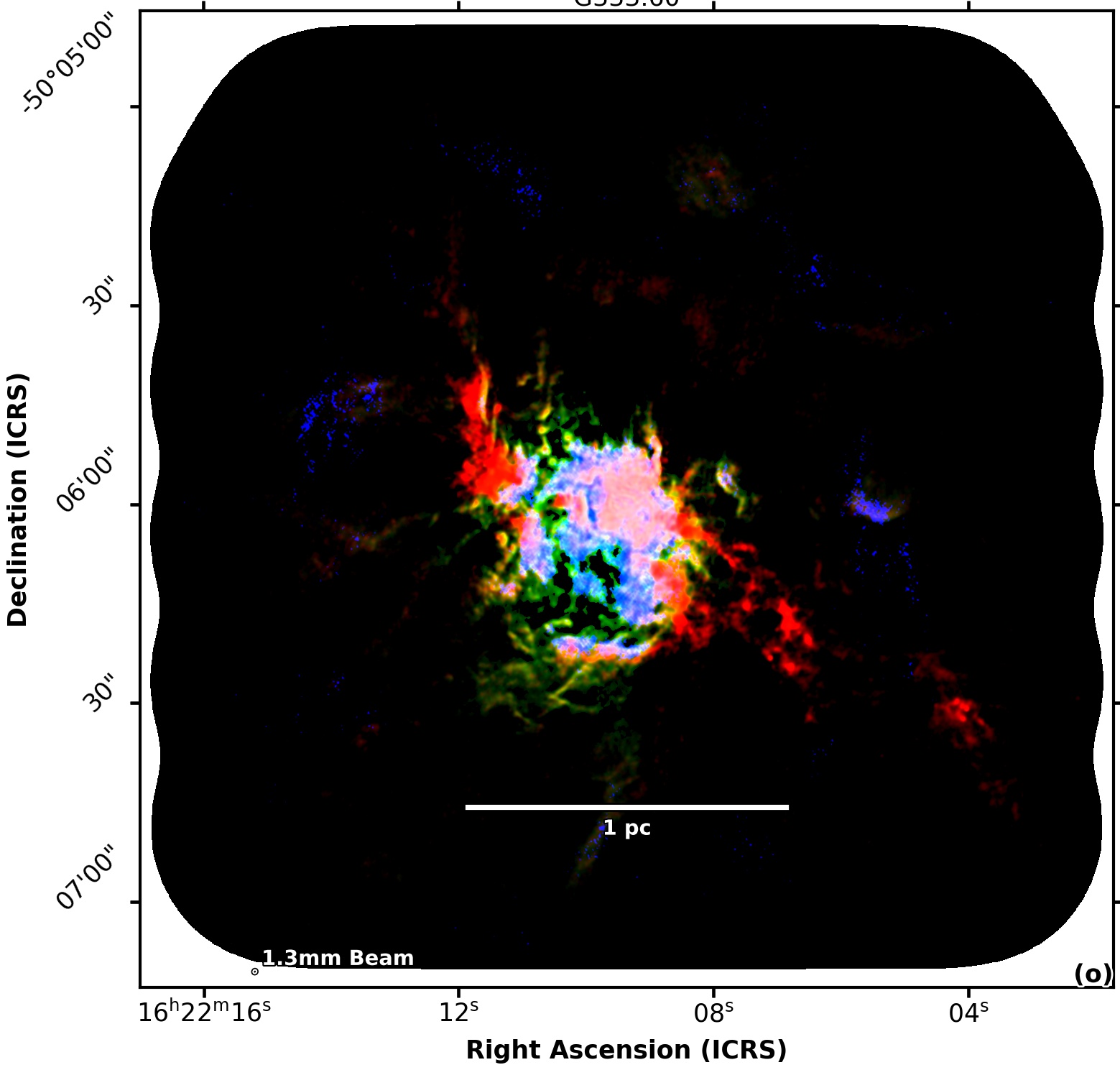}
    \vskip 0.2cm
    \caption{\textbf{(Continued)} Three-color ALMA  12~m array images of the Evolved clouds: G010.62 (in \textsl{l}), W51-IRS2 (in \textsl{m}), G012.80 (in \textsl{n}), and G333.60 (in \textsl{o}). 
    There is an almost complete coincidence between the free-free emission at the H$41\alpha$ recombination line frequency (blue) and the 3~mm continuum (green), which is dominated in \textsl{l--o} by free-free emission. Red filamentary structures represent the few locations where thermal dust emission dominates the continuum millimeter emission.}
\end{figure*}

\section{First results and survey potential}
\label{s:highlights}

We present here the continuum images and preliminary line data set used to refine the evolutionary stage classification of the ALMA-IMF protoclusters (see Sect.~\ref{s:revised}), to characterize their core content (see Sect.~\ref{s:cores}) and the gas concentration from cloud to cores (see Sect.~\ref{s:concentration}), and to discuss the potential of ALMA-IMF data to constrain the gas kinematics and molecular complexity of clouds (see Sects.~\ref{s:kin}--\ref{s:hotcore}). 

\cref{tab:evol} lists the main characteristics of the ALMA-IMF protocluster clouds, among which is their total cloud mass, $M^{\rm cloud}_{\rm 870\,\mu m}$, integrated over the ALMA-IMF 1.3~mm image coverage.
This mass is computed from the $870\,\mu$m flux of the clouds, $S^{\rm cloud}_{\rm 870\,\mu m}$, using \cref{eq:mass870}. Given that the temperatures of the brightest ATLASGAL sources, as measured in NH$_3$, vary (\citealt{wienen12, wienen15}; see also Fig.~7 of
\citealt{csengeri17b}), we assumed $\tdust=20$~K, 25~K and 30~K, for the Young, Intermediate, and Evolved clouds (as defined in Sect.~\ref{s:revised}), respectively.
Assuming a single $\tdust=20$~K temperature would increase the mass of Intermediate and Evolved clouds by 1.3 and 1.75, respectively.

\begin{figure}[htpb!]
    \centering
    \includegraphics[width=0.47\textwidth]{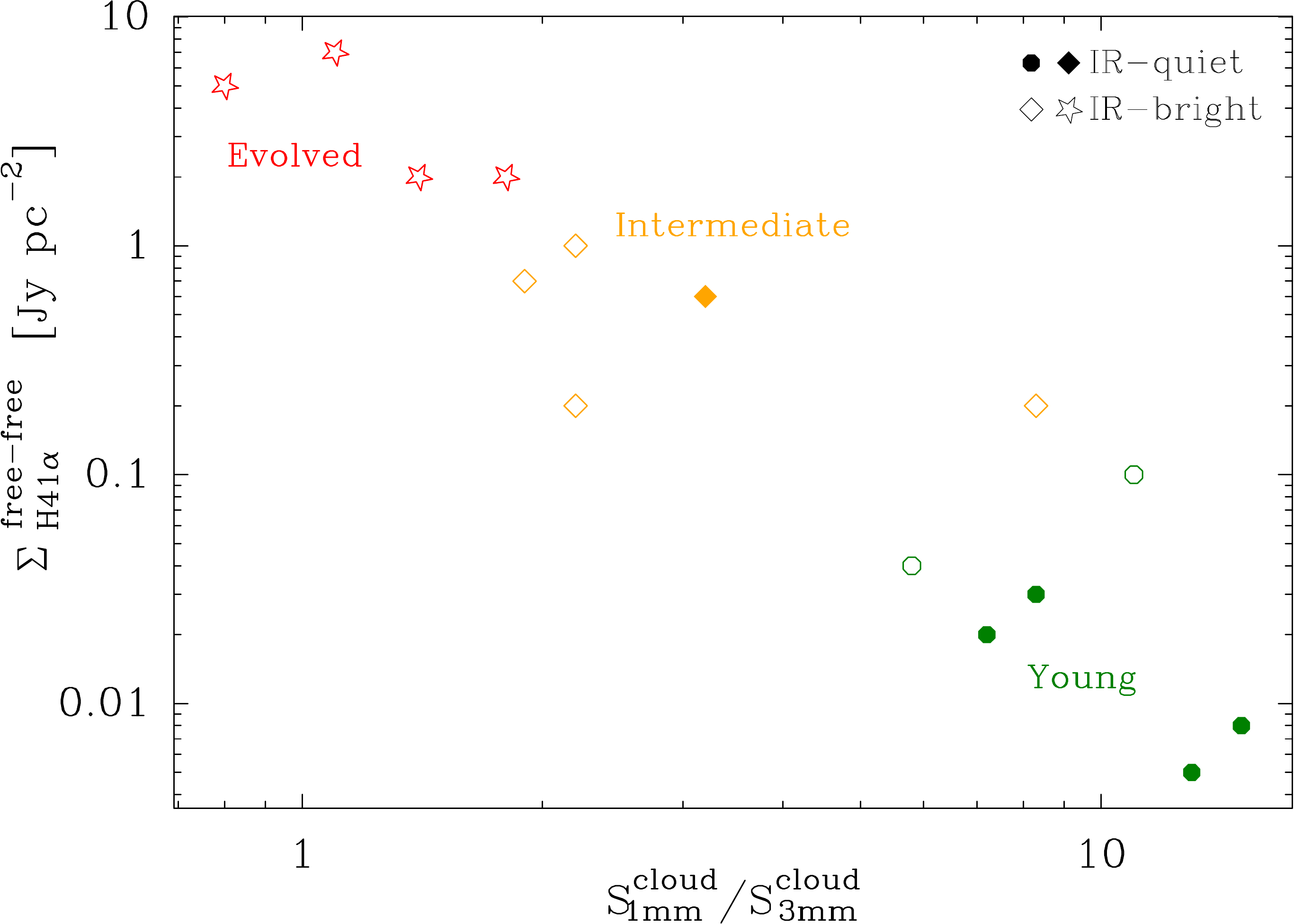}
    \caption{Evolutionary sequence of the ALMA-IMF protoclusters, assuming that free-free emission increases over time. 
    The 1.3~mm-to-3~mm flux ratios decrease when the free-free emission dominates the thermal dust emission at 3~mm. The provisional evolutionary stage of IR-quiet and IR-bright protoclusters is indicated by filled and empty square markers, respectively. Their revised evolutionary stage, from Young, to Intermediate, to Evolved, are shown by green, orange, and red markers, respectively.}
    \label{fig:Evol-Prot}
\end{figure}

\subsection{Evolutionary stage of the ALMA-IMF protoclusters}
\label{s:revised}

We have improved the initial evolutionary stage classification of the 15 protoclusters listed in \cref{tab:sample}, separating them between Young, Intermediate, and Evolved protoclusters. Determining the evolutionary stage of a $\pc^2$-size cloud is nontrivial because its structures are expected to form continuously, concentrate, form stars, heat, be partly ionized, and finally disperse. To this end, we utilize two criteria: the 1.3~mm-to-3~mm flux ratio and the free-free emission at the frequency of the H41$\alpha$ recombination line. They both hinge on the assumption that as high-mass, gas-dominated star-forming protoclusters evolve, they will host more and more \hii regions. Thus, the free-free emission is assumed to increase over time as their \hii bubbles expand and the number of \hii sources increases. We summarize in \cref{fig:Evol-Prot} the evolutionary sequence of the 15 ALMA-IMF protoclusters that is defined using these two criteria and from a visual inspection of \cref{fig:3col}.

We first computed the 1.3~mm-to-3~mm flux ratios of ALMA-IMF clouds using their 1.3~mm and 3~mm 12~m array \bsens images (see \cref{fig:ALMA1mm} and Fig.~1 of \citealt{ginsburg22}). For more evolved clouds, the free-free emission of compact and developed \hii regions (C\hii $\rightarrow$ $\rm \hii$) can dominate at 3~mm, while thermal dust emission would be the major component of the 1.3~mm emission \citep[e.g.,][]{ginsburg20}. The 1.3~mm to 3~mm flux ratio is therefore expected to decrease over time from its value associated with thermal dust emission, to 0.9 for a flat spectral energy distribution \citep[$\alpha (\nu)=-0.1$, e.g.,][]{keto08}. The 1.3~mm ($\nu_{\textrm{B6}}=229.0$~GHz; see Table~D1 in \citealt{ginsburg22}) to 3~mm ($\nu_{\textrm{B3}}=100.78$~GHz) flux ratio of thermal dust emission from clouds at $\tdust=20-30$~K (see \cref{tab:sample}) is expected to be on the order of 20, assuming optically thin emission and with a dust emissivity index of $\beta=1.8$, well suited for clouds \citep{planck11}. We integrated the 1.3~mm and 3~mm fluxes over the common 1.3~mm imaged area (see \cref{tab:evol}) and computed the ratios of integrated fluxes, $S^{\rm cloud}_{\rm 1.3\,mm}/S^{\rm cloud}_{\rm 3\,mm}$. 
In \cref{tab:evol}, the global ratios of the 15 clouds vary from 0.8 to 15, with median values of $\sim$8 and $\sim$2 and $1\,\sigma$ dispersions of $\pm{4}$ and $\pm{3}$ for the IR-quiet and IR-bright cloud populations, respectively.

The evolutionary status derived from the 1.3~mm-to-3~mm flux ratio
is consistent with that derived 
from
the 1.3~mm-to-3~mm spectral index measured both between and within the observed bands 
\citep{ginsburg22}.
The 1.3~mm-to-3~mm flux ratio suggests that, except for three clouds, the IR-quiet/IR-bright classification of protoclusters using their mid-IR flux remains valid. Among the exceptions, the IR-quiet protocluster cloud G008.67 has a low 1.3~mm-to-3~mm flux ratio suggesting that it does not qualify as being Young. Conversely, the IR-bright protocluster clouds G327.29, G337.29, and G351.77 have high ratios that conflict with their previous classification as evolved objects. This tendency is confirmed in \cref{fig:3col}, which presents the three-color ALMA images of the whole ALMA-IMF sample, using the 1.3~mm and 3~mm continuum emission for two of the three color images (red and green). The ALMA imaging indeed separates the cold cloud at the centers of both G327.29 and G337.92 from the developed \hii regions lying at their peripheries (west of G327.29 and north of G337.92). A scaling by the theoretical ratio of thermal dust emission at 20~K shows that thermal dust emission dominates over the whole extent of six protocluster clouds, which we classify as Young (see Figs.~\ref{fig:3col}a--e). Their mean flux ratio, $S^{\rm cloud}_{\rm 1.3\,mm}/S^{\rm cloud}_{\rm 3\,mm} \simeq 11$ with a $1\,\sigma$ dispersion of $\pm{3}$, is smaller than the theoretical one, $20$. Here diffuse free-free emission is sporadically present at the peripheries of the targeted clouds (as in Figs.~\ref{fig:3col}a, e--f) and the extended emission is on average filtered to scales 1.7 times smaller in the 1.3~mm images than in the 3~mm images (see Sect.~\ref{s:dataset}). 

\begin{table*}[htbp!]
\centering
\small
\begin{threeparttable}[c]
\caption{Gas mass distribution within ALMA-IMF protoclusters, from parsec-size clouds to structures with typical sizes of 0.1~pc and finally to 2100~au cores.}
\label{tab:concentration}
 \begin{tabular}{lcrrccrrcc}
\hline \noalign {\smallskip}
Protocluster  
& Spatial
        & $S^{\rm recovered}_{\rm 1.3\,mm}$  & $M^{\rm recovered}_{\rm 1.3\,mm}$ &  \multicolumn{2}{c}{Number of}   & $M^{\rm cores}_{\rm 1.3\,mm}$ & 
        $M^{\rm recovered}_{\rm 1.3\,mm}$ & $M^{\rm cores}_{\rm 1.3\,mm}$ & $M^{\rm cores}_{\rm 1.3\,mm}$\\
cloud name 
& resolution & within $A_{\rm 1.3\,mm}$ & within $A_{\rm 1.3\,mm}$ & extracted & cores\tnote{3} & within $A_{\rm 1.3\,mm}$ & $/M^{\rm cloud}_{\rm 870\,\mu m}$ & $/M^{\rm recovered}_{\rm 1.3\,mm}$ & $/M^{\rm cloud}_{\rm 870\,\mu m}$ \\
& [au] & [Jy]\tnote{1} & [\msun]\tnote{2} & sources & [\#, \%]  & [\msun] & [\%]\tnote{4} & [\%] & [\%]\\
(1) & (2) & (3) & (4) & (5) & (6) & (7) & (8) & (9) & (10)\\
\hline \noalign {\smallskip}
W43-MM1 
    & 2\,430    & 10.4 & 6\,200 & 57 & 56 (98\%) & 872 & {46\%} & 14\% & 6.5\% \\

W43-MM2                 
    & 2\,540    & 2.9 & 1\,700 & 43 & 38 (88\%) & 294 & 15\% & 17\% & 2.5\% \\

G338.93                 
    & 2\,080    & 3.0 & 890 & 51 & 51 (100\%) & 509 & 13\% & 57\% & 7.2\% \\

G328.25                 
    & 1\,350    & 1.4 & 180 & 18 & 18 (100\%)   & 69 & 7\% & 38\% & 2.7\% \\

G337.92                 
    & 1\,460    & 5.1 & 720 & 38 & 37 (97\%)   & 160 & {{28\%}} & 22\% & 6.3\% \\

G327.29         
    & 1\,650    & $16.9-16.4$ & 2\,010 & 47 & 41 (87\%) & 497 & 41\% & 25\% & 9.7\% \\

\hline

G351.77            
    & 1\,540    & $10.3-10.2$ & 800 & 28 & 26 (93\%) & 277 & {{32\%}}
    & 35\% & 11\% \\ 
G008.67                 
    & 2\,250    & $3.7-2.2$ & 500 & 22 & 21 (95\%) & 126 & 16\%
    & 25\% & 4.1\% \\ 
W43-MM3         
    & 2\,690    & $2.5-1.9$ & 1\,100 & 35 & 34 (97\%) & 164 & 21\%
    & 15\% & 3.1\% \\
W51-E   
    & 1\,640    & $30.1-27.6$ & 16\,000 & 58 & 39 (67\%) & 743 & 48\%
    & 5\% & 2.3\% \\ 
G353.41         
    & 1\,590    & $6.6-5.4$ & 420 & 62  & 59 (95\%) & 132 & {17\%}
    & 31\% & 5.3\% \\

\hline
    
G010.62         
    & 2\,310    & $8.7-3.1$ & 1\,500 & 61 & 47 (77\%) & 181 & 22\%
    & 12\% & 2.7\% \\ 
W51-IRS2        
    & 2\,560    & $22.8-16.9$ & 9\,600 & 117 & 96 (82\%) & 825 & 47\%
    & 9\% & 4.0\%\\ 
G012.80           
    & 2\,110    & $22.3-7.7$ & 860 & 82 & 65 (79\%) & 277 & {{18\%}}
    & 32\% & 6.0\% \\
G333.60    
    & 2\,330    & $34.8-6.8$ & 2\,300 & 118 & 66 (56\%) & 461 & 22\%
    & 20\% & 3.8\% \\ 
\hline \noalign {\smallskip}
\end{tabular}
\begin{tablenotes}
\item[1] Flux recovered by the ALMA 12~m array at 1.3~mm and integrated over the 1.3~mm imaged area. In the case of Intermediate and Evolved protocluster clouds and G327.29, a second value, corrected for free-free contamination, is given by ignoring the 1.3~mm fluxes in areas where the free-free emission dominates (as indicated by the H41$\alpha$ image).
\item[2] Mass recovered by the ALMA 12~m array computed from the total 1.3~mm fluxes corrected for free-free contamination (Col. 3, right value) and using \cref{eq:mass} with $\kappa_{\rm 1.3\,mm}=\rm 0.01~cm^2\,g^{-1}$ and $\tdust=20$~K. 
\item[3] Cores are sources extracted at 1.3~mm (Col.~5), whose emission consists of 
thermal dust emission. The sources of Col.~5, which are not included in Col.~6 are candidate ionization peaks detected through their free-free emission.
\item[4] Assuming the same temperature for all type of clouds when measuring their total cloud mass, the gas mass concentration from 1~pc to 0.1~pc cloud structures would be 
reduced by factors of 1.3 and 1.75 for Intermediate and Evolved clouds, respectively.
\end{tablenotes}
\end{threeparttable}
\end{table*}

Owing to the uncertainties from the overall 1.3~mm-to-3~mm flux ratios, we introduced a second criterion based on estimates of the free-free continuum emission (shown in blue in the three-color images, \cref{fig:3col}) using cubes of the H41$\alpha$ hydrogen recombination line. For each ALMA-IMF cloud, we created an H41$\alpha$  image by integrating its spectral cube over the velocity extent of the H41$\alpha$ line. The cloud H41$\alpha$ emission, $S_\mathrm{H41\alpha}^\mathrm{>5\,\sigma}$, is derived from this image clipped at $5\,\sigma$ and integrated over the cloud area. Assuming local thermodynamical equilibrium, the free-free emission of each ALMA-IMF cloud, $S^{\rm free-free}_{\rm H41\alpha}$, is then computed following
\begin{equation}
 \label{eq:freefree}
S^{\rm free-free}_{\rm H41\alpha} =  1.432\times 10^{-4}\times S_\mathrm{H41\alpha}^\mathrm{>5\sigma}\times \nu_0^{-1.1}\, T_e^{1.15}\, (1 + N_\mathrm{He}/N_\mathrm{H})^{-1},
\end{equation}
where $\nu_0 = 92.034$ GHz is the rest frequency of the H41$\alpha$ line and we assume an electron temperature of $T_e = 8000$~K and a relative abundance of helium to hydrogen of $N_\mathrm{He}/N_\mathrm{H} = 0.08$. We finally converted the free-free emission, $S^{\rm free-free}_{\rm H41\alpha}$, to a flux surface density, $\Sigma^{\rm free-free}_{\rm H41\alpha}=S^{\rm free-free}_{\rm H41\alpha}/A_{\rm 1.3\,mm}$, in Jy\,pc$^{-2}$. \cref{tab:evol} lists $\Sigma^{\rm free-free}_{\rm H41\alpha}$ values, which are a good proxy for the average amount of ionized gas in ALMA-IMF clouds. These averaged surface densities of free-free emission have the advantage of being independent of the area of the imaged cloud and, here, $A_{\rm 1.3\,mm}$ ranges from 1.3~pc$^2$ to 8.4~pc$^2$ (see \cref{tab:evol}). Because we designed the survey to have matched sensitivity across all clouds, for $\lesssim$0.16~pc emissions at the H41$\alpha$ frequency (see Sect.~\ref{s:dataset}), and because most of the free-free emission detected in ALMA 12~m array images arise from \hii regions of small sizes (see \cref{fig:3col}), $\Sigma^{\rm free-free}_{\rm H41\alpha}$ does not have a strong distance dependence.
The values given in \cref{tab:evol} can therefore be compared without strong bias from one cloud to another.

The extrema of the present classification are the most informative as they are more robust. Considering the IR-bright sources, we observe that G333.60, G010.62, W51-IRS2, and G012.80 all have strong flux surface densities, $\Sigma^{\rm free-free}_{\rm H41\alpha} \simeq 4~\rm Jy\,pc^{-2}$ with a $1\,\sigma$ dispersion of $\pm 2~\rm Jy\,pc^{-2}$, as well as complex and extended H$41\alpha$ emission, positionally correlated 3~mm and H$41\alpha$ emission, and low 1.3~mm-to-3~mm flux ratios, $S^{\rm cloud}_{\rm 1.3\,mm}/S^{\rm cloud}_{\rm 3\,mm} \simeq 1.3$ with a dispersion of $\pm{0.4}$.  These features together are consistent with advanced \hii activity compared to the other sources in our sample. These four protoclusters are therefore classified as Evolved. Turning next to the IR-quiet sources in \cite{csengeri17b},
Young protoclusters are barely detected in H$41\alpha$, with flux surface densities of their free-free emission two orders of magnitude smaller than those measured for Evolved clouds, median $\Sigma^{\rm free-free}_{\rm H41\alpha} \simeq 0.05\rm ~Jy\,pc^{-2}$ with a dispersion of $\pm 0.04~\rm Jy\,pc^{-2}$, and with no coherent structure detected (see \cref{tab:evol} and \cref{fig:Evol-Prot}).

In between these extrema, we identify the Intermediate protoclusters, which have properties consistent with different aspects of both the Young and Evolved categories described above. Namely, these protoclusters host both dense filamentary structures traced by their thermal dust emission and a couple of small, localized bubbles of ionized gas (see  \cref{fig:3col}g--k).
The latter are generally qualified as hyper-compact \hii (HC$\rm \hii$) regions when they develop in a dense, $\sim$10$^6$~cm$^{-3}$, medium and their extent is smaller than 0.05~pc \citep{hoare07}. They become ultra-compact \hii (UC$\rm \hii$) when they are more extended and develop in a less dense medium, $\sim$0.1~pc and $\sim$10$^4$~cm$^{-3}$ \citep[e.g.,][]{kurtz00}. These young \hii regions are traced by their free-free emission detected both in the 1.3~mm continuum emission band and by the H41$\alpha$ recombination line. G351.77, W43-MM3, W51-E, and G353.41 are four IR-bright protocluster clouds that present those characteristics (see Figs.~\ref{fig:3col}g, i--k). In addition, the G008.67  protocluster cloud, whose IR-quiet classification was already questionable due to its low 1.3~mm-to-3~mm flux ratio, displays an UC\hii region, which qualifies it as Intermediate (see \cref{fig:3col}h). The flux surface densities of the free-free continuum emission of Intermediate protocluster clouds, median $\Sigma^{\rm free-free}_{\rm H41\alpha}=0.5 \rm ~Jy\,pc^{-2}$ with a $1\,\sigma$ dispersion of $\pm 0.3~\rm Jy\,pc^{-2}$, is about ten times lower than that of Evolved protocluster clouds and ten times higher than that of the Young ones (see \cref{tab:evol}). 

We therefore have set up three groups of protocluster clouds according to their evolutionary stage. Six qualify as Young clouds, five as Intermediate, and four as Evolved (see \cref{tab:evol}). This classification is based on visual inspection of the distribution of free-free emission, continuum emission, and multiwavelength continuum ratios, and thus incorporates a wealth of observational information. We robustly distinguish protoclusters devoid of internal ionizing sources, those which harbor a couple of HC\hii or UC\hii regions, and protoclusters whose structure is intertwined with developed and bright \hii regions.
Figure~\ref{fig:Evol-Prot} displays a good correlation between the two quantitative criteria used here, suggesting that the variation in the spatial filtering within the cloud sample (see Sect.~\ref{s:dataset}) has no significant impact on our classification.

As shown in Tables~\ref{tab:sample} and \ref{tab:evol}, 
the three classes of protoclusters span the same range of distances to the Sun, demonstrating that to first order no distance biases affect our classification. The Young and Intermediate protocluster clouds are however much smaller in size than the Evolved ones, their median values being $\sim$2.1~pc$^2$ and $\sim$2.6~pc$^2$ 
versus $\sim$4.8~pc$^2$. The total protocluster gas masses 
of the Young, Intermediate, and Evolved clouds 
have median values 
within a factor of two from each other: $\sim$$6\times 10^3~\msun$, $\sim$$3\times 10^3~\msun$, and $\sim$$9\times 10^3~\msun$, respectively.

\subsection{Population of cores in the ALMA-IMF protoclusters}
\label{s:cores}

The star-formation activity of the ALMA-IMF massive clouds, and therefore the richness of the ALMA-IMF core database, can be assessed by the number of cores one can detect. In Paper III, \cite{louvet22} extracted sources from
the \cleanest continuum images (see definition in Sect.~\ref{s:dataset}) and identified about 840 compact sources.
Moreover, we showed that using both the \bsens and \cleanest images (see definitions in Sect.~\ref{s:dataset}), the number of robust core detections could further increase by a factor of up to $\sim$2 (Pouteau et al. in prep.). Extracting these additional sources requires careful treatment of the line contamination to exclude emission peaks associated with line rather than continuum emission. The additional sources are predominantly lower-mass cores and the fluxes of common sources are consistent between the two approaches.
One can also gain 
in source detections using images 
where the cirrus noise is reduced by the Multiscale non-Gaussian Segmentation (\textsl{MnGSeg}) technique \citep{robitaille19}. \textsl{MnGSeg} separates these cloud structures, which are incoherent from one scale to another and referred to as  Gaussian, from the filaments and cores, which are coherent structures and are associated with star formation.
Thus, the core database of the ALMA-IMF Large Program can potentially contain about $1\,500$ objects. To focus on cores that are real density peaks, \cite{louvet22} excluded millimeter sources that could correspond to free-free continuum peaks, that is, sources associated with inhomogeneities of \hii regions that develop in the ALMA-IMF protoclusters. This exclusion marginally reduced the number of cores by $\sim$5\% in Young regions, $\sim$13\% in Intermediate regions and reduced it further, by $\sim$27\%, in  Evolved clouds. \cref{tab:concentration} lists, for each of the ALMA-IMF clouds, the number of sources and cores, identified in the \cleanest images \citep[see][]{louvet22}. Sources are emission peaks whose size is limited by their structured background and neighboring sources. We used here the {multiscale, multiwavelength extraction method of sources and filaments} \textsl{getsf} that spatially decomposes the observed images to separate relatively round sources from elongated filaments and their background cloud \citep{menshchikov21}. Cores are \textsl{getsf} sources associated with thermal dust emission. The number of cores per protocluster at a given evolutionary stage, computed as the mean values of Col.~6 of \cref{tab:concentration}, correlates, as expected, with the median protocluster mass at this evolutionary stage, computed as the mean values of Col.~5 of \cref{tab:sample}. Moreover, these provide a roughly homogeneous surface number density of cores, which is with our $\sim$0.15~$\msun$ point mass sensitivity and $\sim$2100~au resolution of
$\sim$12.9~cores per pc$^2$ with a $1\,\sigma$ dispersion of $\pm1.6$~cores per pc$^2$.

The core masses are computed from the 1.3~mm flux, $S^{\rm core\, \it i}_{\rm 1.3\,mm}$, under the assumption of optically thin thermal dust emission. We here adapted \cref{eq:mass870} to measure the cumulative mass of cores in each ALMA-IMF cloud, $M^{\rm cores}_{\rm 1.3\,mm}$, from the 1.3~mm flux of all individual cores. We used the following equation, and provide here a numerical application whose dependence on each physical variable is given, for simplicity, in the Rayleigh-Jeans approximation:
\begin{eqnarray}
 \label{eq:mass}
 M^{\rm cores}_{\rm 1.3\,mm} & =    & \sum_{i = \rm first\, core}^{\rm last\, core} \frac{S^{\rm core\, \it i}_{\rm 1.3\,mm}\; d^2}{ \kappa_{\rm 1.3\,mm}\; B_{\rm 1.3\,mm}(\tdust)} \\
 &\approx& 300 \ \msun \times 
              \sum_{i = \rm first\, core}^{\rm last\, core} \left(\frac {S^{\rm core\, \it i}_{\rm 1.3\,mm}}{\mbox{1~Jy}}\right) \left(\frac {T_{\rm dust}^{\rm core\, \it i}}{\rm 20~K}\right)^{-1}
               \nonumber\\
      &      & \times 
        \left(\frac {d}{\mbox{3.9~kpc}} \right)^2
        \left(\frac {\kappa_{\rm 1.3\,mm}}{\rm 0.01\,cm^2\, g^{-1}}\right)^{-1},
        \nonumber
\end{eqnarray}
where $\kappa_{\rm 1.3\,mm}$ is the dust opacity per unit (gas~$+$~dust) mass at 1.3~mm. We chose $\kappa_{\rm 1.3\,mm}=\rm 0.01\,cm^2\, g^{-1}$ \citep[see][with a gas-to-dust ratio of 100]{OsHe94}, a value adapted to cores, which are generally dense and cold cloud structures. We assumed a mass-averaged dust temperature of $\tdust=20$~K for most of the ALMA-IMF cores. Dust temperatures are indeed expected to range from 15~K for shielded pre-stellar cores to 30~K for low- to intermediate-mass protostellar cores. Our present knowledge of the cores' nature precludes fine-tuning their temperature but future ALMA-IMF articles will address this point in detail. A couple of bright 1.3~mm cores per massive cloud, however, are expected to host hot cores and thus should have much higher dust temperatures estimated with, for example, CH$_3$CN, and CH$_3$CCH lines \citep[e.g.,][see also Sect.~\ref{s:hotcore}]{ginsburg17, bonfand17, gieser21}. From our experience \citep[e.g.,][]{motte18b},
we here assume $\tdust=75$~K for the ten cores with fluxes larger than $400~{\rm mJy} \times \left( \frac{\rm 3.9\, kpc}{d}\right)^2$, which would correspond to $>$120~$\msun$ cores if $\tdust=20$~K is assumed (see \cref{eq:mass}).
According to \cite{motte18b} and Pouteau et al. (in prep.), we estimate that about 20 cores (i.e., the most massive 3\% of the core sample) could be partially optically thick. Therefore, their masses would be underestimated by factors of 15\%--40\% when using \cref{eq:mass} \citep[see, e.g.,][]
{motte18b}.
We estimated the absolute values of the core masses to be uncertain by a factor of a few, and the relative values between cores to be uncertain by $\sim$50\%.

\begin{figure}[htpb!]
    \centering
    \includegraphics[width=0.48\textwidth]{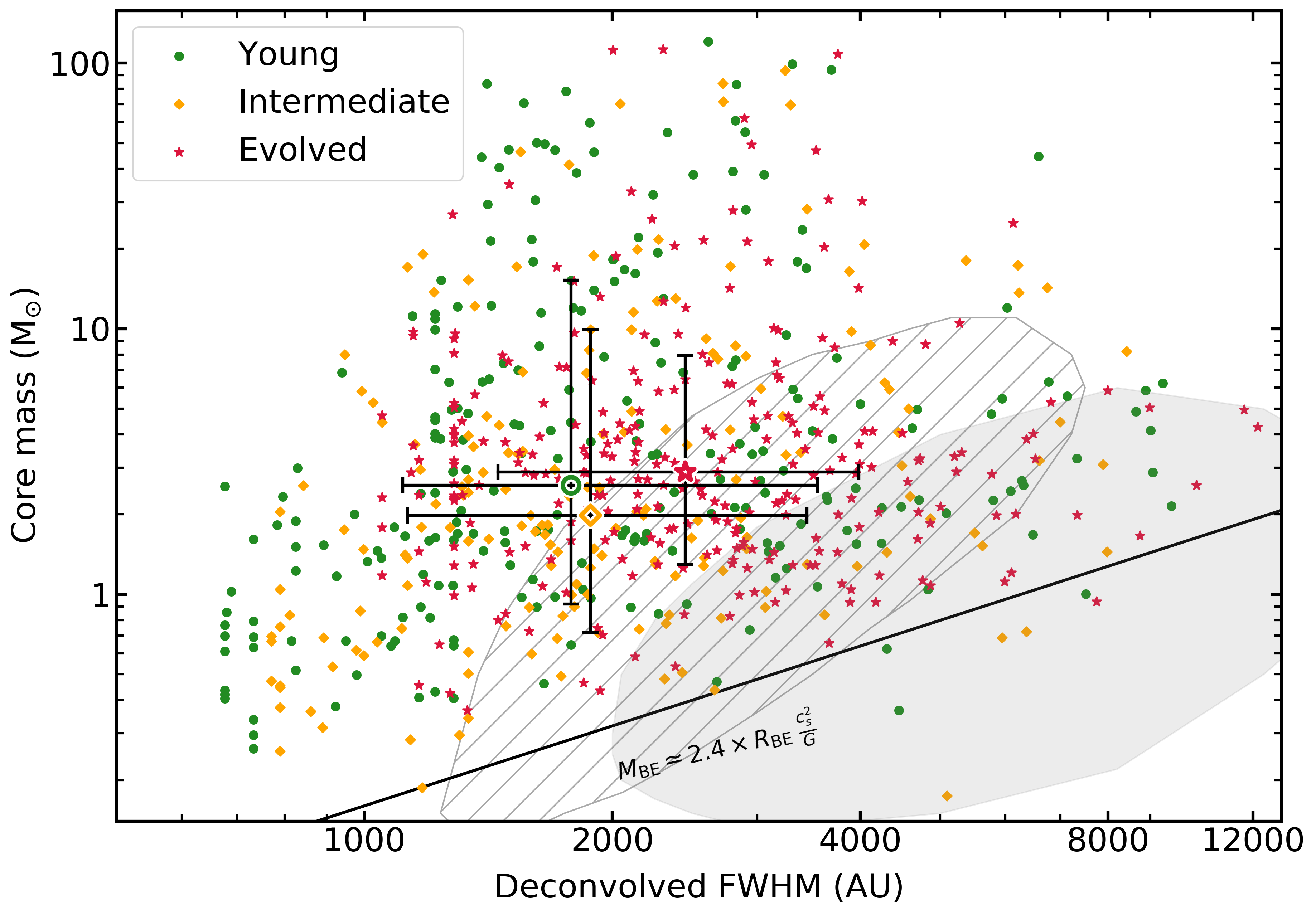}
    \caption{ Mass versus size distribution of the ALMA-IMF cores, compared, for reference, to the mass-size relation of critical Bonnor-Ebert spheres for $\tdust=20$~K (continuous line) and the location of 90\% of the core sample of \cite{konyves15} and \cite{sanhueza19} (dark and light gray shaded areas at the bottom-right corner). The median (open markers) and $1\,\sigma$ dispersion (error bars) of the core distributions for each group of Young (green), Intermediate (orange), and Evolved (red) protocluster clouds are relatively homogeneous. The core FWHMs are deconvolved from the beams,
    where the minimum deconvolved size corresponds to a half-beam angular diameter and the largest cores are not necessarily the most massive. 
    }
    \label{fig:Mass-Size}
\end{figure}

\begin{figure*}[hbtp!]
    \centering
    \includegraphics[width=0.505\textwidth]{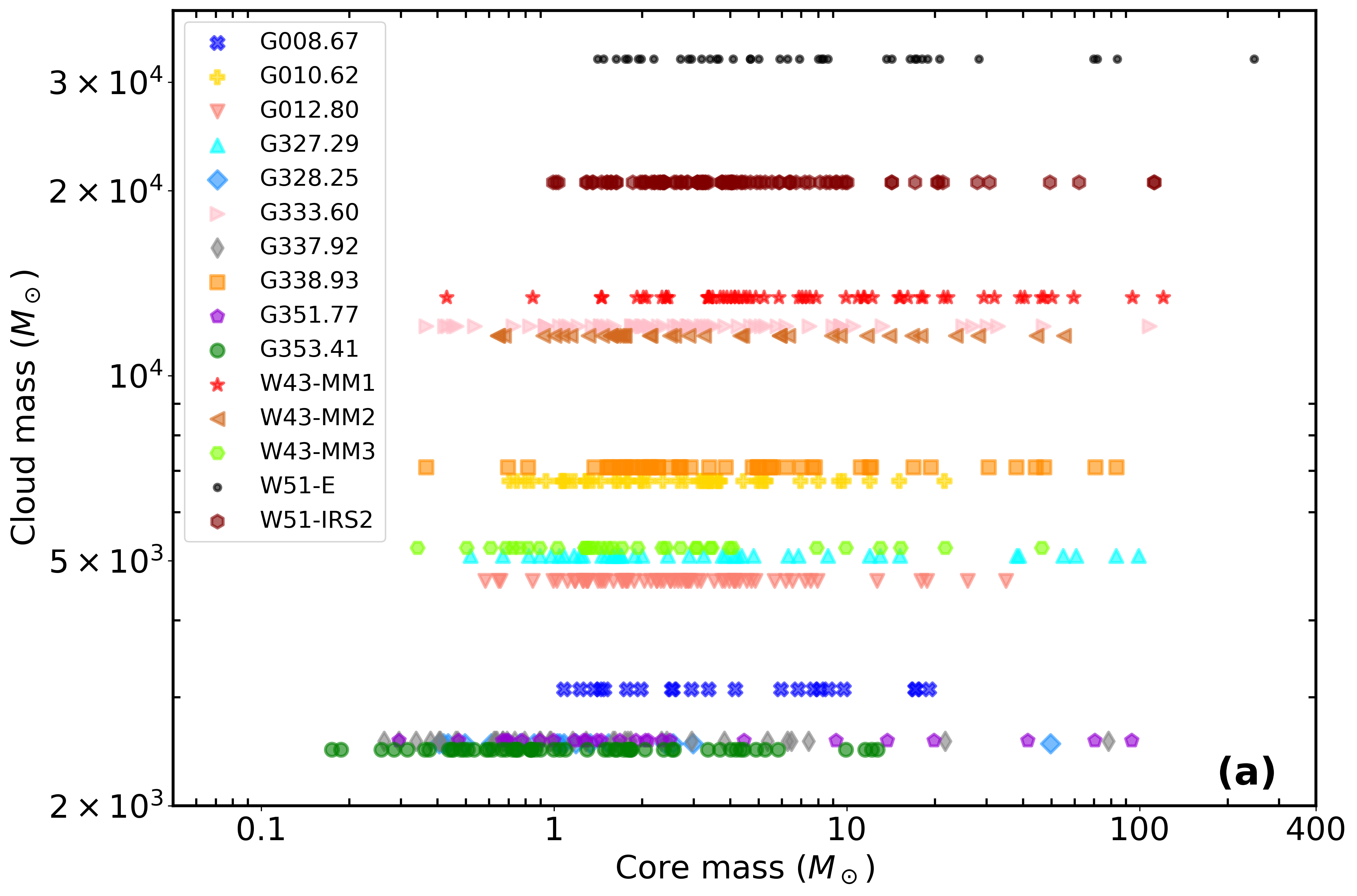}\hskip 0.3cm \includegraphics[width=0.47\textwidth]{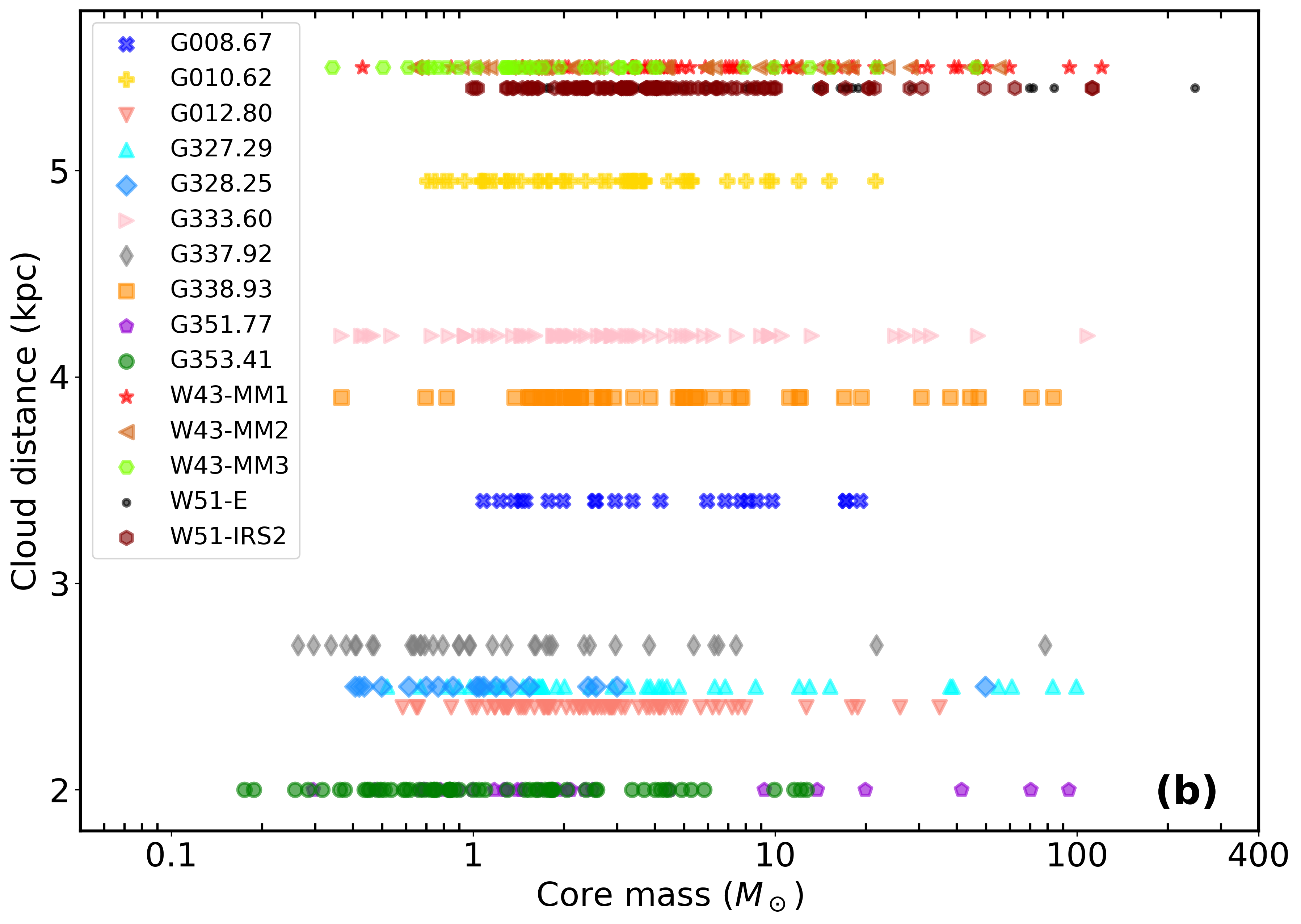}
    \caption{Distribution of the ALMA-IMF core masses as a function of the cloud mass (in \textsl{a}) and distance (in \textsl{b}) of each ALMA-IMF protocluster. The most massive clouds, which also are the farthest, tend to host the most massive cores. The mass sensitivity 
    and, consequently, the lowest-mass core of each core catalog vary by factors of $2.5$. There is no significant bias with the cloud distance to the Sun.}
    \label{fig:CoreContent}
\end{figure*}

Figure~\ref{fig:Mass-Size} presents the mass-to-size diagram of the $\sim$700 cores extracted in the \cleanest continuum images \citep[see catalogs in][and \cref{tab:concentration}]{louvet22}. Cores have deconvolved sizes ranging from $\sim$700~au to $\sim$12\,000~au, with a median value of $\sim$2\,100~au. Their masses span more than three decades, from $\sim$0.15$~\msun$ to $\sim$250~$\msun$ and are $\sim$100 times and $\sim$10 times denser than the cores extracted in low-mass star-forming clouds and infrared-dark clouds, respectively \citep[e.g.,][see \cref{fig:Mass-Size}]{konyves15, sanhueza19}. Naively, \cref{fig:Mass-Size} implies that the ALMA-IMF cores are gravitationally bound as all cores, with the exception of seven, reside above the thermal value of the critical Bonnor-Ebert mass ($M_{\rm BE}$) for a certain radius ($R_{\rm BE}$), assuming $\tdust=20$~K:
\begin{equation}
M_{\rm BE} \simeq 2.4 \times R_{\rm BE} \, \frac{c_{\rm s}^2}{G},
\end{equation}
where $c_{\rm s}$ is the isothermal sound speed and $G$ is the gravitational constant. The dynamical state of the cores, however, requires further scrutiny of nonthermal motions associated with turbulence, magnetic fields, and external pressure via the wealth of spectral lines that we detect in ALMA-IMF.
Core masses are approximately evenly distributed with size, independent of the evolutionary classification of their host clouds. 
As shown in \cref{fig:Mass-Size}, the median sizes and masses of cores in the Young, Intermediate, and Evolved protoclusters are within factors of 14\% and 15\% of each others, respectively. To more robustly compare the core population of the ALMA-IMF clouds, one should
convolve images to the same spatial resolution before extracting cores 
\citep[see, e.g.,][]{louvet21} but this result, derived from images with $2100\pm400$~au resolutions, already suggests a good degree of homogeneity in terms of cores' size and mass ranges within the ALMA-IMF cloud sample.

In \cref{fig:CoreContent}, we plot the mass of each core as a function of the mass and distance of its parental cloud. While cloud masses span a decade, the ranges of the core masses are similar and do not depend much on the cloud distance or evolutionary stage. The lowest-mass cores detected in the ALMA-IMF protoclusters have masses that depend on the spatial resolution and mass sensitivity of the \cleanest images. In the W51 and W43 protocluster clouds, this sensitivity is particularly limited by the small bandwidth used to estimate their line-free continuum emission \citep[][see their Figs.~3--4]{ginsburg22}.
Interestingly, the most massive ALMA-IMF clouds tend to host the most massive cores, even if the mass of the these cores is computed with a dust temperature of $\tdust=75$~K. Based on the Spearman’s rank correlation coefficient \citep{cohen88}, we find a strong correlation ($\rho_s$ = 0.65 and a low probability, $p=0.009$, for no correlation) between the mass of the host clouds and the most massive core in each of them. \cite{lin19} found a similar trend for larger cloud structures, while \cite{sanhueza19} found no such correlation
for cores in a sample of infrared-dark clouds. Given that the latter clouds are five to ten times less massive than the ALMA-IMF clouds, they could either represent earlier stages of protocluster evolution, where the gas is not highly centrally concentrated yet, or they could be progenitors of stellar clusters that are less rich and form stars less massive than the ALMA-IMF protoclusters.
Within the ALMA-IMF clouds, we found 79 cores that have masses larger than 16~$\msun$ and could represent the precursors of high-mass stars, assuming a gas-to-star conversion factor of $50\%$ for these cores. High-mass star precursors could be even more numerous if the core mass reservoirs are fed by inflowing gas from filaments, as suggested by, for example, \cite{motte18a} and observed by, for example, \cite{csengeri11a, olguin21}, or if cores host protostars that have already accreted significant mass. This large number of high-mass star precursors implies, for the ALMA-IMF pc-size clouds, either a high star-formation efficiency ({SFE$\:\simeq22\%$} assuming the IMF of \citealt{kroupa01}) or a lower, more typical, {star-formation efficiency} with a top-heavy IMF.

The ALMA-IMF core catalog that is summarized in \cref{tab:concentration} and will be published 
in the companion paper Paper III \citep{louvet22} is already rich enough to investigate variations in the shape of the CMF. Deeper core extractions will allow individual CMFs to be constructed for all ALMA-IMF protoclusters and even their subregions. 
ALMA-IMF studies will look for varying CMFs that must be related to (1) the evolutionary stages of clouds determined in Sect.~\ref{s:revised} and those of sub-clouds as given by the number of protostars versus pre-stellar cores (see Sects.~\ref{s:kin}--\ref{s:hotcore}), (2) the cloud mass and density structure partly investigated in \cref{fig:CoreContent}a and Sect.~\ref{s:concentration}, and (3) the cloud kinematics illustrated in Sects.~\ref{s:kin}--\ref{s:hotcore}. Due to the strategy chosen to image the ALMA-IMF protocluster clouds (see Sect.~\ref{s:strategy}), there should not be significant bias in spatial resolution, mass sensitivity, and evolutionary stage with cloud distance (see \cref{fig:CoreContent}b).

\subsection{Distribution of gas mass from cloud to cores}
\label{s:concentration}

The distribution of gas mass from the scale of clouds to the scale of cores provides insight into the ability of clouds to form stars, the {star-formation efficiency} \citep[e.g.,][]{louvet14, nony21}. Known to depend on the cloud environment in the Milky Way \citep{nguyen11a, veneziani13, kruijssen14}, the {star-formation efficiency} also depends on the relationship between stars and their cloud gas reservoir used for protostellar accretion, and therefore between the IMF and the CMF. Gas concentration can be investigated measuring density gradients \citep[e.g.,][]{didelon15, StGo16, alvarez21}, power spectra of the coherent cloud structures associated with star formation using multi-scale, multi-fractal techniques such as \textsl{MnGSeg} or 
{the reduced wavelet scattering transform}
(\textsl{RWST}) \citep{robitaille19, allys19}, 
or using less informative functions such as 
probability distribution functions \citep[e.g.,][]{kainulainen13,schneider15, StKa15}. Due to the low spatial dynamic range (i.e., one to two orders of magnitude) of the ALMA-IMF continuum images currently available (12~m array; see Sect.~\ref{s:dataset}), we limit our structural study here to ratios of masses measured at three different physical scales. 

The mass of each ALMA-IMF cloud and the cumulative mass of their hosted cores are given in Tables~\ref{tab:evol}--\ref{tab:concentration}, respectively. In addition to these 1~pc- and 0.01~pc-size cloud and core scale structures, the intermediate-scale structures recovered in the ALMA 1.3~mm 12~m array images mostly consist of  clumps {or} filaments with characteristic sizes close to the largest angular scale discussed in Sect.~\ref{s:dataset} (i.e., $\sim$0.1~pc). For each protocluster, the 1.3~mm fluxes, $S^{\rm recovered}_{\rm 1.3\,mm}$, are integrated over the entire cloud extent, as imaged at 1.3~mm. In the cases of Intermediate and Evolved protocluster clouds, thermal dust emission fluxes are estimated ignoring the 1.3~mm fluxes in areas where the free-free emission dominates (as indicated by H41$\alpha$). The masses of cloud structures recovered by ALMA at 1.3~mm, $M^{\rm recovered}_{\rm 1.3\,mm}$, are then computed from these fluxes, using \cref{eq:mass} and assuming a temperature of $\tdust=20$~K whatever the evolutionary stage of the clouds. 

\cref{tab:concentration} lists the three mass ratios that quantify gas mass concentrations from its cloud to the clumps {or} filaments (1~pc to 0.1~pc), from clumps {or} filaments to cores (0.1~pc to 0.01~pc), and from the cloud to its cores. Their absolute values are uncertain, such as mass estimates, by a factor of a few but their relative values possibly by less than 50\%. The $M^{\rm recovered}_{\rm 1.3\,mm}/M^{\rm cloud}_{\rm 870\,\mu m}$ ratio measures the percentage of the cloud mass preferentially located within dense filaments and cores. This ratio covers a wide range of values with a median of $\simeq$22\% and a $1\,\sigma$ dispersion of $\pm13\%$. Similar values for the median and dispersion are observed for the concentration of clump {or} filament mass within cores: $M^{\rm cores}_{\rm 1.3\,mm}/M^{\rm recovered}_{\rm 1.3\,mm} \simeq 22\%$ and $1\,\sigma \simeq \pm13\%$. 
The large relative dispersion observed for these two ratios of gas mass concentration, $\frac{13\%}{22\%} \simeq 0.6$, partly arises from differences in the interferometric filtering (see \cref{tab:sensitivity_scale_overview}).
However, since the theoretical filtering presents a $1\,\sigma$ relative dispersion of 0.3 (see Sect.~\ref{s:dataset}), it cannot be the sole cause of the 0.6 dispersions of the $M^{\rm recovered}_{\rm 1.3\,mm}/M^{\rm cloud}_{\rm 870\,\mu m}$ and $M^{\rm cores}_{\rm 1.3\,mm}/M^{\rm recovered}_{\rm 1.3\,mm}$ mass ratios.
Moreover the concentration of the gas mass of clouds into cores, independent of any interferometric filtering, displays values similarly dispersed around its median: $M^{\rm cores}_{\rm 1.3\,mm}/M^{\rm cloud}_{\rm 870\,\mu m}\simeq4.1\%$ and $1\sigma\simeq \pm2.6\%$, corresponding to a relative dispersion of 0.6. The variations in gas mass concentration from 1~pc to 0.1~pc and from 0.1~pc to 0.01~pc could therefore trace intrinsic physical variations from one cloud to another. These variations in cloud gas concentration will be investigated in the future in the context of the mass and spatial distributions of cores, and in conjunction with other cloud physical properties (mass, density, turbulence level, and kinematics).
When we compare the three subgroups of clouds at different evolutionary stages, we find no significant change in gas mass concentration over time. If confirmed, this result suggests that the cloud structure remains the same in young clouds and parts of evolved clouds outside \hii regions. If the spatial filtering does not change too much our mass ratio measurements, the feedback effects of stars on clouds would therefore remain very limited as long as their \hii regions do not extend over the entire cloud.

\begin{figure*}[htbp!]
    \centering
    \hskip -1.cm \subfloat{\includegraphics[width=1.05\textwidth]{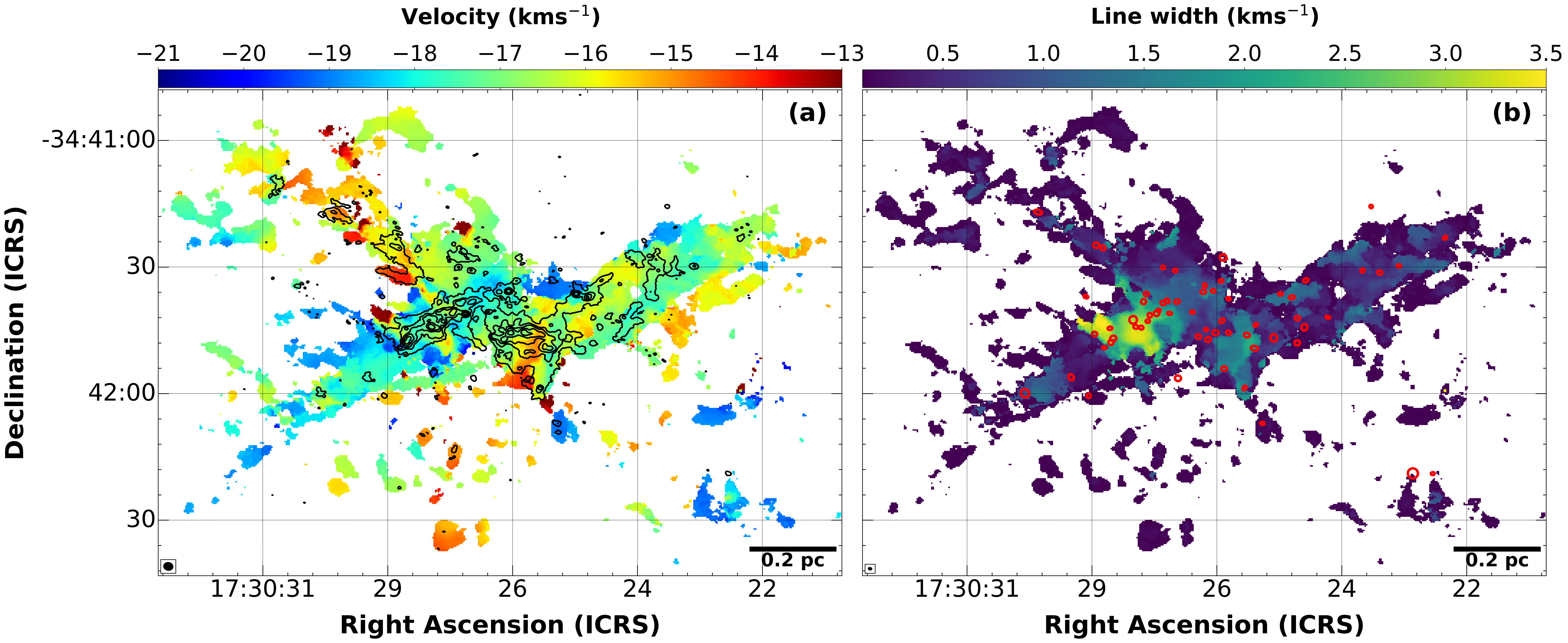}}
    \caption{Network of filaments, which interact at the center of the G353.41 protocluster. First-moment (in \textsl{a}) and second-moment (in \textsl{b}) 12~m array images of the N$_2$H$^+$(1-0) isolated hyperfine satellite, corresponding to the centroid velocity and line width.
    The 1.3~mm continuum emission is overlaid in \textsl{(a)} in black contours, which are 4, 20, 40, 80, and 160 in units of $\sigma =0.3$~mJy\,beam$^{-1}$.
    Red ellipses in \textsl{b} denote the ALMA-IMF core locations. 
    Ellipses in the lower-left corners represent the angular resolutions of the N$_2$H$^+$(1-0) in \textsl{a} and 1.3~mm continuum images in \textsl{b}.
    }
    \label{fig:dynamics}
\end{figure*}

\subsection{Gas kinematics in massive molecular clouds}
\label{s:kin}

In dynamical scenarios, molecular clouds are moving structures from birth to death \citep[e.g.,][]{ballesteros20}. They form by the collision of H\mbox{\sc ~i} gas streams or cloud-cloud collision, they constantly evolve and eventually disperse, notably due to feedback effects such as protostellar outflows, \hii regions, and supernovae. In quasi-static scenarios, on the other hand, clouds are formed by the slow agglomeration of gas, and they evolve little until their dispersion under the effects of stellar feedback \citep{MKOs07}. Observational studies from the past decade have reported intense gas motions on and around filaments: global infall of the gas surrounding filaments \citep[e.g.,][]{williams02, peretto07, schneider10, galvan10, jackson19, Bonne2020}; oscillations and rotation of filaments \citep[e.g.,][]{StGo16, alvarez21}; hub-type sub-filament accretion \citep[e.g.,][]{galvan13, peretto13, nakamura14, trevino19, chen19}, braiding of sub-filaments (also called fibers) into filaments or ridges \citep{schneider10, hennemann12, henshaw14, hacar18, GoLoSt19}; and local gas infall toward cores \citep{galvan09, csengeri11a, olguin21}. A few studies have also reported large-scale low-velocity shocks in high-mass star-forming regions, traced by SiO molecular line emissions \citep{jimenez10, sanhueza13, nguyen13, louvet16}. Such emission could emerge as a consequence of the formation of high-mass star-forming regions through cloud-cloud collision.
 
The ALMA-IMF line data cubes contain several emission lines necessary to trace the gas kinematics (velocity gradients, infall, rotation, turbulence level, and shocks) of cloud structures from the average $\rm 10^4~cm^{-3}$ gas density of massive clouds up to the $\rm 10^7~cm^{-3}$ density of cores. The Large Program setup indeed notably covers, with increasing critical density ($n_{\rm crit} = 1\times 10^4-5\times 10^6~\rm cm^{-3}$), the C$^{18}$O~(2-1), N$_2$H$^+$~(1-0), DCO$^+$~(3-2), DCN~(3-2), 
N$_2$D$^+$~(3-2), and  $^{13}$CS~(5-4) emission lines (see \cref{tab:lines}). 
The constraints obtained for these different lines, along with shock lines such as
SiO~(5-4), will be combined in future work to quantify the kinematics of molecular clouds and protoclusters over two orders of magnitude in physical scale, from one parsec to one hundredth of a parsec. The multi-scale kinematics of each ALMA-IMF cloud can then be confronted with theories of global infall, hierarchical collapse, or inertial inflow \citep[e.g.,][]{smith09,vazquez19, padoan20} and observational models of rotation, oscillation, or clump-fed accretion \citep[e.g.,][]{braine20, stutz18, motte18a}.

The first ALMA-IMF studies focus on the N$_2$H$^+$~(1-0) line, which traces gas filaments crossing the protocluster clouds.
Figure~\ref{fig:dynamics}a displays, for the G353.41 protocluster, the line centroid velocity of the isolated hyperfine satellite of the N$_2$H$^+$~(1-0) multiplet ($J\,F_1\,F =1\,0\,1\rightarrow 0\,1\,2$, at $\sim$93.176~GHz). With this single piece of information, one can already identify three
possible velocity components, separated by $\sim$2~$\kms$. The first one, at $-18~\kms$, crosses the cloud from southeast to northwest. The second one, at $-16~\kms$, presents a ``V'' shape from northeast to the center region and finally northwest and the last one, at $-14~\kms$, extends from northeast to center.
These velocity components are globally filamentary and may represent filaments that interact at the central hub of G353.41, which hosts four intermediate-mass cores ($>$10~\msun; see \cref{fig:CoreContent}) including an UC\hii region (see Figs.~\ref{fig:ALMA1mm}l and \ref{fig:3col}l; see also \cref{tab:evol}).These filamentary components exhibit significant velocity differences corresponding to a crossing time on the order of $2.5\times 10^5\,$Myr (1~pc at 4 km\,s$^{-1}$); they may therefore play an important role in the building up and continued growth of the dense structures of the G353.41 cloud.

Figure~\ref{fig:dynamics}b presents, for the same isolated hyperfine satellite of the N$_2$H$^+$~(1-0) multiplet, the velocity width of the line consisting of all velocity components and fitted by a single Gaussian. While the line generally appears to have width on  the order of $0.5-1.5~\kms$, the multiple velocity components observed on the line of sight of the overlapping areas of these filamentary components mimic line broadening with velocity widths larger than 3~$\kms$. Disentangling the multiple velocity components is necessary to properly study the complex dynamics of the region.
Overall, the filament network of \cref{fig:dynamics} is reminiscent of the fan morphology 
observed for other intermediate- to high-mass hubs \citep[e.g.,][]{peretto13, nakamura14, lu18, trevino19, GoLoSt19}. 
As shown in \cref{fig:dynamics}b, these dense
filaments host many continuum cores, with the most massive ones at their potential connection-hub. 

A major challenge of the ALMA-IMF project is to quantify the mass growth of cores as a function of environment. The mass of the gas channeled through these filaments should therefore be measured to estimate the mass inflow accretion rates toward each of the detected cores. 
We started investigating the nature of the cores detected by \citet[][see also Sect.~\ref{s:cores}]{louvet22}. An important goal is to compare pre-stellar and protostellar CMFs, as was done for example by \cite{HaFu08}. To this end, we use the CO~(2-1) and SiO~(5-4) lines to search for outflows driven by continuum cores; cores with and without outflows are therefore called protostellar and pre-stellar cores, respectively.
In parallel, a detailed characterization of the outflows through their CO~(2-1) and SiO~(5-4) emission will allow us to reach three objectives: reveal the episodicity of the accretion-ejection processes, as done by, e.g., \cite{nony20}, constrain the turbulence level injected by the outflows into ambient molecular gas on cloud scales \citep[e.g.,][]{li20} at the different evolutionary stages, and search for molecular outflows of peculiar morphology \citep[e.g.,][]{tafoya21}.

Due to the 
higher sensitivity of future measurements using the wealth of spectral lines (see \cref{tab:lines} and Sect.~\ref{s:hotcore}), including the entire N$_2$H$^+$~(1-0) line multiplet in combined 12~m plus 7~m plus Total Power data cubes, 
the conditions (optical depth, excitation temperature), radial and line-of-sight velocities, and line widths in the ALMA-IMF protoclusters will be determined in detail in future work. In turn, these measurements will permit the evaluation of potential changes in dense gas properties with evolutionary stage (see Sect.~\ref{s:revised}) and core properties (see Sects.~\ref{s:cores} and \ref{s:hotcore}).

\begin{figure*}
    \centering
    \includegraphics[width=1\textwidth]{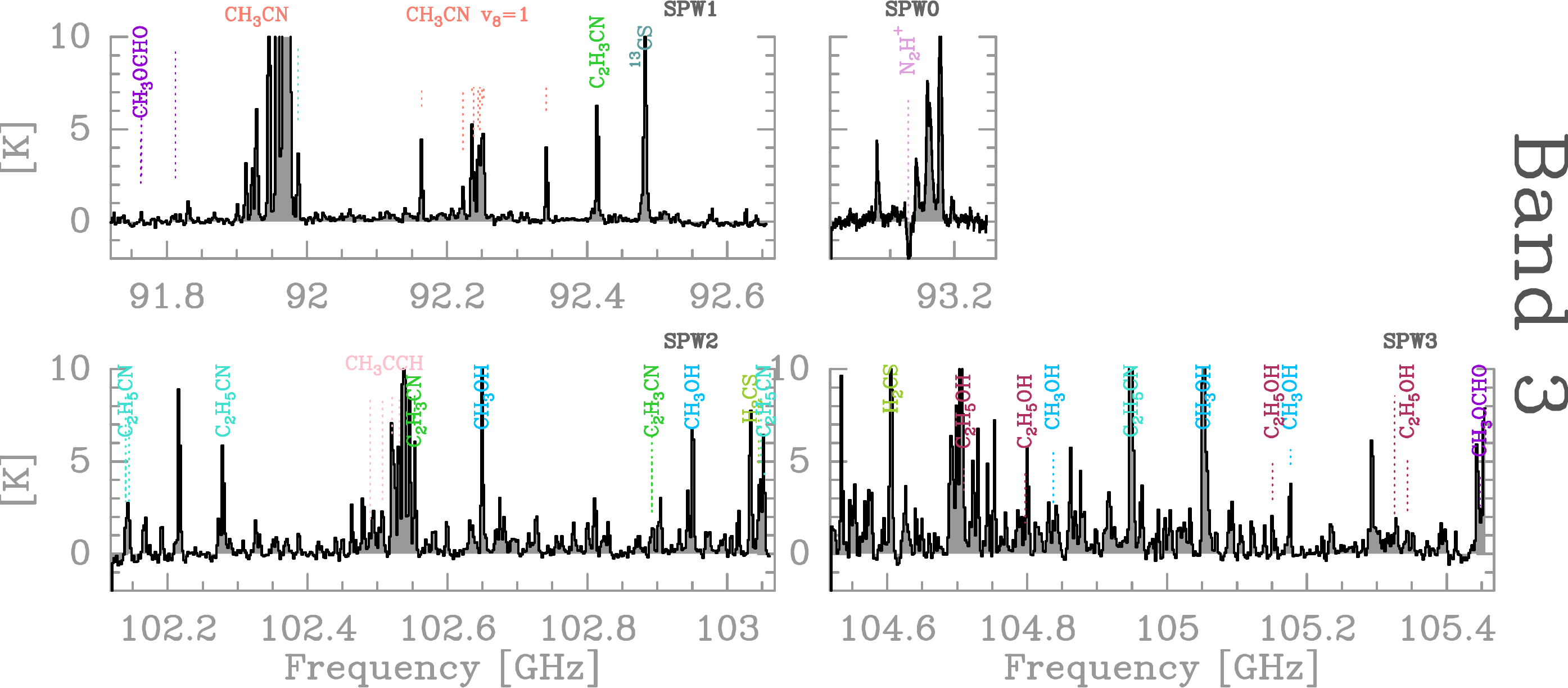}
    \includegraphics[width=1\textwidth]{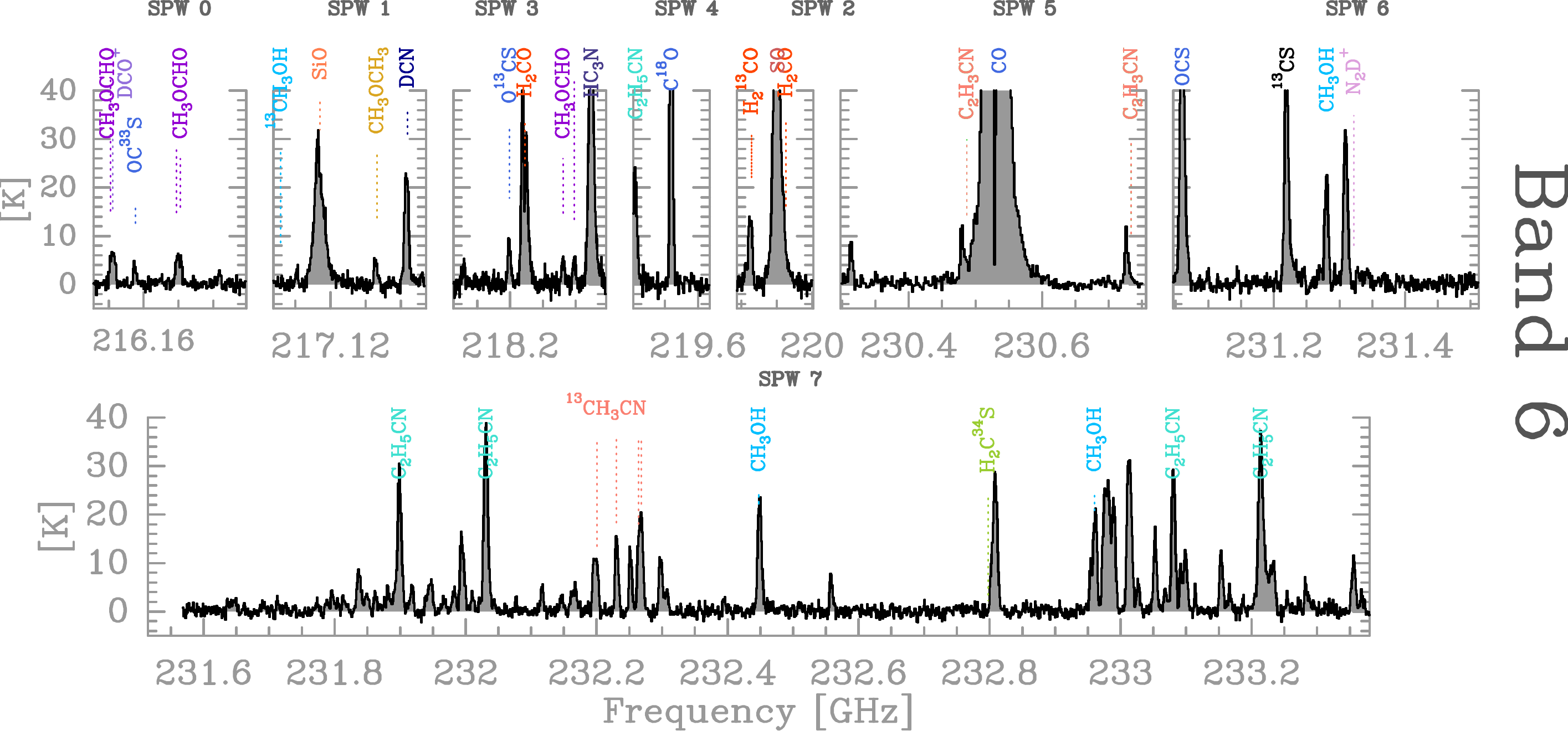}
    \caption{ALMA-IMF spectrum, consisting of the 12 spectral windows of the 12~m array data cubes, observed for the brightest 1.3~mm source of the G351.77 protocluster. Located at (RA, Dec)$_{\rm{ICRS}}$=($+17^h$26$^m$42${\rlap{.}{^s}}51$, $-36^\circ09^{\prime}17.53^{\prime\prime}$), this hot core is associated with a core that has a $\sim$15--30~$\msun$ mass assuming optically thin thermal dust emission at $\tdust=50-100$~K. The spectra have been converted to a brightness temperature scale in kelvins using the respective synthesized beam values. The brightest spectral lines are identified with colored labels.}
    \label{fig:hotcore}
\end{figure*}

\subsection{Hot cores and molecular complexity in high- and low-mass protostars}
\label{s:hotcore}

Emission lines of COMs originating from bright hot cores or chemically rich protostars are present over multiple, if not all, ALMA bands. With a 6.4~GHz noncontinuous bandwidth, corresponding to 2.9~GHz covered by four spectral windows at 3~mm plus 3.5~GHz covered in eight spectral windows at 1.3~mm, the ALMA-IMF data will reveal the molecular content of a large population of objects (see Sect.~\ref{s:dataset} and transitions in \cref{fig:hotcore} and \cref{tab:lines}). Cores extracted in the ALMA-IMF protoclusters cover a broad range of masses, $0.15-250~\msun$, and all evolutionary stages from pre-stellar cores, to protostars, to HC\hii regions (see Sect.~\ref{s:cores} and \cref{fig:Mass-Size}). The brightest cores frequently display line-rich spectra associated with hot molecular gas (see, e.g, \cref{fig:hotcore}). Indeed, several well-known hot cores are within the ALMA-IMF images, especially in all the Evolved clouds, G010.62 \citep{Liu2010, law21}, W51-IRS2 \citep[e.g.,][]{henkel13}, G012.80 \citep{immer14}, and G333.6 \citep{Lo2015}. Other well-studied hot cores include those in the Intermediate clouds G351.77 \citep[e.g.,][]{leurini08}, G008.67 \citep[e.g.,][]{hernandez14}, and W51-E \citep[e.g.,][]{Zhang1998,Rivilla2017} and one in the Young G327.29 cloud \citep{Wyrowski2008, Bisschop2013}.

A first-look analysis of ALMA-IMF data cubes concerning hot molecular gas indicates that all the targeted protoclusters, even the youngest ones, contain at least one core with line-rich spectra. In particular, the 10 brightest continuum sources from \cref{tab:concentration}, corresponding to cores with masses $>$25~$\msun$ assuming $\tdust=75$~K (see Sect.~\ref{s:cores}), all exhibit numerous transitions from COMs in all spectral bands and thus are potential hot cores. This trend suggests that the hot core phase would appear very early in the evolutionary sequence of the formation of high-mass stars and massive clusters. 
While most protoclusters classified as Intermediate or Evolved are associated with known hot cores, several of the youngest ALMA-IMF protoclusters host previously unrecognized (e.g., G337.92 and W43-MM2) or recently identified hot cores (W43-MM1, \citealp{molet19}), or a hot core precursor \citep[G328.25][]{csengeri19}.
Figure~\ref{fig:hotcore} displays,
the complete ALMA-IMF spectrum of the well-known hot core located in the G351.77 protocluster cloud, also known as IRAS~17233$-$3606
\citep[e.g.,][]{leurini08}. A large number of transitions from COMs commonly observed toward hot cores are clearly detected, the brightest of which originate from the CH$_3$OH, CH$_3$CN, CH$_3$CCH, CH$_3$OCHO, C$_2$H$_5$CN, and CH$_3$CHO molecules (see \cref{fig:hotcore}). These lines will be used to assess and statistically compare the molecular content of the thousand cores detected by ALMA-IMF (see \cref{tab:concentration}). Among these transitions, those of the CH$_3$OH, CH$_3$CN, and CH$_3$CCH molecules can serve as excellent probes of the physical conditions of hot molecular gas \citep[e.g.,][]{gieser21}. They can be used to estimate kinetic temperatures,
column densities and molecular abundances. Temperature estimates based on molecular tracers will put additional constraints on the average dust temperatures, and hence will allow us to better characterize the properties of protostars and improve their mass estimates. Heavier COMs and rotational transitions from vibrationally excited states such as CH$_3$CN  ($v_8=1$) in \cref{fig:hotcore} should be detected toward some of the most massive ALMA-IMF cores, thus providing further constraints on their excitation conditions. 

The sensitivity of the ALMA-IMF data cubes at 1.3~mm reaches about 0.5--1~K at 1~\kms\ resolution (see Sect.~\ref{s:dataset} and \cref{fig:hotcore}). This sensitivity is largely sufficient to detect, in the ALMA-IMF clouds, a hot core such as Orion-KL that exhibits spectral lines with peak brightness temperatures of $\sim$20-30\,K 
in a 2000~au beam \citep[e.g.,][]{Brouillet2015, pagani17}. Archetypical hot corinos, such as IRAS~16293 or IRAS~4A with sizes of at most a few 100~au \citep{Jorgensen2016, belloche20},
would themselves remain undetected in our survey because of the large beam dilution.
Preliminary investigations of the ALMA-IMF data cubes suggest, however, that line transitions typical of hot cores are detected toward cores of low to intermediate masses (i.e., $2-4~\msun$; e.g., \citealt{motte18b}).
Altogether, based on a first look analysis of the detection rates of COMs toward cores of the W43-MM1, W43-MM2, and W43-MM3 protoclusters (Brouillet et al. in prep.), we expect to identify one to five cores with line-rich spectra in each ALMA-IMF cloud, giving statistics of $15-75$ high-mass and intermediate-mass hot cores.

The advantage of the ALMA-IMF survey to study the molecular complexity is that it provides a large sample of hot cores and various cloud environments undergoing dynamical events (gas inflow and protostellar outflows; see Sect.~\ref{s:kin}), all studied with the same physical scale, sensitivity, and spectral coverage. Recent studies of molecular complexity show that, in addition to the classical radiative heating corresponding to hot cores, a range of physical processes may lead to the emergence of emission from COMs. For example, significant amounts of COMs are released in the gas phase through shocks created by protostellar outflows (such as L1157-B1 and IRAS 4A, e.g., \citealt{Lefloch2017, DeSimone20}), in externally heated regions (such as Orion-KL and {photodissociation regions}; \citealt{favre11, pagani19, leGal2017}), and through  accretion shocks \citep[as observed toward G328.25,][]{csengeri19}. Beyond the emergence of COMs, ALMA-IMF also covers several deuterated molecules as well as isotopologues from abundant species such as CH$_3$CN, CH$_3$OH, OCS, and H$_2$CO. Measurements of the deuteration and isotopic fractionation provide indications of physical conditions that may impact the chemistry. Therefore, the physical processes and the chemistry driving the emergence of molecular complexity in high-mass protostars and dynamically evolving clouds could be statistically established with ALMA-IMF and then further compared to chemical models \citep[see][]{garrod2006, ruaud2016, holdship2017}. This goal is particularly important for emerging hot cores, as they offer excellent laboratories to study the early warm-up phase chemistry.


\section{Conclusion} \label{s:conc}

ALMA-IMF\footnote{\#2017.1.01355.L, PIs: Motte, Ginsburg, Louvet, Sanhueza; see \url{http://www.almaimf.com}}
is a Cycle~5 Large Program carried out with ALMA 12~m, 7~m, and Total Power arrays. Its potential and first highlights based on 12~m array data can be summarized as follows:
\begin{enumerate}
\item We selected 15 massive ($2-33\times 10^3~\msun$), nearby ($2.5-5.5$~kpc) protoclusters that span early protocluster evolutionary stages (see Sect.~\ref{s:target} and \cref{tab:sample}). ALMA mosaics cover total noncontiguous areas of $\sim$53~pc$^2$ at 1.3~mm and $\sim$122~pc$^2$ at 3~mm in the typical yet extreme environments of the Milky Way clusters in formation (see \cref{fig:overview}). The ALMA-IMF spectral setup was carefully designed to focus on lines tracing gas motions from clouds to cores. In addition, the ALMA-IMF setup covers a 6.4~GHz noncontinuous bandwidth at 3~mm (Band~3, 99.66~GHz) and 1~mm (Band 6, 228.4~GHz) used to survey lines from COMs.
\item The ALMA-IMF data set is homogeneous, with approximately matched point mass sensitivity and physical resolution across the protocluster sample that spans a factor of $\sim$3 in distance. Hence, the key feature of our approach is the lack of significant distance bias, which enables robust, synergistic science on the emergence of the IMF and star clusters. 
 The ALMA-IMF database consists of 1~mm and 3~mm continuum images that are sensitive to $\sim$0.18~$\msun$ and $\sim$0.6~$\msun$ point-like cores, respectively, at a matched spatial resolution of $\sim$2\,100$\pm400$~au (see, e.g., \cref{fig:3col}). Moreover, the ALMA-IMF data set contains many emission lines that trace dense molecular gas, outflows, shocks, {COMs probing hot cores, and} recombination lines that trace the ionized gas (see Sect.~\ref{s:obs} and \cref{tab:lines}). The whole ALMA-IMF data set is processed with a pipeline\footnote{see \url{https://github.com/ALMA-IMF/reduction}} 
described in Paper II \citep{ginsburg22}, in which we have carried out a homogeneous, repeatable, and high-quality reduction. 
\item We improved the evolutionary stage classification of the 15 protoclusters based on visual inspection and quantitative measurements of the distribution of free-free and thermal emission (see Sect.~\ref{s:revised} and \cref{tab:evol}). Four protoclusters are classified as Evolved based on their advanced \hii activity compared to the other sources in our sample. They present strong, complex, and extended free-free emission traced by the H41$\alpha$ {line} and 3~mm continuum (see Figs.~\ref{fig:3col}l--o and \ref{fig:Evol-Prot}). Six protoclusters are classified as Young based on being devoid of internal ionizing sources (see Figs.~\ref{fig:3col}a--f and \ref{fig:Evol-Prot}), with free-free emission two orders of magnitude smaller than those measured for Evolved clouds. In between these extrema, Intermediate protoclusters host both dense filamentary structures traced by their thermal dust emission and small, localized bubbles of ionized gas (see  Figs.~\ref{fig:3col}g--k and \ref{fig:Evol-Prot}).
\item 
The ALMA-IMF core catalog contains $\sim$700 cores spanning $\sim$0.15~$\msun$ to $\sim$250~$\msun$, with a median size of $\sim$2\,100~au (see \cref{fig:Mass-Size}, \cref{tab:concentration}, and Sect.~\ref{s:cores}). This core sample, published in Paper III \citep{louvet22}, has no significant bias with cloud distance or cloud evolutionary stage (see Figs.~\ref{fig:Mass-Size}-\ref{fig:CoreContent}). Within the ALMA-IMF clouds, we found 79 cores that have masses larger than 16~$\msun$, which could represent the precursors of high-mass stars, assuming a gas-to-star conversion factor of $50\%$ for these cores. The most massive protocluster clouds tend to host the most massive cores, even if the masses of such cores are computed with a dust temperature of $\tdust=75$~K (see \cref{fig:CoreContent}). Core catalogs of \cite{louvet22} will be used to build CMFs 
and study their variations with cloud characteristics and evolutionary stage.
\item ALMA-IMF has the ability to constrain the distribution of gas mass from the scale of clouds to the scale of cores and thus provide insight into the {star-formation efficiency}. Due to the current dynamic range, however, we limit our structural analysis to  mass ratios that quantify the gas mass concentration from cloud to clumps {or} filaments (1~pc to 0.1~pc), from clumps {or} filaments to cores (0.1~pc to 0.01~pc), and from cloud to cores (see \cref{tab:concentration}). Initial results on the concentration of cloud gas into cores suggest that stellar feedback has little effect on the structure development of high-density gas (see Sect.~\ref{s:concentration}).
\item The ALMA-IMF line data cubes contain all the emission lines necessary to trace the gas kinematics (velocity gradients, infall, rotation, turbulence level, and shocks) of cloud structures from the average $\rm 10^4~cm^{-3}$ gas density of massive clouds up to the $\rm 10^7~cm^{-3}$ density of cores (see \cref{tab:lines} and Sect.~\ref{s:kin}). The constraints obtained for these different lines, along with shock tracing lines such as SiO~(5-4), will be combined to quantify the kinematics of molecular clouds and protoclusters over two orders of magnitude in physical scale, from one parsec to one-hundredth of a parsec. In particular, the N$_2$H$^+$~(1-0) line shows  
networks of filaments that may trace
inflow gas streamers (see, e.g., \cref{fig:dynamics}). They will be traced down to the scale of cores to potentially quantify the growth of core masses and the evolution over time of the shape of the CMF. 
\item 
ALMA-IMF has the potential to identify several tens of cores that exhibit line-rich spectra potentially corresponding to hot cores (see, e.g., \cref{fig:hotcore} and Sect.~\ref{s:hotcore}). Beyond the well-known hot cores hosted in Evolved protoclusters, we also cover several regions where completely new, bright hot cores can be recognized. The detection of COMs toward the brightest sources suggests that the hot core phase appears early in the emergence of high-mass protostars. Thanks to the similar sensitivity and spatial resolution toward each protocluster, we will be able to perform a homogeneous characterization of their molecular content and use spectroscopic tracers to constrain the emerging molecular complexity in protostars of high to intermediate mass.
\end{enumerate}

The ultimate objective of ALMA-IMF is to push forward our understanding of the IMF of stars and stimulate improvements to star-formation models, taking the effects of cloud characteristics and evolution into account. To this end, we will provide the community with a high-legacy database of protocluster clouds, filaments, cores, hot cores, outflows, and inflows at matched sensitivity. We emphasize that ALMA-IMF spans a Milky Way-relevant sample that captures the range in gas mass and evolutionary stages necessary to achieve this legacy value.


\begin{acknowledgements}
This paper makes use of the following ALMA data: ADS/JAO.ALMA\#2017.1.01355.L, \#2013.1.01365.S, and \#2015.1.01273.S. ALMA is a partnership of ESO (representing its member states), NSF (USA) and NINS (Japan), together with NRC (Canada), MOST and ASIAA (Taiwan), and KASI (Republic of Korea), in cooperation with the Republic of Chile. The Joint ALMA Observatory is operated by ESO, AUI/NRAO and NAOJ. 
This project has received funding from the European Research Council (ERC) via the ERC Synergy Grant \textsl{ECOGAL} (grant 855130), from the French Agence Nationale de la Recherche (ANR) through the project \textsl{COSMHIC} (ANR-20-CE31-0009), and the French Programme National de Physique Stellaire and Physique et Chimie du Milieu Interstellaire (PNPS and PCMI) of CNRS/INSU (with INC/INP/IN2P3).
SB acknowledges support from the French Agence Nationale de la Recherche (ANR) through the project \textsl{GENESIS} (ANR-16-CE92-0035-01).
TCs and MB have received financial support from the French State in the framework of the IdEx Universit\'e de Bordeaux Investments for the future Program.
YP, ALS, GB, and BL acknowledge funding from the European Research Council (ERC) under the European Union’s Horizon 2020 research and innovation programme, for the Project “The Dawn of Organic Chemistry ” (DOC), grant agreement No 741002.
FL acknowledges the support of the Marie Curie Action of the European Union (project \textsl{MagiKStar}, Grant agreement number 841276).
AS gratefully acknowledges funding support through Fondecyt Regular (project code 1180350) and from the Chilean Centro de Excelencia en Astrofísica y Tecnologías Afines (CATA) Basal grant AFB-170002.
RGM acknowledges support from UNAM-PAPIIT project IN104319 and from CONACyT Ciencia de Frontera project ID  86372. Part of this work was performed at the high-performance computers at IRyA-UNAM. We acknowledge the investment over the years from CONACyT and UNAM, as well as the work from the IT staff at this institute.
AG acknowledges support from the National Science Foundation under grant No. 2008101. 
GB also acknowledge funding from the State Agency for Research (AEI) of the Spanish MCIU through the AYA2017-84390-C2-2-R grant.
PS and BW were supported by a Grant-in-Aid for Scientific Research (KAKENHI Number 18H01259) of the Japan Society for the Promotion of Science (JSPS). P.S. and H.-L.L. gratefully acknowledge the support from the NAOJ Visiting Fellow Program to visit the National Astronomical Observatory of Japan in 2019, February.
RAG gratefully acknowledges support from ANID Beca Doctorado Nacional 21200897.
TB acknowledges the support from S. N. Bose National Centre for Basic Sciences under the Department of Science and Technology, Govt. of India.
GB also acknowledges funding from the State Agency for Research (AEI) of the Spanish MCIU through the AYA2017-84390-C2-2-R grant.
CB gratefully acknowledges support from the National Science Foundation under Award No. 1816715.
LB acknowledges support from ANID BASAL grant AFB-170002.
DW gratefully acknowledges support from the National Science Foundation under Award No. 1816715.
\end{acknowledgements}

\bibliographystyle{aa}  
\bibliography{41677_astroph}


\begin{appendix}
\section{The most massive ATLASGAL clumps at 2-5.5~kpc}
\label{s:appendix}

\begin{figure}
    \centering
    \includegraphics[width=0.53\textwidth]{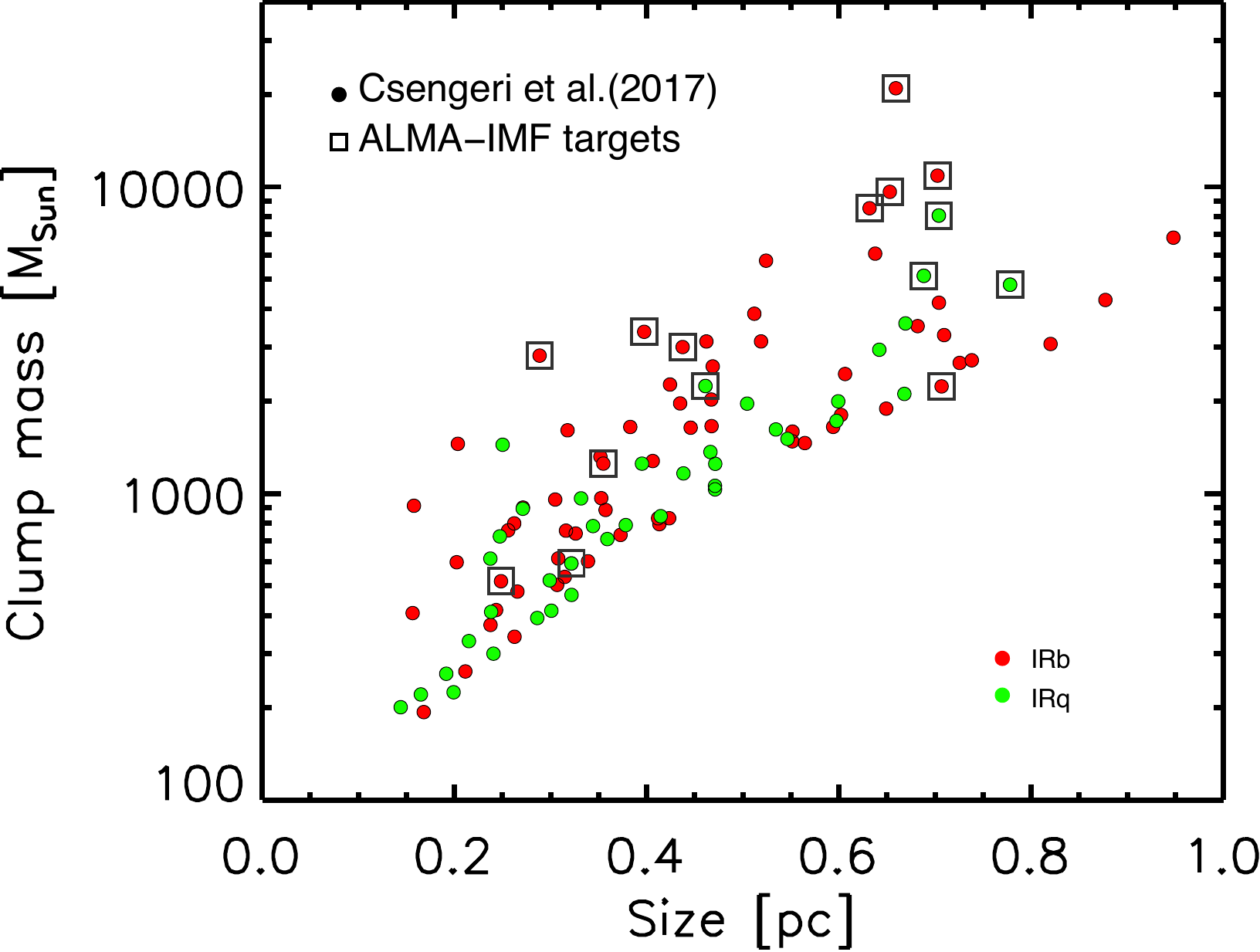}
    \caption{Mass versus size distribution of the ATLASGAL clumps selected for the ALMA-IMF survey (highlighted with squares; see \cref{fig:overview}), compared to the rest of the \citet{csengeri17b} sample at distances between 1~kpc and 5.5~kpc. ATLASGAL clumps targeted by ALMA-IMF are among the brightest and most concentrated. For more details on the selection criteria, see Sect.~\ref{s:target}.
    }
    \label{fig:sample}
\end{figure}

Figure~\ref{fig:sample} presents the basic characteristics of the ALMA-IMF clumps, taken from the catalog of \cite{csengeri17b}.
\cref{appendixtab:timea-table} lists the most extreme clumps from the \citet{csengeri17b} catalog. This list served as a first step in our source selection. In addition to the source position and association to molecular complexes and IRAS sources, we also list projects from the ALMA archive that cover these positions.

{\renewcommand{\arraystretch}{1.5}
\begin{table*}[ht]
\centering
\tiny
\begin{threeparttable}[c]
\caption{Massive ATLASGAL clumps that have $S^{\rm int}_{\rm 870\mu m}>25$~Jy fluxes and are located between 2~kpc and 5.5~kpc from the Sun.}
\label{appendixtab:timea-table}
\begin{tabular}{lcccccccc}
\hline\hline
ATLASGAL \& common\tnote{1} & $d$ & Ref.\tnote{2} & Evolutionary & $S^{\rm int}_{\rm 870\mu m}$\tnote{3} & ALMA & Molecular \\ 
names & [kpc] & $d$ & stage\tnote{3} & [Jy] & projects \tnote{4} & complex\tnote{5} \\ 
\hline\hline
              
G049.4888-0.3882  /    \underline{W51-E}    / W51 Main          & 5.4$\pm 0.3$  & (1) & IR-bright   & 113.2           
& (1), (8a,b), (9) & W51 \\ 
&&&&&(10), (11), (12) &&&\\

G333.6036-0.2130 / \underline{G333.60} / IRAS 16183-4958                                     & 4.2$\pm 0.7$   & (2)       & IR-bright     & 97.4         
& (1) & RCW106 \\
G012.8057-0.1994 / \underline{G012.80}           / W33-IRS3                                  & 2.4$\pm 0.2$   & (3)           & IR-bright   & {92.4}         
& (1), (5a) & W33 \\
G351.7747-0.5369 / \underline{G351.77}  /  IRAS 17233-3606                             & 2.0$\pm 0.7$   & (4)     & IR-bright   & 81.8          
& (1), (3), (13), (14) & G353 \\ 
G327.2921-0.5792 / \underline{G327.29}                / RCW97                           & 2.5$\pm 0.5$   & (2)       & IR-bright & 71.2       
& (1), (15) & G327 \\
G353.4102-0.3611 / \underline{G353.41}          / IRAS 17271-3439                       & 2.0$\pm 0.7$   & (4)       &  IR-bright  & 68.4       
& (1) , (3)& G353 \\ 

G010.6237-0.3833 / \underline{G010.62}       / W31                                     & 4.95$\pm 0.5$  & (5)       & IR-bright   & 54.9         
& (1), (5a) , (14b) & W31 \\ 
&&&&&(16), (17) &&&\\

G333.1341-0.4314 / IRAS 16172-5028                            & 4.2$\pm 0.7$   & (2)       & IR-bright & 54.2      
& (2), (3), (4) & RCW106 \\
G338.9249+0.5539 / \underline{G338.92}                                                 & 3.9$\pm 1.0$   & (2)       & IR-quiet     & 53.3
& (1), (2), (18) & \\ 
G049.4897-0.3697 / \underline{W51-IRS2} /       IRAS 19213+1424
                 & 5.4$\pm 0.3$  & (1)            & IR-bright   & 52.0       
& (1), (8a,b), (10) & W51 \\ 
G332.8262-0.5493 / IRAS 16164-5046                            & 4.2$\pm 0.7$   & (2)       & IR-bright & 51.4
& (2), (5a), (3) & RCW106 \\
G348.7260-1.0391 / IRAS 17167-3854                                      & 3.4$\pm 0.3$ & (6)        & IR-bright & 43.4
& (2) & RCW122\\ 
G030.8166-0.0561 / \underline{W43-MM1}                         & 5.5$\pm 0.4$  & (7)       & IR-quiet     & {42.3}
& (1), (19a,b) & W43 \\ 
&&&&& (20a,b), (21) &&&\\

G328.8087+0.6328  / IRAS 15520-5234            & 2.5$\pm 0.5$   & (2)       & IR-bright   & 40.6
& & G327\\ 
G333.2841-0.3868                                                       & 4.2$\pm 0.7$   & (2)       & IR-bright     & 37.5
& (2), (4) & RCW106 \\ 
G049.4908-0.3792 / W51-Main & 5.4$\pm 0.3$  & (1)            & IR-bright   & 36.9
& (1) in W51-E field & W51\\
G337.9154-0.4773 / \underline{G337.92}        / IRAS16274-4701                         & 2.7$\pm 0.7$   & (4)       & {IR-bright}     & 36.6         
& (1), (5a), (5b), (6),  & G337\\ 
G301.1365-0.2256 / IRAS 12326-6245                          & 4.2$\pm 0.7$   & (2)       & IR-bright & 34.5   
& (2), (5), (3) & \\
G327.3005-0.5509 /  IRAS 15492-5426                                                 & 2.5$\pm 0.5$   & (2)       & IR-bright   & 32.3         
& & G327\\ 
G337.4052-0.4024 / IRAS 16351-4722     & 2.7$\pm 0.7$   & (5)       & IR-bright     & 31.7
& (2), (5b), (3), (6) & G337 \\ 
G008.6702-0.3557  / \underline{G008.67}                                                 & 3.4$\pm 0.3$   & (2)       & IR-quiet     & 30.6
& (1), (5b), (22) & \\ 
G322.1581+0.6354 / RCW92                             & 3.2$\pm 1.1$   & (2)       & IR-bright     & 30.3          
&  & \\ 
G329.0303-0.2022   / IRAS 15566-5304                                                        & 2.5$\pm 0.5$   & (2)       & IR-quiet     & 29.4          
& (3) & G327 \\ 
G330.8788-0.3681 / IRAS 16065-5158                                              & 4.2$\pm 0.7$   & (2)       & IR-quiet & 28.1          
& & RCW106\\ 
G019.6084-0.2346 / IRAS 18248-1158                                              & 3.6$\pm 0.8$   & (2)       & IR-bright   & 27.6         
& (7) & \\ 
G326.6577+0.5941 / IRAS 15408-5356 / RCW95         & 2.5$\pm 0.5$   & (2)       & IR-bright & 26.2          
& & G327\\
G305.2083+0.2063 / IRAS 13079-6218                     & 3.5$\pm 2.0$   & (8)     & IR-bright     & 26.1           
& (2), (3), (4) &\\ 
G030.7016-0.0672  / \underline{W43-MM2}                          & 5.5$\pm 0.4$  & (7)       & IR-quiet     & 25.1          
& (1), (20c) & W43\\
\hline
G328.2551-0.5321 / \underline{G328.25}                                                & 2.5$\pm 0.5$   & (2)       & IR-quiet     & 
15.0          
& (1), (23) & G327 \\
G030.7173-0.0822 / \underline{W43-MM3}                         & 5.5$\pm 0.4$  & (7)       & IR-bright   & 11.7          
& (1), (2), (20c) & W43\\ 
\hline
\end{tabular}
\begin{tablenotes}
\item[1] ATLASGAL name (underlying their Galactic coordinates) from \cite{csengeri17b} together with adopted name for ALMA-IMF (underlined names) as well as other most common names.
\item[2] References for the distance to the Sun: (1) \cite{sato10}; (2) \cite{csengeri17b}; (3) \cite{immer13}; (4) This paper ; (5) \cite{sanna14}; (6) \cite{reid14}; (7) \cite{zhang14}; (8) \cite{russeil03}.
\item[3] Evolutionary stage and 870~$\mu$m integrated fluxes, taken from \cite{csengeri17b}.
\item[4] Observed as part of the: (1) ALMA-IMF Large Program \#2017.1.01355.L, by Motte, Ginsburg, Louvet, Sanhueza et al. (B6+B3 mosaics); (2) Large Program \#2019.1.00195.L (B6) by Molinari, Schilke et al.; (3) Program by Liu et al. \#2019.1.00685.S (B3); (4) Program by Barnes et al. \#2019.1.01031.S (B3, mosaic); (5) Programs by Leurini et al. a) \#2016.1.01347.S, b) \#2017.1.00377.S (B6+B3, mosaics); (6) Program by Hacar et al. \#2018.1.00697.S (B3, mosaic); (7) Program by Se-Hyung et al. \#2013.1.00266.S (B3); (8) Programs by Ginsburg et al. a) \#2013.1.00308.S (B6, mosaic) and b) \#2017.1.00293.S (B3); (9) Program by Kim et al. \#2015.1.01571.S (B6); (10) Program by Goddi et al. \#2015.1.01596.S (B6); (11) Program by Su et al. \#2016.1.00268.S (B3); (12) Program by Rivilla et al. \#2016.1.01071.S (B3); (13) Program by Beuther et al. \#2015.1.00496.S (B6); (14) Program by Sanhueza et al. a) \#2017.1.00237.S and b) \#2016.1.01036.S (B6); (15) Program by Schilke et al. \#2016.1.00168.S (B6, mosaic); (16) Program by Gerin et al. \#2013.1.01194.S (B3); (17) Program by Zhang et al. \#2015.1.00106.S (B6); (18) Program by Fuller et al. \#2015.1.01312.S (B6); (19) Program by Motte et al. a) \#2013.1.01365.S (B6, mosaic) and b) \#2015.1.01273.S (B6, mosaic); (20) Program by Louvet et al. a) \#2015.1.01020.S (B6), b) \#2018.1.01787.S (B3) and c) \#2017.1.00226; (21) Program by Kim et al. \#2018.1.01288.S (B6); (22) Program by Shirley et al. \#2017.1.01116.S (B3, B6); (23) Program by Csengeri et al. \#2019.2.00093.S (B3).
\item[5] Name of the molecular cloud complex hosting the ATLASGAL clump.
\end{tablenotes}
\end{threeparttable}
\end{table*}
}

\end{appendix}

\end{document}